\documentclass[conference, 10pt]{IEEEtran}
\IEEEoverridecommandlockouts

\usepackage{cite}
\usepackage{amsmath,amssymb,amsfonts}
\usepackage{algorithmic}
\usepackage{graphicx}
\usepackage{textcomp}
\usepackage{xcolor}
\def\BibTeX{{\rm B\kern-.05em{\sc i\kern-.025em b}\kern-.08em
    T\kern-.1667em\lower.7ex\hbox{E}\kern-.125emX}}

\usepackage{bm}
\usepackage{stmaryrd}
\usepackage{mathtools}
\usepackage{macros}
\usepackage{xspace}
\usepackage{caption}
\usepackage{subcaption}
\usepackage{tikz}
\usetikzlibrary{petri,calc,backgrounds,shapes,decorations,automata,patterns}
\usepackage{xcolor}
\usepackage{pgfplots}
\pgfdeclarelayer{bg}    
\pgfsetlayers{bg,main}
\usetikzlibrary{decorations.pathreplacing,patterns,petri,calc}
\usetikzlibrary{backgrounds}
\usepackage{array}
\usepackage{balance}
\usepackage{booktabs}
\usepackage{enumitem}
\usepackage{cleveref}
\usepackage{thm-restate}
\usepackage{nicefrac}
\pgfdeclarelayer{bg}
\pgfsetlayers{bg,main}

\begin{document}

\title{Verifying linear temporal specifications of constant-rate
  multi-mode systems
\thanks{M.~Blondin was
    supported by a Discovery Grant from the Natural Sciences and
    Engineering Research Council of Canada (NSERC), and by the Fonds
    de recherche du Qu\'{e}bec -- Nature et technologies
    (FRQNT). P.~Offtermatt is now at Informal Systems, Munich, Germany.}
}

\author{\IEEEauthorblockN{Michael Blondin}
\IEEEauthorblockA{\textit{Universit\'{e} de Sherbrooke}\\
  Sherbrooke, Canada \\
  michael.blondin@usherbrooke.ca}
\and
\IEEEauthorblockN{Philip Offtermatt}
\IEEEauthorblockA{\textit{Universit\'{e} de Sherbrooke}\\
  Sherbrooke, Canada \\
  \textit{University of Warsaw} \\
  Warsaw, Poland \\
  philip.offtermatt@usherbrooke.ca}
\and
\IEEEauthorblockN{Alex Sansfa\c{c}on-Buchanan}
\IEEEauthorblockA{\textit{Universit\'{e} de Sherbrooke}\\
  Sherbrooke, Canada \\
  alex.sansfacon-buchanan@usherbrooke.ca}
}

\maketitle

\begin{abstract}
  Constant-rate multi-mode systems (MMS) are hybrid systems with
  finitely many modes and real-valued variables that evolve over
  continuous time according to mode-specific constant rates. We
  introduce a variant of linear temporal logic (LTL) for MMS, and we
  investigate the complexity of the model-checking problem for
  syntactic fragments of LTL. We obtain a complexity landscape where
  each fragment is either P-complete, NP-complete or
  undecidable. These results generalize and unify several results on
  MMS and continuous counter systems.
\end{abstract}

\section{Introduction}
\label{sec:introduction}
Constant-rate multi-mode systems (MMS) are hybrid systems with
finitely many modes and a finite number of real-valued variables that
evolve over continuous time according to mode-specific constant rates.

MMS were originally introduced by Alur \etal\ to model, \eg, problems
related to green scheduling and reducing energy peak consumption of
systems~\cite{alur2012mms}. There, they consider the problems of safe
schedulability and safe reachability with respect to zones defined as
bounded convex polytopes.

\emph{Safe schedulability} asks whether a given MMS admits a
non-Zeno\footnote{Informally, this means that there cannot be
  infinitely many mode switches within a finite amount of time.}
infinite execution that remains within a given safety zone. \emph{Safe
  reachability} asks whether a given MMS has a finite execution that
reaches a target point, while staying within a given safety zone along
the way. Both problems were shown to be solvable in polynomial
time~\cite{alur2012mms}.

A similar problem was studied by Krishna \etal\ in the context of
motion planning~\cite{reachavoid2017krishna}. There, the authors are
interested in the \emph{reach-avoid problem}. In the latter, the goal
is to reach a given target point without ever entering any of the
given obstacles. The authors of~\cite{reachavoid2017krishna} consider
obstacles specified as convex polytopes. They show that the
reach-avoid problem is decidable if the obstacles are closed, and is
undecidable in general. They further provide an implementation of
their procedure which is benchmarked positively against the \emph{Open
  Motion Planning Library}.

\paragraph*{Contribution}

The aforementioned problems were solved with ad hoc
approaches. Moreover, many natural problems cannot be expressed in
these existing frameworks. One such problem is \emph{safe repeated
  reachability}, where the goal is to find a non-Zeno infinite
execution that remains within a safety zone and visits a finite set of
zones infinitely often.

We propose a framework that encompasses all of the above. More
precisely, we introduce a linear temporal logic (LTL) for MMS. Our
variant uses bounded convex polytopes as atomic propositions. We omit
the next operator $\X$ which is ill-suited for the continuous behavior
of MMS. Moreover, we use a strict-future interpretation of the until
temporal operator $\U$, inspired from metric temporal
logic~\cite{koymans1990specifying} (more precisely from
$\text{MITL}_{0, \infty})$. In particular, our logic can express
\begin{itemize}
\item Safe schedulability: $\G\, Z_\text{safe}$;

\item Safe reachability: $Z_\text{safe} \U \{\vec{x}_\text{target}\}$;

\item Reach-avoid: $(\neg O_1 \land \cdots \land \neg O_n) \U
  \{\vec{x}_\text{target}\}$; and

\item Safe repeated reachability: $(\G\, Z_\text{safe}) \land
  \bigwedge_{i=1}^n (\G\F\, Z_i)$.
\end{itemize}

We investigate the computational complexity of LTL model checking,
which asks, given an MMS $\M$, a starting point $\vec{x}$ and an LTL
formula $\varphi$, whether there is a non-Zeno infinite execution of
$\M$ that satisfies $\varphi$ from $\vec{x}$, denoted $\vec{x}
\models_\M \varphi$. We consider the syntactic fragments obtained by
(dis)allowing operators from $\{\U, \F, \G, \land, \lor, \neg\}$ and
allowing at least one temporal operator\footnote{Without any temporal
  operator, the logic has nothing to do with MMS; it becomes
  quantifier-free linear arithmetic.}. We establish the computational
complexity of \emph{all} of the $2^6 - 2^3 = 56$ fragments: Each one
is either P-complete, NP-complete or undecidable.

Our work is also closely related to the study of counter systems like
vector addition systems (VAS) and Petri nets. These models have
countless applications ranging from program verification and
synthesis, to the formal analysis of chemical, biological and business
processes (\eg, see~\cite{GS92,EGLM17,HGD08,Aal98}). Moreover, the
continuous relaxation of counter systems has been successfully
employed in practice to alleviate their tremendous computational
complexity (\eg, see~\cite{ELMMN14,BHO21}).

The behavior of an MMS amounts to continuous pseudo-reachability of
VAS and Petri nets, \ie\ where the effect of transitions can be scaled
by positive real values, and without the requirement that counters
must remain non-negative. The latter requirement can be regained in
our logic. While we do not investigate unbounded zones in their full
generality, we consider semi-bounded linear formulas, which include
formulas of the form $(\G\, Z) \land \cdots$ or $Z \U \cdots$, where
$Z$ is unbounded, and so can be set to $Z \defeq \Rnon^d$. In
particular, our results imply the known fact that continuous
reachability, \ie\ checking $\Rnon^d \U \{\vec{x}_\text{target}\}$,
can be done in polynomial time~\cite{FH15}. Moreover, we establish the
decidability of richer properties. Thus, our work can be seen as a
unifying and more general framework for MMS and continuous VAS/Petri nets.

\paragraph*{Results}

\begin{figure}[h]
  \begin{center}
    \begin{tikzpicture}[auto, semithick, transform shape, scale=0.85]  
  \colorlet{colP}{colA}
  \colorlet{colNP}{colB}
  \colorlet{colUndec}{colC}
  
  \fill[pattern color=colP!75, postaction={pattern=north east lines},
    colP, fill opacity=0.4] (-3.75, -1.5) rectangle (5.30, 1.35);
  
  \fill[pattern color=colNP!85!black, postaction={pattern=crosshatch dots},
    colNP, fill opacity=0.4] (-3.75, 3.60) rectangle (5.30, 1.35);

  \fill[pattern color=colUndec!50, postaction={pattern=crosshatch},
    colUndec, fill opacity=0.4] (-3.75, 7.85) rectangle (5.30, 3.60);
  
  \node[rotate=90, black!70!colP]     at (5.05, -0.1) {P-complete};
  \node[rotate=90, black!70!colNP]    at (5.05, 2.45) {NP-complete};
  \node[rotate=90, black!70!colUndec] at (5.05, 5.65) {Undecidable};

  \node (FG) {$\{\F, \G\}$};

  \node[below  left=10pt and 0cm of FG] (F) {$\{\F\}$};
  \node[below right=10pt and 0cm of FG] (G) {$\{\G\}$};

  \node[above=10pt of FG, xshift=2cm, yshift=-20pt, font=\small] (FGn) {
    \begin{tabular}{l}
      $\{\F, \neg\} \equiv {}$ \\
      $\{\F, \G, \neg\} \equiv {}$ \\
      $\{\G, \neg\}$
    \end{tabular}
  }; 
  \node[above=-15pt of FGn, xshift=10pt] (FGnphantom) {};
      
  \node[left=2.5cm of FGn] (Fo) {$\{\F, \lor\}$};
  \node[right=10pt of FGn] (Ga) {$\{\G, \land\}$};
  
  \node[above=2.5cm of F]  (Fa)  {$\{\F, \land\}$};
  \node[above=10pt of Fa]  (FGa) {$\{\F, \G, \land\}$};
  \node[left =5pt  of FGa] (Fao) {$\{\F, \land, \lor\}$};
  
  \node[above=2cm of FGa] (FGao) {$\{\F, \G, \land, \lor\}$};
    
  \node[above=1.25cm of FGao] (all) {$\{\U, \F, \G, \land, \lor, \neg\}$};
  
  \node[above=15pt of FGa, xshift=2.75cm] (Go) {$\{\G, \lor\}$};
  \node[above=5pt  of Go,  xshift=-1cm]  (FGo) {$\{\F, \G, \lor\}$};
  \node[above=5pt  of Go,  xshift=1cm]   (Gao) {$\{\G, \land, \lor\}$};

  \node[left=3.5cm of Go] (U) {$\{\U, \ldots\}$};
  
  \node[above=1cm of Gao, xshift=-0.5cm, font=\tiny] (FGaon) {
    \begin{tabular}{l}
      $\mathmakebox[38pt][r]{\{\F, \lor, \neg\} \equiv {}}
      \mathmakebox[26pt][l]{\{\F, \land, \neg\}} \equiv
      \{\F, \land, \lor, \neg\} \equiv {}$ \\
      
      $\mathmakebox[38pt][r]{\{\G, \lor, \neg\} \equiv {}}
      \mathmakebox[26pt][l]{\{\G, \land, \neg\}} \equiv
      \{\G, \land, \lor, \neg\} \equiv {}$ \\
      
      $\mathmakebox[38pt][r]{\{\F, \G, \lor, \neg\} \equiv {}} \{\F, \G,
      \land, \neg\} \equiv \{\F, \G, \land, \lor, \neg\}$
    \end{tabular}
  };
  
  \path[->]
  (FG) edge[out=45, in=180] node {} (FGn)
  (F)  edge node {} (FG)
  (G)  edge node {} (FG)
  (F)  edge node {} (Fo)
  (G)  edge[out=0, in=-145] node {} (Ga)
  
  (F)  edge node {} (Fa)
  (Fo) edge[bend left=10] node {} (Fao)
  (Ga) edge[out=90, in=0] node {} (FGa)
  
  (Fo) edge[out=15, in=-100, looseness=1] node {} (FGo)
  (G)  edge[out=0, in=0,  looseness=1]  node {} (Go)
  (FG) edge[out=90, in=-60, looseness=0.5] node {} (FGo)  
  (FGnphantom) edge[out=45, in=-25, looseness=0.5] node {} (FGaon)
  (Ga) edge[out=90, in=-35] node {} (Gao)
  
  (Fa) edge node {} (FGa)
  (Fa) edge node {} (Fao)
  (FG) edge[out=110, in=-35] node {} (FGa)
  
  (Fao) edge[bend left=10] node {} (FGao)
  (FGa) edge node {} (FGao)
  
  (Go)  edge node {} (FGo)
  (Go)  edge node {} (Gao)
  (FGo) edge node {} (FGao)
  (Gao) edge[bend right=10] node {} (FGao)
  
  (U)    edge[bend left=20] node {} (all)
  
  (FGao)  edge[bend right=10] node {} (FGaon)
  (FGaon) edge[out=150, in=0] node {} (all)
  ;

  \node[below=-5pt of F, font=\footnotesize, black!60!colP] {
    Thm.~\ref{thm:f:g:Phard}
  };
  
  \node[below=-5pt of G, font=\footnotesize, black!60!colP] {
    Thm.~\ref{thm:f:g:Phard}
  };

  \node[below=-5pt of Fo, xshift=-12pt, font=\footnotesize, black!60!colP] {
    Thm.~\ref{thm:fgn:for:P}
  };

  \node[below=-5pt of Ga, xshift=2pt, font=\footnotesize, black!60!colP] {
    Thm.~\ref{thm:fgn:for:P}
  };

  \node[below=-5pt of FGn, xshift=-5pt, font=\footnotesize, black!50!colP] {
    Thm.~\ref{thm:fgn:for:P}
  };

  \node[below=-5pt of Fa, xshift=-18pt, font=\footnotesize, black!50!colNP] {
    Thm.~\ref{thm:fa:NPhard}
  };

  \node[above=-5pt of FGa, xshift=18pt, font=\footnotesize, black!50!colNP] {
    Thm.~\ref{thm:fga:NP}
  };

  \node[above=-5pt of Fao, xshift=-12pt, font=\footnotesize, black!50!colNP] {
    Thm.~\ref{thm:fao:NP}
  };

  \node[below=-5pt of U, xshift=-2pt, font=\footnotesize, black!50!colUndec] {
    Thm.~\ref{thm:u:go:undec}
  };

  \node[below=-5pt of Go, font=\footnotesize, black!50!colUndec] {
    Thm.~\ref{thm:u:go:undec}
  };
\end{tikzpicture}\hspace*{-2pt}
  \end{center}

  \caption{Complexity landscape of LTL model checking for MMS. An edge
    from $X$ to $Y$ indicates that any formula from \fragmentBx{X} is
    equivalent to some formula from \fragmentBx{Y}. Each expression
    ``$X \equiv Y$'' stands for $X \leftrightarrow Y$, \ie, an edge
    from $X$ to $Y$ and an edge from $Y$ to $X$. Node $\{\U, \ldots\}$
    stands for any LTL fragment that contains
    $\U$.}\label{fig:complexity}
\end{figure}
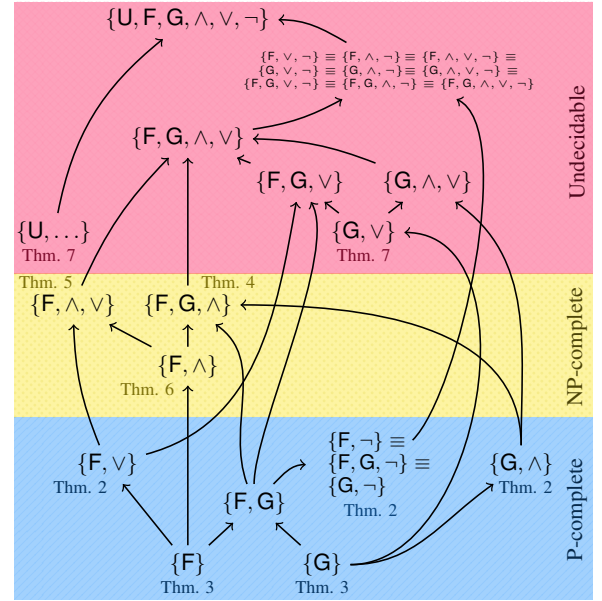

Let us write \fragmentBx{X} to denote the set of LTL formulas using
only operators from $X$, and \fragmentx{X} for the same fragment but
with zones possibly unbounded. We obtain the full complexity landscape
depicted in \Cref{fig:complexity}. Our contribution is summarized by
the following three points.

\paragraph*{I)}

We show that \fragmentB{\F, \G, \land} is in NP, and hence
that \fragmentB{\F, \land, \lor} is as well. More precisely, we prove
that:
\begin{enumerate}
\item Formulas from this fragment can be put in a normal form, coined
  as \emph{flat formulas}, where the nesting of temporal operators is
  restricted;

\item Flat formulas can be translated into generalized
  B\"uchi automata with transition-based acceptance, no cycles except
  for self-loops (``almost acyclic'') and linear width;

\item Testing whether an MMS $\M$ satisfies a specification given by
  such an automaton $\A$ can be done in NP by guessing a so-called
  linear path scheme $S$ of $\A$; constructing a so-called linear
  formula $\psi$ equivalent to $S$, and testing whether
  $\vec{x} \models_\M \psi$ in polynomial time.
\end{enumerate}

Step~2 is inspired by the work of K\v{r}et\'{i}nsk\'{y} and
Esparza~\cite{KE12} on deterministic Muller automata for classical LTL
restricted to $\{\F, \G, \land, \lor, \neg\}$. Our construction also
deals with classical LTL, restricted to $\{\F, \G, \land\}$, and is
thus an indirect contribution to logic and automata independent of
MMS.

In particular, in Step~3 we establish a polynomial-time LTL fragment for
MMS, namely \emph{semi-bounded linear LTL formulas}. We do so by using a
polynomial-time fragment of existential linear arithmetic, introduced
in~\cite{blondin2017logics} for the purpose of characterizing
reachability sets of continuous Petri nets. In particular, we show how
to translate LTL formulas of the form $\psi = (\G Z_0) \land
\bigwedge_{i=1}^n \G \F Z_i$, with $Z_0$ unbounded, into the logic
of~\cite{blondin2017logics}. This is challenging, in contrast to
simply handling $\G Z_0$, with $Z_0$ bounded, as done
in~\cite{alur2012mms}. It involves a technical characterization of MMS
and points that satisfy $\psi$, which, in particular, goes through a
careful use of Farkas' lemma.

As a corollary of Step~3, we show that \fragmentB{\F, \G, \neg},
\fragment{\F, \lor} and \fragment{\G, \land} are solvable in
polynomial time. These fragments include safe schedulability and safe
reachability, which generalizes their membership in
P~\cite{alur2012mms}.

\paragraph*{II)}

We show the NP-hardness of \fragmentB{\F, \land} by reducing from
SUBSET-SUM. With the previous results, this shows that
\fragmentB{\F, \G, \land} and \fragmentB{\F, \land, \lor} are NP-complete.

\paragraph*{III)}

We show that \fragmentB{\U} and \fragmentB{\G, \lor} are both
undecidable, by reducing from the reachability problem for Petri nets
with inhibitor arcs. This ``generalizes'' the
undecidability of the reach-avoid problem established
in~\cite{reachavoid2017krishna}. Their proof indirectly shows that the
model checking problem is undecidable for formulas of the form
$(Z_1 \lor \cdots \lor Z_n) \U \{\vec{x}_\text{target}\}$ where each
$Z_i$ is a possibly unbounded zone. We strengthen this result by using
bounded zones only.

\paragraph*{Further related work}

MMS are related to hybrid
automata~\cite{henzinger2000theory}. Contrary to MMS, however, the
latter allow for a finite control structure, and modes of non-constant
rates. Their immense modelling power leads to the undecidability of
most problems, including reachability, \ie\ formulas of the form $\F\,
\{\vec{x}_\text{target}\}$. Yet, some researchers have investigated
decision procedures for temporal specification languages such as
signal temporal logic (\eg, see~\cite{BL19}).

Timed automata~\cite{alur1999timed} form another related type of
hybrid system. In this model, all variables (known as clocks)
increase at the \emph{same} constant rate, as opposed to the case of
MMS. On the other hand, timed automata are equipped with a finite
control structure, which is not the case of MMS.

Bounded-rate multi-mode systems generalize MMS~\cite{bhave2015bounded,
  alur2017bounded}. In this model, the mode-dependent rates are given as
bounded convex polytopes. The setting can be seen as a two-player
game. Player~1 chooses a mode and a duration, and Player~2 chooses the
rates from the set for that mode. The system evolves according to the
rates chosen by Player~2. In this context, ``schedulability'' is a
strategy for Player~1 that never leaves the safety zone, no matter the
choices of Player~2.

Small fragments of classical LTL have been investigated in the
literature, \eg\ see~\cite[Table~1]{DS02}. In
particular, \fragment{\F, \G, \allowbreak \land, \allowbreak \lor, \allowbreak \neg}
has been studied in~\cite{SC85,KE12} under the names $\mathrm{L}(\F)$
and $(\F, \G)$, and \fragment{\F, \land} has been studied
in~\cite{AT04} under the name $\textsc{LTL}_{+}(\Diamond, \land)$. The
authors of~\cite{AT04} show that a fragment, called
$\textsc{LTL}^\text{PODB}$, and which is incomparable
to \fragment{\F, \G, \land}, admits partially-ordered deterministic
B\"{u}chi automata of exponential size and linear width. To the best
of our knowledge, there is no work dedicated
to \fragment{\F, \G, \land}, and in particular to its translation into
automata of linear width.

\paragraph*{Organization}

In \Cref{sec:preliminaries}, we introduce basic definitions, MMS and
LTL. We further relate LTL over MMS with classical LTL over infinite
words. In \Cref{sec:buechi}, we show that any formula from
classical \fragment{\F, \G, \land} translates into a specific type of
$\omega$-automaton, which amounts to a disjunction of so-called linear
LTL formulas. In \Cref{sec:ptime}, we show that linear LTL formulas
over MMS can be model-checked in polynomial time. From this, we
establish the P-completeness of some syntactic
fragments. In \Cref{sec:np} and \Cref{sec:undecidable}, we
respectively prove the NP-completeness and undecidability of the other
fragments. We conclude in \Cref{sec:conlusion}. Due to space
limitation, many proofs are deferred to the full version which is
freely available on arXiv.

\section{Preliminaries}
\label{sec:preliminaries}
We write $\N$ to denote $\{0, 1, \ldots\}$, $\Z$ to denote the integers, and $\R$ to denote the
reals. We use subscripts to restrict these sets, \eg\ $\Rpos \defeq \{x
\in \R : x > 0\}$. We write $[\alpha, \beta] \defeq \{x \in \R :
\alpha \leq x \leq \beta\}$ and $[a..b] \defeq \{i \in \N : a \leq i
\leq b\}$. We also use (semi-)open intervals, \eg\ $(1, 2] = [1, 2]
  \setminus \{1\}$.

Let $I$ be a set of indices and let $X \subseteq \R^I$. We write
$\vec{e}_i \in \R^I$ for the vector with $\vec{e}_i(i) = 1$ and
$\vec{e}_i(j) = 0$ for all $j \neq i$, and $\vec{0}$ to denote the
vector such that $\vec{0}(i) = 0$ for all $i \in I$. Let
$\norm{\vec{x}} \defeq \max\{|\vec{x}(i)| : i \in I\}$ and $\norm{X}
\defeq \sup\{\norm{\vec{x}} : \vec{x} \in X\}$. We say that $X$ is
\emph{convex} if $\lambda \vec{x} + (1 - \lambda) \vec{y} \in X$ for
all $\lambda \in [0, 1]$ and $\vec{x}, \vec{y} \in X$, and
\emph{bounded} if $\norm{X} \leq b$ for some $b \in \Rnon$.

We write $2^\Sigma$ for the powerset of $\Sigma$. Given a nonempty
finite sequence $w$, let $w^\omega \defeq ww
\cdots$. Let $\Sigma^\omega \defeq \{w_0 w_1 \cdots : w_i \in
\Sigma\}$ be the set of infinite sequences with elements from
$\Sigma$.

\subsection{Constant-rate multi-mode systems}

A \emph{$d$-dimensional constant-rate multi-mode system}
\emph{(MMS)}, with $d \in \N_{\geq 1}$, is a finite set $\M \subseteq
\R^d$ whose elements are called \emph{modes}. A \emph{schedule} is a
(finite or infinite) sequence $\pi = (\alpha_1, \vec{m}_1) (\alpha_2,
\vec{m}_2) \cdots$, where each $(\alpha_i, \vec{m}_i) \in \Rpos \times
\M$. To ease the notation, we often write, \eg, $\vec{m}\, \frac{1}{2}
\vec{m}'$ rather than $(1, \vec{m}) (1/2, \vec{m}')$. Given $\lambda
\in \Rpos$, we define the schedule $\lambda \pi$ as $\pi$ with each
$\alpha_i$ replaced by $\lambda \alpha_i$. The \emph{size} of $\pi$,
denoted $\abs{\pi}$, is its number of pairs. The \emph{effect} of
$\pi$ is $\effect{\pi} \defeq \sum_{i} \alpha_i \vec{m}_i$. The
\emph{support} of $\pi$ is $\supp{\pi} \defeq \{\vec{m}_1, \vec{m}_2,
\ldots\}$. Let $\weight{\pi}{\vec{m}} \defeq \sum_{i : \vec{m}_i =
  \vec{m}} \alpha_i$ and $\tweight{\pi} \defeq \sum_{\vec{m} \in \M}
\weight{\pi}{\vec{m}}$. We say that an infinite schedule $\pi$ is
\emph{non-Zeno} if $\tweight{\pi} = \infty$. The \emph{Parikh image}
of a finite schedule $\pi$ is denoted $\parikh{\pi} \in \Rnon^\M$,
\ie\ $\parikh{\pi}(\vec{m}) \defeq \weight{\pi}{\vec{m}}$. We say that
two finite schedules are \emph{equivalent}, denoted with $\equiv$, if
they are equal after merging consecutive equal modes, \ie\ using the
rule $\pi (\alpha, \vec{m}) (\beta, \vec{m}) \pi' \equiv \pi (\alpha +
\beta, \vec{m}) \pi'$. Let $\FromTo{\pi}{\tau}{\tau'}$ be the
schedule obtained from $\pi$ starting where time $\tau$ has elapsed,
and ending where time $\tau'$ has elapsed; \eg, for $\pi = (2,
\vec{m}_1)(0.5, \vec{m}_2)(1, \vec{m}_3)^\omega$, we have
$\FromTo{\pi}{0}{1} = (1, \vec{m}_1)$, $\FromTo{\pi}{0.5}{2.25} =
(1.5, \vec{m}_1)(0.25, \vec{m}_2)$ and $\FromTo{\pi}{3}{} = (0.5,
\vec{m}_3) (1, \vec{m}_3)^\omega$.

An \emph{execution} is a (finite or infinite) sequence $\sigma =
\vec{x}_0 I_0 \vec{x}_1 \allowbreak I_1 \allowbreak \vec{x}_2 \cdots$
where $\vec{x}_0, \vec{x}_1, \ldots \in \R^d$, $I_0, I_1, \ldots
\subseteq \Rnon$ are closed intervals with distinct endpoints, $\min
I_0 = 0$, and $\min I_j = \max I_{j-1}$ for all $j \in \N_{> 0}$. Let
$\dom \sigma \defeq I_0 \cup I_1 \cup \cdots$. For every $\tau \in
\dom \sigma$, with $\tau \in I_j$, let
\[
\sigma(\tau) \defeq
\vec{x}_j +
\frac{\tau - \min I_j}{\max I_j - \min I_j} \cdot (\vec{x}_{j+1} - \vec{x}_j).
\]
We define $\FromTo{\sigma}{\tau}{\tau'}$, where $[\tau, \tau']
\subseteq \dom \sigma$, as the execution $\sigma'$ that satisfies
$\sigma'(\alpha) = \sigma(\tau + \alpha)$ for every $\alpha \in [0,
  \tau' - \tau]$.

A schedule $\pi = (\alpha_1, \vec{m}_1)(\alpha_2, \vec{m}_2) \cdots$,
together with a point $\vec{x}_0$, gives rise to an execution
$\induced{\pi}{\vec{x}_0} \defeq \vec{x}_0 I_0 \vec{x}_1 \cdots$ where
$I_0 \defeq [0, \alpha_1]$, $I_j \defeq [\max I_{j-1}, \max I_{j-1} +
  \alpha_{j+1}]$ and $\vec{x}_j \defeq \vec{x}_{j-1} + \alpha_j
\vec{m}_j$. We use the notation $\vec{x} \trans{\pi} \vec{y}$ to
denote the fact that $\pi$ is a schedule that, from $\vec{x}$, gives
rise to an execution ending in $\vec{y}$. If we only care about the
existence of such a schedule, we may write $\vec{x} \trans{*}
\vec{y}$, or write $\vec{x} \trans{+} \vec{y}$ to denote that there is
such a nonempty schedule. We sometimes omit either of the two
endpoints if its value is irrelevant, \eg\ $\vec{x} \trans{\pi} {}$
stands for $\vec{x} \trans{\pi} \vec{x} + \effect{\pi}$. Given a set
$Z \subseteq \R^d$, we write $\vec{x} \mreach{\pi}{Z}$ to denote that
the execution never leaves $Z$, \ie\ $\sigma(\tau) \in Z$ for all
$\tau \in \dom \sigma$, where
$\sigma \defeq \induced{\pi}{\vec{x}}$. We extend this notation to any set of
sets $\mathcal{X}$, requiring that, for all $\tau \in \dom \sigma$,
there exists $Z \in \mathcal{X}$ such that $\sigma(\tau)
\in Z$.

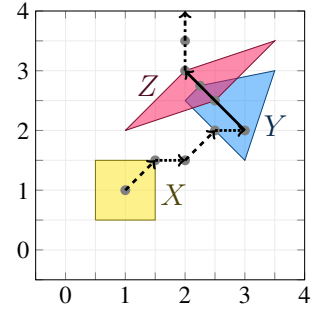
\begin{figure}
  \begin{center}
    \begin{tikzpicture}[auto, scale=0.66, transform shape]
  \tikzstyle{every node} = [font=\Large]
  
  \begin{axis}[
      xmin = -0.5, ymin = -0.5,
      xmax = 4, ymax = 4, 
      xlabel = {}, ylabel = {},
      xtick distance=0.5, ytick distance=0.5,
      xticklabels = {,, 0,, 1,, 2,, 3,, 4},
      yticklabels = {,, 0,, 1,, 2,, 3,, 4},
      width = 7cm, height = 7cm,
      grid = both,
      grid style = {line width = 0.1pt, draw = gray!15},
    ]
    
    \addplot[colB!50!black, fill=colB, fill opacity=0.5]
    coordinates {
      (0.5, 0.5) (0.5, 1.5) (1.5, 1.5) (1.5, 0.5) (0.5, 0.5)
    }
    node[
      above, xshift=45pt, yshift=5pt,
      colB!30!black, fill opacity=1.0, font=\LARGE
    ] {$X$};
    
    \addplot[colA!50!black, fill=colA, fill opacity=0.5]
    coordinates {
      (2, 2.5) (2.25, 2.75) (3.5, 3) (3, 1.5) (2, 2.5)
    }
    node[
      below, xshift=52pt, yshift=-5pt,
      colA!30!black, fill opacity=1.0, font=\LARGE
    ] {$Y$};
    
    \addplot[colC!50!black, fill=colC, fill opacity=0.5]
    coordinates {
      (1, 2) (2.5, 2.5) (3.5, 3.5) (2, 3) (1, 2)
    }
    node[
      below, xshift=13pt, yshift=35pt,
      colC!30!black, fill opacity=1.0, font=\LARGE
    ] {$Z$};          
    
    \tikzstyle{cmark} = [circle, fill, black!50, inner sep=2pt];
    
    \node[cmark] at (axis cs:1, 1)       {};
    \node[cmark] at (axis cs:1.5, 1.5)   {};
    \node[cmark] at (axis cs:2, 1.5)     {};
    \node[cmark] at (axis cs:2.5, 2)     {};
    \node[cmark] at (axis cs:3, 2)       {};
    \node[cmark] at (axis cs:2.5, 2.5)   {};
    \node[cmark] at (axis cs:2.25, 2.75) {};
    \node[cmark] at (axis cs:2, 3)       {};
    \node[cmark] at (axis cs:2, 3.5)     {};
    
    \addplot[ultra thick, ->, dashed] coordinates {
      (1, 1)
      (1.5, 1.5)
    }
    node[] {};
    
    \addplot[ultra thick, ->, densely dotted] coordinates {
      (1.5, 1.5)
      (2, 1.5)
    } node[] {};
    
    \addplot[ultra thick, ->, dashed] coordinates {
      (2, 1.5)
      (2.5, 2)
    } node[] {};
    
    \addplot[ultra thick, ->, densely dotted] coordinates {
      (2.5, 2)
      (3, 2)
    } node[] {};
    
    \addplot[ultra thick, ->] coordinates {
      (3, 2)
      (2, 3)
    } node[] {};
    
    \addplot[ultra thick, ->, dashdotted] coordinates {
      (2, 3)
      (2, 4)
    } node[] {};
  \end{axis}
\end{tikzpicture}
  \end{center}
  \caption{An execution $\sigma$ is depicted as a directed
    path. Three bounded zones $X, Y, Z$ are depicted as filled colored
    polygons. A trace is obtained from $\sigma$ from the marked
    points.}\label{fig:exec}
\end{figure}

\begin{example}\label{ex:mms}
  Let $\M \defeq \{(0, 1), (1, 0), (1, 1), (-1, 1)\}$. Let $\pi \defeq
  \frac{1}{2}(1, 1)\ \frac{1}{2}(1, 0)\ \frac{1}{2}(1,
  1)\ \frac{1}{2}(1, 0)\ (-1, 1)\ (0, 1)\ \cdots$ be a schedule. The
  execution $\sigma \defeq \induced{\pi}{(1, 1)}$ is depicted in
  \Cref{fig:exec} as a directed path, with distinct styles to
  distinguish the modes (ignore the circular marks and colored
  polygons):
  \begin{multline*}
    (1, 1)\,      \textcolor{black!60}{[0, 0.5]}\,
    (1.5, 1.5)\,  \textcolor{black!60}{[0.5, 1]}\,
    (2, 1.5)\,    \textcolor{black!60}{[1, 1.5]}\,
    (2.5, 2)\, \\ \textcolor{black!60}{[1.5, 2]}\,
    (3, 2)\,      \textcolor{black!60}{[2, 3]}\,
    (2, 3)\,      \textcolor{black!60}{[3, 4]}\,
    (2, 4)\,      \textcolor{black!60}{\cdots}. \tag*{\qed}
  \end{multline*}
\end{example}

\subsection{A linear temporal logic for MMS}\label{ssec:ltl:mms}

A \emph{zone} $Z \subseteq \R^d$ is a convex polytope represented as
the intersection of finitely many closed half-spaces, \ie\ $Z
= \{\vec{x} \in \R^d : \mat{A}\vec{x} \leq \vec{b}\}$ for some
$\mat{A} \in \Z^{k \times d}$ and $\vec{b} \in \Z^k$.

\emph{Linear temporal logic (LTL)}, over a finite set of zones $\AP$,
has the following syntax:
\[
\varphi ::=
\true \mid Z \mid \neg \varphi \mid \varphi \land \varphi \mid \varphi
\lor \varphi \mid \F \varphi \mid \G \varphi \mid \varphi \U \varphi,
\]
where $Z \in \AP$. Let \fragmentx{X} denote the set of LTL
formulas (syntactically) using only the operators from $X$. For
example, \fragment{\F, \G, \land}\ describes the LTL formulas using
only operators $\F$, $\G$ and $\land$. We write $\text{LTL}_\text{B}$
to indicate that all zones from $\AP$ must be bounded. We say that an
LTL formula is \emph{negation-free} if it contains no occurrence of
$\neg$.

We define the semantics over infinite executions:
\begin{alignat*}{3}
  \sigma, \tau &\models \true
  &&\iff \true, \\
  \sigma, \tau &\models Z
  &&\iff \sigma(\tau) \in Z, \\
  \sigma, \tau &\models \neg \varphi
  &&\iff \neg(\sigma, \tau \models \varphi), \\
  \sigma, \tau &\models \varphi \land \varphi'
  &&\iff (\sigma, \tau \models \varphi) \land
         (\sigma, \tau \models \varphi'), \\
  \sigma, \tau &\models \varphi \lor \varphi'
  &&\iff (\sigma, \tau \models \varphi) \lor
         (\sigma, \tau \models \varphi'), \\
  \sigma, \tau &\models \F \varphi
  &&\iff \exists \tau' \geq \tau : \sigma, \tau' \models \varphi, \\
  \sigma, \tau &\models \G \varphi
  &&\iff \forall \tau' \geq \tau : \sigma, \tau' \models \varphi, \\
  \sigma, \tau &\models \varphi \U \varphi'
  &&\iff \exists \tau' \geq \tau : (\sigma, \tau' \models \varphi') \\
  &&&\hspace{1.5cm} \land (\forall \tau'' \in [\tau, \tau') :
    \sigma, \tau'' \models \varphi).
\end{alignat*}

We write $\sigma \models \varphi$ iff $\sigma, 0 \models \varphi$. We
say that two formulas are \emph{equivalent}, denoted $\varphi \equiv
\varphi'$, if they are satisfied by the same executions. In
particular, $\F \psi \equiv \true \U \psi$ and $\G \psi \equiv \neg \F
\neg \psi$.

Let $\M$ be an MMS and let $\vec{x} \in \R^d$. We say that $\vec{x}
\models_\M \varphi$ iff $\M$ has a non-Zeno infinite schedule $\pi$
such that $\induced{\pi}{\vec{x}} \models \varphi$. The
\emph{model-checking problem} of a fragment \fragmentx{X} asks, given
$\M$, $\vec{x}$ and $\varphi \in \text{\fragmentx{X}}$, whether
$\vec{x} \models_\M \varphi$.

\begin{example}\label{ex:ltl}
  Recall the MMS $\M$ and the schedule $\pi$ from \Cref{ex:mms}. Let $X$,
  $Y$ and $Z$ be the bounded zones colored in \Cref{fig:exec}, \eg\
  $X \defeq \{(x, y) \in \R^2 : 0.5 \leq x \leq 1.5, 0.5 \leq y \leq
  1.5\}$. We have $(1, 1) \models_\M X \land \F((Y \land \neg
  Z) \land \F Z)$. \hfill\qed
\end{example}

\subsection{Connection with classical LTL}\label{ssec:ltl}
\emph{Classical linear temporal logic (LTL)} (without temporal
operator $\X$) has the same syntax as the logic from
\Cref{ssec:ltl:mms}, but is interpreted over infinite words $w \in
(2^\AP)^\omega$:
\begin{alignat*}{3}
  w, i &\models \true,
  &&\iff \true, \\
  w, i &\models a
  &&\iff a \in w(i), \\
  w, i &\models \neg \varphi
  &&\iff \neg(w, i \models \varphi), \\
  w, i &\models \varphi \land \varphi'
  &&\iff (w, i \models \varphi) \land
         (w, i \models \varphi'), \\
  w, i &\models \varphi \lor \varphi'
  &&\iff (w, i \models \varphi) \lor
         (w, i \models \varphi'), \\
  w, i &\models \F \varphi
  &&\iff \exists j \geq i : w, j \models \varphi, \\
  w, i &\models \G \varphi
  &&\iff \forall j \geq i : w, j \models \varphi, \\
  w, i &\models \varphi \U \varphi'
  &&\iff \exists j \geq i : (w, j \models \varphi') \\
  &&&\hspace{1.5cm} \land (\forall k \in [i..j-1] :
    w, k \models \varphi).
\end{alignat*}

We write $w \models \varphi$ iff $w, 0 \models \varphi$. Observe that
$w, i \models \varphi$ holds iff $\FromTo{w}{i}{} \models \varphi$,
where $\FromTo{w}{i}{} \defeq w(i) w(i+1) \cdots$. We write $\varphi
\equiv \varphi'$ if $\varphi$ and $\varphi'$ are satisfied by the
same infinite words. In order to relate LTL over executions with LTL over
infinite words, we introduce the notion of traces. Informally, a trace
captures the zone changes within an execution.

Let $\chi_{\AP} \colon \R^d \to \AP$ be the function that yields the
set of zones a given point lies in: $\indic{\vec{x}} \defeq
\{Z \in \AP : \vec{x} \in Z\}$. Let $\sigma$ be an execution. We say
that word $w$ is a \emph{trace} of $\sigma$ if there exist $\tau_0 < \tau_1
< \cdots \in \Rnon$ such that
\begin{itemize}
\item $\dom \sigma = [\tau_0, \tau_1] \cup [\tau_1, \tau_2] \cup \cdots$,

\item $w(i) = \indic{\sigma(\tau_i)}$ for every $i \in \N$, and

\item for every $i \in \N$, there exists $j \in \{i, i+1\}$ such that:
$\indic{\sigma(\tau')} = \indic{\sigma(\tau_j)}$ for all $\tau' \in
(\tau_i, \tau_{i+1})$.
\end{itemize}

\begin{example}
  Recall execution $\sigma$ from \Cref{ex:mms}. The word $w \defeq
  \{X\} \{X\} \emptyset \{Y\} \{Y\} \{Y, Z\} \{Y, Z\} \{Z\} \emptyset
  \emptyset \cdots$ is a trace of $\sigma$. As depicted with circular
  marks in \Cref{fig:exec}, it is obtained from $\tau_0 \defeq 0,
  \tau_1 \defeq 0.5, \tau_2 \defeq 1, \tau_3 \defeq 1.5, \tau_4 \defeq
  2, \tau_5 \defeq 2.5, \tau_6 \defeq 2.75, \tau_7 \defeq 3, \tau_8
  \defeq 3.5$ and so on. \hfill\qed
\end{example}

With a bit of care, it is possible to prove that any execution
admits a trace. Moreover, in the absence of negations, model-checking
an execution amounts to model-checking any of its traces under the
classical LTL semantics. Thus, equivalences
of negation-free classical LTL also hold under the semantics over
executions.

\begin{restatable}{proposition}{propHasTrace}\label{prop:has:trace}
  Any execution $\sigma$ has a trace.
\end{restatable}

\begin{restatable}{proposition}{propLtlRealDiscrete}\label{prop:ltl:real:discrete}
  Let $\sigma$ be an execution with $\dom \sigma = \Rnon$, let $w$ be
  a trace of $\sigma$, and let $\varphi$ be a negation-free LTL
  formula. It is the case that $\sigma \models \varphi$ iff $w \models
  \varphi$.
\end{restatable}

\section{From \fragment{\F, \G, \land} to linear LTL}
\label{sec:buechi}
This section deals with classical LTL formulas interpreted over
infinite words. We show that any formula
from \fragment{\F, \allowbreak \G, \allowbreak \land} corresponds to
an automaton of a certain shape, which amounts to what we call linear
LTL formulas.

\subsection{From \fragment{\F, \G, \land} to flat formulas}

We say that a formula is \emph{pseudo-atomic} if it is a conjunction
of atomic propositions. By convention, an empty conjunction amounts to
$\true$. An LTL formula $\varphi$ is \emph{flat} if it has this form:
\[
\psi \land \G \psi' \land \bigwedge_{i \in I} \G\F \psi_i'' \land
\bigwedge_{j \in J} \F \varphi_j,
\]
where $\psi$, $\psi'$ and $\psi_i''$ are pseudo-atomic; and
$\varphi_j$ is flat.

Given a formula $\varphi \in \text{\fragment{\F, \G, \land}}$ of this
form:
\[
\varphi = \psi \land \bigwedge_{i \in I} \G \varphi_i \land
\bigwedge_{j \in J} \F \varphi_j,
\]
we define these mappings:
\begin{align*}
  \flat_{\G}(\varphi)
  &\defeq \G \psi \land \bigwedge_{i \in I}
  \flat_{\G}(\varphi_i) \land \bigwedge_{j \in J}
  \flat_{\G\F}(\varphi_j), \\
  \flat_{\G\F}(\varphi)  
  &\defeq \G\F \psi \land
  \bigwedge_{i \in I} \flat_{\F\G}(\varphi_i) \land
  \bigwedge_{j \in J} \flat_{\G\F}(\varphi_j), \\
  \flat_{\F\G}(\varphi)
  &\defeq \F\G \psi \land \bigwedge_{i \in I}
  \flat_{\F\G}(\varphi_i) \land \bigwedge_{j \in J}
  \flat_{\G\F}(\varphi_j), \\
  \flat(\varphi) 
  &\defeq \psi \land \bigwedge_{i \in I} \flat_{\G}(\varphi_i)
  \land \bigwedge_{j \in J} \F\, \flat(\varphi_j).
\end{align*}

As its name suggests, it follows by induction that formula
$\flat(\varphi)$ is flat. Moreover, the following holds.

\begin{restatable}{proposition}{propEquivFlatFormula}\label{prop:equiv:flat:formula}
  It is the case that $\flat(\varphi) \equiv \varphi$.
\end{restatable}

\subsection{From flat formulas to almost acyclic automata}

\newcommand{\unfold}[1]{\mathfrak{U}(#1)}

We say that an automaton is \emph{almost acyclic} if, for every
pair of states
$q \neq r$, it is the case that $q \trans{*} r$ implies $r
\not\trans{*} q$, \ie\ cycles must be self-loops. The \emph{width} of
an almost acyclic automaton is the maximal length among its simple
paths.

We will prove that any formula $\varphi \in \text{\fragment{\F, \G,
    \land}}$ can be translated into an almost acyclic automaton
$\A_\varphi$ of linear width in the size of $\varphi$, and such that
$\A_\varphi$ accepts $w$ iff $w \models \varphi$. We will formally
define the acceptance condition later on, but for readers familiar
with $\omega$-automata: $\A_\varphi$ will be a generalized B\"{u}chi
automaton with accepting transitions.

In order to define $\A_\varphi$, we first provide intermediate
definitions. Let $\mathfrak{U} \colon \text{\fragment{\F, \G, \land}}
\to 2^{\text{\fragment{\F, \G, \land}}}$ be defined by $\unfold{\true}
\defeq \{\true\}$, $\unfold{a} \defeq \{a\}$, $\unfold{\G \varphi}
\defeq \{\G \varphi\}$,
\begin{align*}
  \unfold{\varphi_1 \land \varphi_2}
  &\defeq \{\psi_1 \land \psi_2 : \psi_1 \in \unfold{\varphi_1},
  \psi_2 \in \unfold{\varphi_2}\}, \\
  \unfold{\F \varphi}
  &\defeq \{\F \varphi\} \cup \unfold{\varphi}.
\end{align*}

\begin{example}
  The set $\unfold{\G a \land \F(b \land \F\G c)}$ is equal to
  \[
  \{
  \G a \land \F(b \land \F\G c),
  \G a \land b \land \F\G c,
  \G a \land b \land \G c
  \}. \tag*{\qed}
  \]
\end{example}

\newcommand{\prop}[1]{\mathrm{prop}(#1)}

\noindent Given $A \subseteq \AP$, let $\prop{\bigwedge_{a \in A} a} \defeq
A$. Given $A \subseteq \AP$ and a flat formula $\varphi = \psi \land
\G \psi' \land \bigwedge_{i \in I} \G\F \psi_i'' \land \bigwedge_{j
  \in J} \F \varphi_j$, let
\[
\varphi[A] \defeq
\begin{cases}
  \displaystyle
  \G \psi' \land \bigwedge_{i \in I} \G\F \psi_i'' \land \bigwedge_{j
    \in J} \F \varphi_j
  & \!\!\text{if } \prop{\psi \land \psi'} \subseteq A, \\ \false
  & \!\!\text{otherwise}. \\
\end{cases}
\]
Given $\varphi \in \text{\fragment{\F, \G, \land}}$, the automaton
$A_\varphi \defeq (Q, \allowbreak \Sigma, \allowbreak {\trans{}},
\allowbreak q_0)$ is defined respectively by the following states,
alphabet, transitions and initial state:
\begin{align*}
  Q
  &\defeq \{\psi \in \text{\fragment{\F, \G, \land}} : \psi \text{ is
    flat}\}, \\  
  \Sigma
  &\defeq 2^\AP, \\
  \trans{} &\defeq \{(\psi, A, \psi'):
  \exists \psi'' \in \unfold{\psi} \text{ s.t.\ } \psi' =
  \psi''[A] \neq \false\}, \\
  q_0 &\defeq \flat(\varphi).
\end{align*}

\begin{example}
  Let $\varphi \defeq a \land \F b$, which is flat. We have
  $\unfold{\varphi} = \{a \land \F b, a \land b\}$. The automaton
  $\A_\varphi$ is depicted at the top of \Cref{fig:automaton}. Note
  that $w \models \varphi$ iff there is an infinite path from the
  initial state, labeled with $w$, that visits $\true$.

  Let $\varphi' \defeq \G\F(a \land \G c) \land \F b$. We have
  $\flat(\varphi') = \G\F a \land \F\G c \land \F b$. Hence,
  $\unfold{\flat(\varphi')}$ is equal to
  \[
  \{
  \G\F a \land \F\G c \land \F b,
  \G\F a \land \F\G c \land b,
  \G\F a \land \G c \land \F b,  
  \G\F a \land \G c \land b
  \}.
  \]
  The automaton $\A_{\varphi'}$ is depicted at the bottom of
  \Cref{fig:automaton}. Let $q \defeq \G\F a \land \G c$. Note that $w
  \models \varphi'$ iff there is an infinite path from the initial state,
  labeled with $w$, that visits the set of transitions $\{(q, A, q) :
  A \supseteq \{a, c\}\}$ infinitely often. \hfill\qed
\end{example}

\begin{figure}
  \begin{center}
    \begin{tikzpicture}[auto, node distance=2cm, very thick, initial text=, scale=0.75, transform shape]
  \tikzstyle{cstate} = [state, ellipse, inner sep=0pt];

  \node[cstate, initial above]     (q0) {$a \land \F b$};
  \node[cstate, right=1.5cm of q0] (q1) {$\F b$};        
  \node[cstate, right=1.5cm of q1] (q2) {$\true$};        
  
  \path[->]
  (q0) edge[] node {$\upc{\{a\}}$} (q1)
  (q0) edge[bend right=20] node[swap] {$\upc{\{a, b\}}$} (q2)
  (q1) edge[] node {$\upc{\{b\}}$} (q2)

  (q1) edge[loop above] node {$\upc{\emptyset}$} ()
  (q2) edge[loop right] node {$\upc{\emptyset}$} ()
  ;
  
  \node[cstate, initial above, below=1.25cm of q0] (r0) {
    $\G\F a \land \F\G c \land \F b$
  };
  \node[cstate, below of=r0] (r2) {$\G\F a \land \G c \land \F b$};
  \node[cstate, right of=r0, shift={(3, 0)}] (r1) {$\G\F a \land \F\G c$};
  \node[cstate, right of=r2, shift={(3, 0)}] (r3) {$\G\F a \land \G c$};
  
  \path[->]
  (r0) edge[] node[]     {$\upc{\{b\}}$}    (r1)
  (r1) edge[] node[]     {$\upc{\{c\}}$}    (r3)
  (r0) edge[] node[swap] {$\upc{\{c\}}$}    (r2)
  (r2) edge[] node[swap] {$\upc{\{b, c\}}$} (r3)
  
  (r0) edge[] node[] {$\upc{\{b, c\}}$} (r3)
  
  (r0) edge[out=175, in=185, looseness=5] node[swap] {
    $\upc{\emptyset}$
  } (r0)
  
  (r1) edge[out=-7, in=7, looseness=5] node[swap] {
    $\upc{\emptyset}$
  } (r1)
  
  (r2) edge[out=175, in=185, looseness=5] node[swap] {
    $\upc{\{c\}}$
  } (r2)
  
  (r3) edge[out=-7, in=7, looseness=5] node[swap] {
    $\upc{\{c\}}$
  } (r3);
  
  \node[above=2pt of q0, xshift=-35pt, font=\Large] {$\A_\varphi$:};
  \node[above=2pt of r0, xshift=-35pt, font=\Large] {$\A_{\varphi'}$:};
\end{tikzpicture}      
  \end{center}
  \caption{Automata for $\varphi = a \land \F b$ (top) and $\varphi' =
    \G\F(a \land \G c) \land \F b$ (bottom). Each $\upc{A}$ stands for
    $\{A' \subseteq \AP : A' \supseteq A\}$.}\label{fig:automaton}
\end{figure}

\subsubsection{Shape of automaton $\A_\varphi$}

We first seek to prove that $\A_\varphi$ is almost acyclic. For every
$\varphi \in \text{\fragment{\F, \G, \land}}$, let $|\true| = |a|
\defeq 1$, $|\varphi_1 \wedge \varphi_2| \defeq |\varphi_1| + 1 +
|\varphi_2|$ and $|\G\varphi| = |\F\varphi| \defeq 1 +
|\varphi|$. Moreover, let $\Fcount{\true} = \Fcount{a} = \Fcount{\G
  \varphi} \defeq 0$, $\Fcount{\varphi_1 \wedge \varphi_2} \defeq
\Fcount{\varphi_1} + \Fcount{\varphi_2}$ and $\Fcount{\F\varphi}
\defeq 1 + \Fcount{\varphi}$. The properties below follow by
induction.

\begin{restatable}{proposition}{propFCount}\label{prop:F:count}
  Let $\varphi \in \text{\fragment{\F, \G, \land}}$. This holds:
  \begin{enumerate}
  \item if $\varphi$ is flat, then $\Fcount{\varphi} >
    \Fcount{\varphi'}$ for all $\varphi' \in \unfold{\varphi}
    \setminus \{\varphi\}$,\label{itm:greater:Foperator}

  \item if $\varphi$ is flat, then $\Fcount{\varphi} \geq
    \Fcount{\varphi[A]}$ for all $A \subseteq
    \AP$,\label{itm:bracketoperator:Fcount}

  \item $|\varphi| \geq
    \Fcount{\mathrm{flat}(\varphi)}$.\label{itm:Formulasize:Foperator}
  \end{enumerate}
\end{restatable}

This proposition follows from \Cref{prop:F:count}:

\begin{restatable}{proposition}{propAutomataFcountLowering}\label{prop:automata:FcountLowering}
  Let $r_0 \trans{A_1} r_1 \trans{A_2} \cdots \trans{A_n} r_n$ be a
  simple path of $\A_\varphi$. It is the case that $\Fcount{r_1} >
  \cdots > \Fcount{r_n}$.
\end{restatable}

\begin{proposition}\label{prop:automata:nocycle}
  Let $\varphi \in \text{\fragment{\F, \G, \land}}$. Automaton
  $\A_\varphi$ is almost acyclic and its width belongs to
  $\bigO(|\varphi|)$.
\end{proposition}

\begin{IEEEproof}
  Let us first prove almost acyclicity. For the sake of contradiction,
  suppose that $\A_\varphi$ has a simple cycle $q \trans{u} r
  \trans{v} q$ where $q \neq r$. Since $q \trans{*} r$, it follows
  from \Cref{itm:greater:Foperator,itm:bracketoperator:Fcount} of
  \Cref{prop:F:count} that $\Fcount{q} \geq \Fcount{r}$. Since $q \neq
  r$, we have $|u| > 1$ and $|v| > 1$. Thus,
  \Cref{prop:automata:FcountLowering} yields $\Fcount{r} >
  \Fcount{q}$, which is a contradiction.

  Let us now bound the width $n$ of $A_\varphi$. Let $q_0 \trans{A_1}
  q_1 \trans{A_2} \cdots \trans{A_n} q_n$ be a simple path of
  $A_\varphi$. We have
  \begin{align*}
    n
    &\leq \Fcount{q_1} + 1
    && \text{(by \Cref{prop:automata:FcountLowering})} \\
    &\leq \Fcount{q_0} + 1
    && \text{(by
      \Cref{itm:greater:Foperator,itm:bracketoperator:Fcount} of
      \Cref{prop:F:count})} \\
    &= \Fcount{\flat(\varphi)} + 1
    && \text{(by def.\ of $q_0$)} \\
    &\leq |\varphi| + 1
    && \text{(by
      \Cref{itm:Formulasize:Foperator} of \Cref{prop:F:count})}. \tag*{\qedhere}
  \end{align*}
\end{IEEEproof}

\subsubsection{Language of $\A_\varphi$}

Let us define the acceptance condition of automaton $\A_\varphi$. Let
$F \defeq \{q \in Q : q_0 \trans{+} q \land \Fcount{q} =
0\}$. By definition, each state $q \in F$ is of the form $\G \psi
\land \bigwedge_{j \in J} \G\F \psi_j'$. Given such a state $q$, we
define \[T_{q, j} \defeq \{(q, A, q) \in {\trans{}}: \prop{\psi_j'}
\subseteq A\}.\] We say that word $w \in (2^\AP)^\omega$ is
\emph{accepted} by $\A_\varphi$, denoted $w \in \lang{\A_\varphi}$,
iff there exist $q \in F$ and an infinite path from $q_0$ that visits
$q$ and, for each $j \in J$, the set $T_{q, j}$ infinitely often. In
the remainder, we prove that $w \in \lang{\A_\varphi}$ iff $w \models
\varphi$.

\begin{restatable}{lemma}{lemConsume}\label{lem:consume}
  Let $\varphi \in \text{\fragment{\F, \G, \land}}$ be a flat
  formula. These two properties are equivalent to $w \models \varphi$:
  \begin{enumerate}
  \item there exists $\varphi'$ such that $\varphi \trans{w(0)}
    \varphi'$ and $\FromTo{w}{1}{} \models \varphi'$;

  \item there exist $i \in \N$ and $\varphi'$ such that $\varphi
    \trans{w(0) \cdots w(i-1)} \varphi'$, $\Fcount{\varphi'} = 0$ and
    $\FromTo{w}{i}{} \models \varphi'$.
  \end{enumerate}
\end{restatable}

\begin{proposition}\label{prop:equiv:automata:formula}
  Let $\varphi \in \text{\fragment{\F, \G, \land}}$. It is the case
  that $w \models \varphi$ iff $w \in \lang{\A_\varphi}$.
\end{proposition}

\begin{IEEEproof}
  $\Rightarrow$) By \Cref{lem:consume}(2), there are $k \in \N$ and
  $\varphi'$ with
  \[
  \flat(\varphi) \trans{w(0) \cdots w(k-1)} \varphi',
  \Fcount{\varphi'} = 0 \text{ and }
  \FromTo{w}{k}{} \models \varphi'.
  \]
  As $\Fcount{\varphi'} = 0$, we have $\unfold{\varphi'} =
  \{\varphi'\}$. So, \Cref{lem:consume}(1) yields
  \[
  \varphi' \trans{w(k)} \varphi' \trans{w(k+1)} \varphi'
  \trans{w(k+2)} \cdots.
  \]
  As $\varphi' \in F$, it has the form $\G \psi \land \bigwedge_{j \in
    J} \G\F \psi_j'$. In particular, this means that $\FromTo{w}{k}{} \models
  \bigwedge_{j \in J} \G\F \psi_j'$. Recall that $T_{\varphi', j} =
  \{(\varphi', A, \varphi') \in {\trans{}}: \prop{\psi_j'} \subseteq
  A\}$. So, for each $j \in J$, the set $T_{\varphi', j}$ is visited
  infinitely often.
  
  $\Leftarrow$) By $w \in \lang{A_\varphi}$, there exist $q \in F$ and
  $k \in \N$ such that
  \begin{itemize}
  \item $q = \G \psi \land \bigwedge_{j \in J} \G\F \psi_j'$,

  \item $q_0 \trans{w(0) \cdots w(k-1)} q$, and

  \item some infinite path $q \trans{\FromTo{w}{k}{}}$ visits, for each $j \in
    J$, the set $T_{q, j}$ infinitely often.
  \end{itemize}
  Recall that $T_{q, j} = \{(q, A, q) \in {\trans{}}: \prop{\psi_j'}
  \subseteq A\}$. Since $q \trans{\FromTo{w}{k}{}}$ visits each $T_{q,
    j}$ infinitely often, we have $\FromTo{w}{k}{} \models
  \bigwedge_{j \in J} \G\F \psi_j'$. By $\unfold{q} = \{q\}$ and by
  definition of ${\trans{}}$, we have $\FromTo{w}{k}{} \models \G
  \psi$. So, $\FromTo{w}{k}{} \models q$. By repeated applications of
  \Cref{lem:consume}(1), this implies $w \models q_0 = \flat(\varphi)
  \equiv \varphi$.
\end{IEEEproof}

\subsection{From almost acyclic automata to linear LTL}

In this subsection, we show that almost acyclic automata are
equivalent to finite sets of so-called linear LTL formulas, with the
goal of showing that \fragmentB{\F, \G, \land} belongs to NP
in the forthcoming \Cref{subsection:membership}.

For every $A \subseteq \AP$, let $\upc{A} \defeq \{A' \subseteq \AP :
A' \supseteq A\}$. We say that $X \subseteq 2^\AP$ is \emph{simple} if
$X = \upc{A}$ for some $A \subseteq \AP$.

\begin{example}\label{ex:paths:struct}
  Consider the bottom automaton of \Cref{fig:automaton}. Its infinite
  paths are captured by these three expressions:
  \begin{itemize}
  \item $\upc{\emptyset}^*\ \upc{\{b\}}\
    \upc{\emptyset}^*\ \upc{\{c\}}\ \upc{\{c\}}^\omega$,

  \item $\upc{\emptyset}^*\ \upc{\{b, c\}}\ \upc{\{c\}}^\omega$,

  \item $\upc{\emptyset}^*\ \upc{\{c\}}\ \upc{\{c\}}^*\
    \upc{\{b, c\}}\ \upc{\{c\}}^\omega$,
  \end{itemize}
  which respectively amount to these LTL formulas:  
  \begin{itemize}
  \item $\true \U (b \land (\true \U (c \land \G c)))$,

  \item $\true \U ((b \land c) \land \G c)$,

  \item $\true \U (c \land (c \U ((b \land c) \land \G c)))$.
  \end{itemize}
  Taking into account the acceptance condition of the automaton on its
  bottom-right transition, we obtain these LTL formulas:
  \begin{itemize}
  \item $\true \U (b \land (\true \U (c \land (\G c \land \G\F a))))$,

  \item $\true \U ((b \land c) \land (\G c \land \G\F a))$,

  \item $\true \U (c \land (c \U ((b \land c) \land (\G c \land \G\F
    a))))$. \hfill\qed
  \end{itemize}
\end{example}

For every $q, r \in Q$, let $X_{q, r} \defeq \{A \subseteq \AP : q
\trans{A} r\}$. In general, the paths of $\A_\varphi$ can always be
captured in the fashion of \Cref{ex:paths:struct} due to the following
structure of $A_\varphi$.

\begin{restatable}{proposition}{propAutTransProp}\label{prop:aut:trans:prop}
  Let $q, r \in Q$. It is the case that
  \begin{enumerate}
  \item $X_{q, r}$ is either empty or simple,

  \item if $X_{q, r} \neq \emptyset$, then $X_{r, r} \neq \emptyset$,

  \item if $X_{q, q} \neq \emptyset$, then $X_{q, q} \supseteq X_{q,
    r}$.
  \end{enumerate}
  Moreover, given $\theta \in \unfold{q}$ and $A \subseteq \AP$ such
  that $r = \theta[A]$, the representation of $X_{q, r}$ can be
  obtained in polynomial time.
\end{restatable}

A \emph{linear path scheme (LPS)} of $\A_\varphi$ is a simple path
$r_0 \trans{} \allowbreak r_1 \trans{} \allowbreak \cdots \trans{} \allowbreak r_n$ of $\A_\varphi$ such
that $r_n \in F$. A word $w \in (2^\AP)^\omega$ is \emph{accepted} by
such an LPS $S$, denoted $w \in \lang{S}$, iff $\A_\varphi$ has an
accepting path starting in $r_0$ and visiting $r_1, \ldots, r_n$
(possibly many times). For example, the LPS $(a \land \F b) \trans{}
\F b \trans{} \true$, from the top of \Cref{fig:automaton}, accepts $w
\defeq \{a\}\ \emptyset \{a\}\ \{b\}\ (\emptyset \{a\})^\omega$.

\newcommand{\goals}[1]{\mathrm{goals}(#1)}

We say that an LTL formula is \emph{linear} if it can be derived from
$\psi$ in this grammar:
\begin{align*}
  \psi  &::= A \land \psi' \mid \psi', \\[-5pt]
  \psi' &::= B \U (B' \land \psi') \mid (\G C_0) \land \bigwedge_{i=1}^n
  \G\F C_i,
\end{align*}
where $A, B, B', C_0, \ldots, C_n \subseteq \AP$, $\upc{B} \supseteq
\upc{B'}$, and each subset $D$ stands for formula $\bigwedge_{d \in D} d$. We
prove that any LPS is equivalent to a linear formula.

\begin{proposition}\label{prop:lps:to:formula}
  Given an LPS $S$ of $\A_\varphi$, one can construct, in polynomial
  time, a linear formula $\psi$ s.t.\ $w \in \lang{S}$ iff $w \models
  \psi$.
\end{proposition}

\begin{IEEEproof}
  Let $r_0 \trans{} r_1 \trans{} \cdots \trans{} r_n \in F$ be the
  simple path given by $S$. We inductively construct a formula derived
  from $\psi'$ if $X_{r_0, r_0} \neq \emptyset$, and from $\psi$
  otherwise.

  If $n = 0$, then we have $r_0 = r_n$. As $r_0 \in F$, it is of the
  form $r_0 = \G \psi \land \bigwedge_{j \in J} \G\F \psi_j'$. Recall
  that $T_{r_0, j} \defeq \{(r_0, A, r_0) \in {\trans{}}: A \in
  \upc{\prop{\psi_j'}}\}$. Note that $X_{r_0, r_0} \neq \emptyset$ as
  $\prop{\psi} \in X_{r_0, r_0}$. So, by \Cref{prop:aut:trans:prop},
  it is the case that $X_{r_0, r_0} = \upc{C_0}$ for some $C_0
  \subseteq \AP$. Let $\psi' \defeq \G C_0 \land \bigwedge_{j \in J}
  \G\F C_j$, where $C_j \defeq \prop{\psi_j'}$. By definition, $w \in
  \lang{S}$ iff $w \models \psi'$.

  Assume $n > 0$. Let $S'$ be the LPS $r_1 \trans{} \cdots \trans{}
  r_n \in F$. By \Cref{prop:aut:trans:prop}, since $X_{r_0, r_1} \neq
  \emptyset$, we have $X_{r_1, r_1} \neq \emptyset$. By induction
  hypothesis, there is a linear formula $\psi'$ such that $w \in
  \lang{S'}$ iff $w \models \psi'$. By \Cref{prop:aut:trans:prop},
  there exists $B \subseteq \AP$ such that $X_{r_0, r_1} =
  \upc{B}$. If $X_{r_0, r_0} = \emptyset$, then we set $\psi \defeq B
  \land \psi'$. Otherwise, by \Cref{prop:aut:trans:prop}, there exists
  $A \subseteq \AP$ such that $X_{r_0, r_0} = \upc{A} \supseteq
  \upc{B}$. Thus, we set $\psi \defeq A \U (B \land \psi')$. By
  definition of acceptance, $w \in \lang{S}$ iff $w \models \psi$.
\end{IEEEproof}

\section{P-complete fragments}
\label{sec:ptime}
Let $\goals{A \land \varphi} \defeq \goals{\varphi}$, $\goals{B \U
    (B' \land \varphi)} \defeq \goals{\varphi}$, and $\goals{(\G
    C_0) \land \bigwedge_{i=1}^n \G\F C_i} \defeq \{C_1, \ldots, C_n\}$.
We say that a linear LTL formula $\varphi$, interpreted over
executions, is \emph{semi-bounded} if each zone of $\goals{\varphi}$
is bounded. We will establish the following theorem by
translating linear LTL formulas into a polynomial-time logic introduced
in~\cite{blondin2017logics}.

\begin{theorem}\label{thm:linear:P}
  The model-checking problem for semi-bounded linear LTL formulas is
  in P.
\end{theorem}

Before proving the above theorem, we use it to prove the previously announced
P-completeness results.

\begin{restatable}{theorem}{thmFGNforP}\label{thm:fgn:for:P}
  The model-checking problem is in P for these fragments:
  \fragmentB{\F, \G, \neg}, \fragment{\F, \lor} and \fragment{\G,
    \land}.
\end{restatable}

\begin{IEEEproof}
  Consider a formula from \fragmentB{\F, \G, \neg}. Note that
  $\neg \F \varphi \equiv \G \neg \varphi$ and $\neg\neg \varphi
  \equiv \varphi$. Thus, negations can be pushed inwards. Afterwards,
  we can simplify using $\F \G \F \varphi \equiv \G \F \varphi$ and
  $\G \F \G \varphi \equiv \F \G \varphi$. If the resulting formula is
  negation-free, then it is of the form $\F Z$, $\G Z$, $\G\F Z$ or
  $\F\G Z$. These are all linear as $\F Z \equiv \R^d \U Z$ and $\F\G
  Z \equiv \R^d \U (\G Z)$. Thus, we are done by \Cref{thm:linear:P}. If
  the resulting formula has a negation, then there are four forms to consider:
  (1)~$\F \neg Z$, (2)~$\G \neg Z$, (3)~$\G\F \neg Z$ and (4)~$\F\G
  \neg Z$. These are easy to handle:

  \begin{itemize}
  \item We have $\vec{x} \models_\M \F \neg Z$ iff $\vec{x} \models_\M
    \G\F \neg Z$ iff $\vec{x} \models_\M \F\G \neg Z$ iff $\vec{x}
    \not \in Z$ or there is a mode $\vec{m} \in \M$ such that $\vec{m}
    \neq \vec{0}$.

  \item We have $\vec{x} \models_\M \G \neg Z$ iff $\vec{x} \not \in
    Z$ and there exists a mode $\vec{m} \in \M$ such that for all
    $\alpha \in \Rpos$: $\vec{x} + \alpha \vec{m} \not \in Z$.
  \end{itemize}

  Since $\F\F \varphi \equiv \F \varphi$ and $\F (\varphi \lor \psi)
  \equiv (\F \varphi) \lor (\F \psi)$, any formula from \fragment{\F,
    \lor} can be turned into a disjunction of atomic propositions and
  formulas from \fragment{\F}. So, it suffices to check each
  disjunct in polynomial time.

  As $\G \G \varphi \equiv \G \varphi$ and $(\G \varphi) \land (\G
  \psi) \equiv \G (\varphi \land \psi)$, we can transform formulas from
  \fragment{\G, \land} into the form $\psi \land \G \psi'$,
  where $\psi, \psi'$ are pseudo-atomic. The latter is linear,
  and hence can be model-checked in polynomial time.
\end{IEEEproof}

\begin{restatable}{theorem}{thmFGPhard}\label{thm:f:g:Phard}
  The model-checking problem is P-hard for both \fragmentB{\F}
  and \fragmentB{\G}.
\end{restatable}

\begin{IEEEproof}
  It follows by simple reductions from feasibility of linear programs
  and the monotone circuit-value problem.
\end{IEEEproof}

\subsection{A polynomial-time first-order logic}
\label{ssec:poly:logic}

We recall a first-order logic over the reals introduced
in~\cite{blondin2017logics}. It allows for conjunctions of
\emph{convex semi-linear Horn formulas}, \ie\ formulas of this form:
\[
  \sum_{i=1}^d \vec{a}(i) \cdot \vec{x}(i) \sim c \lor
  \bigvee_{i \in I} \bigwedge_{j \in J_i} \vec{x}(j) > 0,
\]
where $\vec{a} \in \Z^d$, $c \in \Z$, ${\sim} \in \{<, \leq, =, \geq,
  >\}$, and $I$ and each $J_i$ is a finite set of indices. The problem
of determining, given a formula $\varphi$ from this logic, whether
there exists $\vec{x} \in \Rnon^d$ such that $\varphi(\vec{x})$ holds,
can be solved in polynomial time~\cite{blondin2017logics}.

This result extends easily to solutions where $\vec{x}(j) \in \R$ is
allowed, provided that $\vec{x}(j)$ is never used in
disjuncts. Indeed, it suffices to introduce two variables $y, z \in
  \Rnon$ and replace each occurrence of $\vec{x}(j)$ with $y - z$.

Given $\vec{x}(i), \vec{x}(j) \in \Rnon$, we will use $\vec{x}(i) > 0
\rightarrow \vec{x}(j) > 0$ as short for $\vec{x}(i) = 0 \lor
\vec{x}(j) > 0$.

\subsection{Expressing ${\mreach{*}{Z}}$ in first-order logic}
\label{ssec:transu}

We first seek to build a formula $\varphi_Z$ from the
aforementioned logic such that $\varphi_Z(\vec{x}, \vec{\lambda},
\vec{y})$ holds iff there is a finite schedule $\pi$ with $\vec{x}
\mreach{\pi}{Z} \vec{y}$ and $\parikh{\pi} = \vec{\lambda}$. Let us
fix zone $Z$ and modes $\M = \{\vec{m}_1, \ldots, \vec{m}_n\}$. We
take inspiration from the characterization of continuous Petri nets
reachability of~\cite[Thm.~20]{FH15}, which is equivalent to finding a
Parikh image that (1)~admits the right effect, (2)~is forward fireable,
and (3)~is backward fireable. The forthcoming
\Cref{prop:Zsafe:LineConstraints} similarly characterizes $\vec{x}
\mreach{\pi}{Z} \vec{y}$.

\begin{proposition}\label{prop:reduc:Zsafe:exec}
  Let $\vec{x} \mreach{\pi}{Z} \vec{y}$. There exist $\pi_x$ and
  $\pi_y$ such that $\vec{x} \mreach{\pi_x}{Z} {}$,
  ${} \mreach{\pi_y}{Z} \vec{y}$, $\supp{\pi_x} = \supp{\pi_y}
  = \supp{\pi}$ and $|\pi_x| = |\pi_y| = |\supp{\pi}|$.
\end{proposition}

\begin{restatable}{lemma}{lemaModeZSafeOneMode}\label{lema:mode:Zsafe:oneMode}
  Let $\rho(\alpha, \vec{m})\rho'$ be a schedule. This holds:
  \begin{itemize}
  \item If $\vec{x} \mreach{\rho(\alpha, \vec{m})\rho'}{Z}$, then
    $\vec{x} \mreach{\rho(\frac{\alpha}{2},
    \vec{m})\frac{1}{2}\rho'(\frac{\alpha}{2}, \vec{m})}{Z}$,

  \item If ${} \mreach{\rho'(\alpha, \vec{m})\rho}{Z} \vec{y}$, then
    ${} \mreach{(\frac{\alpha}{2},
    \vec{m})\frac{1}{2}\rho'(\frac{\alpha}{2}, \vec{m})\rho}{Z}
    \vec{y}$.
  \end{itemize}
\end{restatable}

\begin{restatable}{proposition}{propModeZSafeExec}\label{prop:mode:Zsafe:exec} 
  Let $\vec{x} \mreach{\pi}{Z} \vec{y}$. There exist
  $\beta \in \N_{\geq 1}$, $\vec{x} \mreach{\pi'}{Z} \vec{y}_Z$ and
  $\vec{x}_Z \mreach{\pi''}{Z} \vec{y}$ such that $|\pi| = |\pi'| =
  |\pi''|$, $\supp{\pi} = \supp{\pi'} = \supp{\pi''}$, and, for every
  $\vec{m} \in \supp{\pi}$, it is the case that $\vec{x}_Z \mreach{(1
  / \beta) \vec{m}}{Z} {}$ and ${} \mreach{(1
  / \beta) \vec{m}}{Z} \vec{y}_Z$.
\end{restatable}

\begin{restatable}{proposition}{propZSafeLineConstruction}\label{prop:Zsafe:LineConstruction}
  Let $\vec{x} \mreach{\pi}{} \vec{y}$, $k \defeq |\pi|$ and $\beta
  \in \N_{\geq 1}$ be such that $\vec{x} \mreach{(1 / \beta)\pi(i)}{Z}
      {}$ and ${} \mreach{(1 / \beta)\pi(i)}{Z} \vec{y}$ hold for all
      $i \in [1..k]$. It is the case that $\vec{x} \mreach{\pi'}{Z}
      \vec{y}$, where $\pi' \defeq ((1 / (\beta k))\pi)^{\beta k}$.
\end{restatable}

\begin{proposition}\label{prop:Zsafe:LineConstraints}
  It is the case that $\vec{x} \mreach{\pi}{Z} \vec{y}$ iff there exist
  $\pi', \pi_\text{fwd}, \pi_\text{bwd}$ with
  \begin{itemize}
  \item $\supp{\pi'} = \supp{\pi_\text{fwd}} = \supp{\pi_\text{bwd}} =
    \supp{\pi}$,

  \item $\vec{x} \mreach{\pi'}{} \vec{y}$, $\vec{x}
    \mreach{\pi_\text{fwd}}{Z}$ and ${} \mreach{\pi_\text{bwd}}{Z}
    \vec{y}$.
  \end{itemize}
\end{proposition}

\begin{IEEEproof}
  $\Rightarrow$) It suffices to take $\pi' = \pi_\text{fwd} =
  \pi_\text{bwd} \defeq \pi$.

  $\Leftarrow$) Let $\beta$ and the following be given by
  \Cref{prop:mode:Zsafe:exec}:
  \[
  \vec{x} \mreach{\pi'_\text{fwd}}{Z} \vec{x}_Z \text{ and } \vec{y}_Z
  \mreach{\pi'_\text{bwd}}{Z} \vec{y}.
  \]
  Let $\gamma \in \N_{\geq 1}$ be sufficiently large so that
  $\parikh{\pi'} \geq \frac{1}{\gamma}(\parikh{\pi'_\text{fwd}} +
  \parikh{\pi'_\text{bwd}})$. Such a $\gamma$ exists as
  $\supp{\pi'_\text{fwd}} = \supp{\pi'_\text{bwd}} = \supp{\pi'}$. Let
  $\pi''$ be any schedule with $\parikh{\pi''} = \parikh{\pi'} -
  \frac{1}{\gamma}(\parikh{\pi'_{\text{fwd}}} +
  \parikh{\pi'_{\text{bwd}}})$. We have
  \[
  \vec{x}
  \mreach{\frac{1}{\gamma}\pi'_\text{fwd}}{Z} \vec{x'}_Z
  \mreach{\pi''}{} \vec{y'}_Z
  \mreach{\frac{1}{\gamma}\pi'_\text{bwd}}{Z}
  \vec{y}.
  \]
  So, by invoking \Cref{prop:Zsafe:LineConstruction} with $\beta \cdot
  \gamma \cdot \lceil \tweight{\pi''} \rceil \in \N_{\geq 1}$, we
  obtain $\pi$ with $\vec{x} \mreach{\pi}{Z} \vec{y}$.
\end{IEEEproof}

We may now conclude this subsection by building a suitable first-order
formula. First, we define
\begin{alignat*}{3}
    \theta_Z(\vec{x}, \parikh{s}, \vec{y})
    &\defeq\ &&
    \exists \{\vec{z}_{i, j} \in Z\}_{i \in [1..n], j \in [0..n]} \\
    &&&
    \exists \{\lambda_{i, j} \in \Rnon\}_{i, j \in [1..n]}\ \\
    &&&
    \exists \vec{\alpha} \in \Rnon^n :
    (\vec{x} = \vec{z}_{1, 0}) \land (\vec{y} = \vec{z}_{n, n}) \land {} \\
    &&&
    \bigwedge_{i \in [1..n]} \bigwedge_{j \in [1..n]} (\vec{z}_{i, j} =
    \vec{z}_{i, j-1} + \lambda_{i, j} \cdot \vec{m}_j) \land {} \\
    &&&
    \bigwedge_{i \in [2..n]} (\vec{z}_{i, 0} = \vec{z}_{i-1, n}) \land {} \\
    &&&
    \bigwedge_{j \in [1..n]} (\vec{\alpha}(j) = \sum_{\mathclap{i \in [1..n]}} \lambda_{i, j})\, \land {} \\
    &&&
    \bigwedge_{j \in [1..n]} (\parikh{s}(j) > 0 \leftrightarrow \vec{\alpha}(j) > 0).
\end{alignat*}
Let $\varphi_Z(\vec{x}, \parikh{\lambda}, \vec{y}) \defeq \psi(\vec{x},
\parikh{\lambda}, \vec{y}) \land \psi_\text{fwd}(\vec{x}, \parikh{\lambda})
\land \psi_\text{bwd}(\parikh{\lambda}, \vec{y})$ where
\begin{align*}
    \psi(\vec{x}, \parikh{\lambda}, \vec{y})
    &\defeq
    (\vec{y} = \vec{x} +
    \parikh{\lambda}(1) \cdot \vec{m}_1 + \ldots +
    \parikh{\lambda}(n) \cdot \vec{m}_n), \\
    \psi_\text{fwd}(\vec{x}, \parikh{s})
    &\defeq
    \exists \vec{x}' \in \R^d : \theta_Z(\vec{x}, \parikh{s}, \vec{x}'), \\
    \psi_\text{bwd}(\parikh{s}, \vec{y})
    &\defeq
    \exists \vec{y}' \in \R^d : \theta_Z(\vec{y}', \parikh{s}, \vec{y}).
\end{align*}

\begin{proposition}
  It is the case that $\varphi_Z(\vec{x}, \vec{\lambda}, \vec{y})$
  holds iff $\vec{x} \mreach{\pi}{Z} \vec{y}$ for some finite schedule
  $\pi$ such that $\parikh{\pi} = \vec{\lambda}$.
\end{proposition}

\begin{IEEEproof}
  First note that formula $\theta_Z(\vec{u}, \vec{s}, \vec{v})$
  guesses a schedule of size at most $n^2$ from $\vec{u}$ to $\vec{v}$
  that remains within $Z$ using precisely the modes from $\{\vec{m}_j
  : j \in [1..n], \vec{s}(j) > 0\}$. The reason $\theta_Z$ uses a
  schedule of size $n^2$, rather than $n$, is to guess
  the order in which modes are first used.
  
  $\Rightarrow$) Suppose that $\psi(\vec{x}, \parikh{\lambda},
  \vec{y}) \land \psi_\text{fwd}(\vec{x}, \parikh{\lambda}) \land
  \psi_\text{bwd}(\parikh{\lambda}, \vec{y})$ holds. Let $\pi' \defeq
  \prod_{i = 1}^n \vec{\lambda}(i) \vec{m}_i$, 
  \[
  \pi_\text{fwd} \defeq \prod_{i = 1}^n \prod_{j = 1}^n
  \lambda^\text{fwd}_{i,j} \vec{m}_j
  \text{ and }
  \pi_\text{bwd} \defeq \prod_{i = 1}^n \prod_{j = 1}^n
  \lambda^\text{bwd}_{i,j}\vec{m}_j,
  \]
  with the convention that $0 \cdot \vec{m}_j$ stands for the empty
  schedule. We clearly have $\vec{x} \mreach{\pi'}{} \vec{y}$. By the
  above observation on $\theta_Z$, we further have $\vec{x}
  \trans{\pi_\text{fwd}} {}$ and ${} \trans{\pi_\text{bwd}}
  \vec{y}$. Moreover, $\supp{\pi'} = \supp{\pi_\text{fwd}} =
  \supp{\pi_\text{bwd}} = \vec{\lambda}$. By
  \Cref{prop:Zsafe:LineConstraints}, we obtain $\vec{x}
  \mreach{\pi}{Z} \vec{y}$ for some $\pi$ with $\parikh{\pi} =
  \vec{\lambda}$.

  $\Leftarrow$) Let $\vec{x} \mreach{\pi}{Z} \vec{y}$ and $\vec{\lambda}
  \defeq \parikh{\pi}$. By \Cref{prop:reduc:Zsafe:exec}, there are
  $\pi_x$ and $\pi_y$ such that $\vec{x} \mreach{\pi_x}{Z} {}$, ${}
  \mreach{\pi_y}{Z} \vec{y}$, $\supp{\pi_x} = \supp{\pi_y} =
  \supp{\pi}$ and $|\pi_x| = |\pi_y| = |\supp{\pi}|$. Thus, we can use
  $\pi$, $\pi_x$ and $\pi_y$ to satisfy $\psi(\vec{x},
  \parikh{\lambda}, \vec{y})$, $\psi_\text{fwd}(\vec{x},
  \parikh{\lambda})$ and $\psi_\text{bwd}(\parikh{\lambda}, \vec{y})$.
\end{IEEEproof}

\subsection{Expressing $\G Z \land \G\F X \land \G\F Y$ in first-order logic}
\label{ssec:lasso}

In this subsection, we build a formula $\varphi_{\G Z \land \G\F
X \land \G\F Y}$ from the logic of \Cref{ssec:poly:logic} such that
$\vec{x} \models_\M \G Z \land \G\F X \land \G\F Y$ iff $\varphi_{\G
Z \land \G\F X \land \G\F Y}(\vec{x})$ holds. Let us fix an MMS $\M$.

\begin{restatable}{proposition}{propZSafeExecOnLine}\label{prop:Zsafe:execOnLine}
  Let $Z$ be a zone, let $\pi$ be a schedule, let $\vec{x}, \vec{x}', \vec{y}
  \in Z$ and let $\beta \in (0, 1]$. Let $\vec{z} \defeq \beta
    \vec{x} + (1-\beta) \vec{y}$ and $\vec{z}' \defeq \beta \vec{x}'
    + (1-\beta) \vec{y}$. If $\vec{x} \mreach{\pi}{Z} \vec{x}'$
    holds, then $\vec{z} \mreach{\beta \pi}{Z} \vec{z}'$.
\end{restatable}

\begin{proposition}\label{prop:Dual}
  Let $X, Y, Z$ be zones where at least one of the three zones is
  bounded. Let $\vec{z} \models_\M \G Z \land \G\F X \land \G\F
  Y$. There exist $\vec{x}_f \in X \cap Z$, $\vec{y}_f \in Y \cap Z$
  and finite schedules $\pi, \pi'$ such that $\vec{z} \mreach{*}{}
  \vec{x}_f \mreach{\pi}{} \vec{y}_f \mreach{\pi'}{} \vec{x}_f$ and
  $\norm{\parikh{\pi} + \parikh{\pi}'} \geq 1$.
\end{proposition}

\begin{IEEEproof}
  Let $X' \defeq X \cap Z$ and $Y' \defeq Y \cap Z$. By assumption,
  there exist $\vec{x}_0, \vec{x}_1, \ldots \in X'$ and $\vec{y}_0,
  \vec{y}_1, \ldots \in Y'$ such that
  \[
  \vec{z}   \mreach{*}{Z}
  \vec{x}_0 \mreach{\pi_0}{Z} \vec{y}_0 \mreach{\pi_0'}{Z}
  \vec{x}_1 \mreach{\pi_1}{Z} \vec{y}_1 \mreach{\pi_1'}{Z}
  \cdots,
  \]
  and $\norm{\parikh{\pi_i} + \parikh{\pi_i'}} \geq 1$ for all $i \in
  \N$. Note that the latter follows from non-Zenoness.

  Let $\vec{A}_1 \vec{\ell} \leq \vec{b}_1$ and $\vec{A}_2 \vec{\ell}
  \leq \vec{b}_2$ be the systems of inequalities that respectively
  represent zones $X'$ and $Y'$. We define $\mat{M}$ as the matrix
  such that each column is a mode from $\M$. Let $\mathcal{S}$ denote
  the following system:
  \begin{align*}
    &\exists \vec{u}_1, \vec{u}_2, \vec{u}_3 \geq \vec{0}:\\
    &\begin{bmatrix}
      \mat{A}_1\mat{M} & \mat{0}          & \mat{0}    \\
      \mat{A}_2\mat{M} & \mat{A}_2\mat{M} & \mat{0}    \\
      \mat{0}          & \mat{M}          & \mat{M}    \\
      \mat{0}          & -\mat{M}         & -\mat{M}   \\
      \vec{0}^T        & \vec{-1}^T       & \vec{-1}^T \\
    \end{bmatrix}
    \begin{bmatrix}
      \vec{u}_1 \\ \vec{u}_2 \\ \vec{u}_3 \\
    \end{bmatrix}
    \leq
    \begin{bmatrix}
      \vec{b}_1 - \mat{A}_1\vec{z} \\
      \vec{b}_2 - \mat{A}_2\vec{z} \\
      \vec{0}                      \\
      \vec{0}                      \\
      -1
    \end{bmatrix}.
  \end{align*}
  Observe that $\mathcal{S}$ is equivalent to the existence of
  $\vec{x}_f \in X', \vec{y}_f \in Y'$ and $\pi, \pi'$ such that
  \[
  \vec{z} \mreach{*}{} \vec{x}_f \mreach{\pi}{} \vec{y}_f
  \mreach{\pi'}{} \vec{x}_f \text{ and } \norm{\parikh{\pi} + \parikh{\pi'}}
  \geq 1.
  \]  
  For the sake of contradiction, suppose that $\mathcal{S}$ has no
  solution. By Farkas' lemma, the following system $\mathcal{S}'$ has
  a solution:
  \begin{align*}
    \exists
    \vec{v}_1, \vec{v}_2 \in \Rnon^d,
    \vec{v}_3 \in \R^d, v_4 \in \Rnon: \\
    \begin{bmatrix}
      \mat{M}^T\mat{A}_1^T & \mat{M­}^T\mat{A}_2^T  & \mat{0}   & \vec{0} \\
      \mat{0}              & \mat{M}^T\mat{A}_2^T  & \mat{M}^T & \vec{-1} \\
      \mat{0}              & \mat{0}               & \mat{M}^T & \vec{-1}
    \end{bmatrix}
    \begin{bmatrix}
      \vec{v}_1 \\ \vec{v}_2 \\ \vec{v}_3 \\ v_4
    \end{bmatrix}
    & \geq \vec{0}, \\
    \begin{bmatrix}
      (\vec{b}_1 - \mat{A}_1\vec{z})^T &
      (\vec{b}_2 - \mat{A}_2\vec{z})^T & -1
    \end{bmatrix}
    \begin{bmatrix}
      \vec{v}_1 \\ \vec{v}_2 \\ v_4
    \end{bmatrix}
    & < 0.
  \end{align*}

  Using the above, we will construct linear functions $g$ and $h$ such
  that $\lim_{i \to \infty} g(\vec{x}_i) = \lim_{i \to \infty}
  h(\vec{y}_i) = \infty$. Since either $X'$ or $Y'$ is bounded, this
  yields a contradiction. We make a case distinction on the value of
  $v_4$.

  \medskip\noindent\emph{Case $v_4 > 0$}. We have $\vec{v}_3^T\mat{M}
  \geq \vec{1}^T v_4 > 0$. Let $g(\vec{x}_i) \defeq \vec{v}_3^T
  (\vec{x}_i - \vec{x}_0)$ and $h(\vec{y}_i) \defeq \vec{v}_3^T
  (\vec{y}_i - \vec{y}_0)$. For every $i \geq 1$,
  \begin{align*}
    g(\vec{x}_i)
    &= \vec{v}_3^T (\vec{x}_i - \vec{x}_0) \\
    &= \vec{v}_3^T (\vec{x}_{i-1} - \vec{x}_0) +
    \vec{v}_3^T (\vec{x}_i - \vec{x}_{i-1}) \\
    &= g(\vec{x}_{i-1}) + \vec{v}_3^T \mat{M}
    (\parikh{\pi_{i-1}} + \parikh{\pi_{i-1}'}) \\
    &\geq g(\vec{x}_{i-1}) + v_4 \vec{1}^T
    (\parikh{\pi_{i-1}} + \parikh{\pi_{i-1}'}) \\
    &\geq g(\vec{x}_{i-1}) + v_4
    && \hspace*{-55pt}\text{(by $\norm{\parikh{\pi_{i-1}} +
        \parikh{\pi_{i-1}'}} \geq 1$)}.
  \end{align*}
  By the above, we conclude that $\lim_{i \to \infty} g(\vec{x}_i) =
  \infty$. The proof for function $h$ is symmetric.

  \medskip\noindent\emph{Case $v_4 = 0$}. In this case, $\mathcal{S}'$
  amounts to:
  \begin{align}
    \vec{v}_1^T\mat{A}_1\mat{M} + \vec{v}_2^T\mat{A}_2\mat{M} &\geq
    \vec{0}^T \label{eq:farkas:1} \\
    \vec{v}_2^T\mat{A}_2\mat{M} + \vec{v}_3^T\mat{M} &\geq \vec{0}^T
    \label{eq:farkas:2} \\
    \vec{v}_3^T\mat{M} &\geq \vec{0}^T \label{eq:farkas:3} \\
    \vec{v}_1^T(\vec{b}_1 - \vec{A}_1\vec{z}) + \vec{v}_2^T(\vec{b}_2
    - \mat{A}_2\vec{z}) &< 0. \label{eq:farkas:4}
  \end{align}
  Let $\lambda \defeq -(\vec{v}_1^T(\vec{b}_1 - \mat{A}_1\vec{z}) +
  \vec{v}_2^T(\vec{b}_2 - \mat{A}_2\vec{z}))$. By~\eqref{eq:farkas:4},
  we have $\lambda > 0$. Let
  \begin{alignat*}{3}
    f'(\vec{x}, \vec{y})
    &\defeq\ && -(\vec{v}_1^T\mat{A}_1(\vec{x} - \vec{z}) +
    \vec{v}_2^T\mat{A}_2(\vec{y} - \vec{z})) + {} \\
    &&&
    \hspace{20pt}
    \vec{v}_2^T\mat{A}_2(\vec{y} - \vec{x}) +
    \vec{v}_3^T(\vec{y}-\vec{x}), \\
    f(\vec{x}, \vec{y})
    &\defeq\ && \vec{v}_3^T(\vec{y} - \vec{x}).
  \end{alignat*}

  Using \eqref{eq:farkas:1} and~\eqref{eq:farkas:2}, it can be shown
  that $f(\vec{x}, \vec{y}) \geq \lambda$. From this,
  we can conclude. Let $g(\vec{x}_i) \defeq f(\vec{x}_0, \vec{x}_i)$
  and $h(\vec{y}_i) \defeq f(\vec{x}_0, \vec{y}_i)$ for all
  $i \in \N$. For all $i \geq 1$, we have
  \begin{alignat*}{4} &&&
  f(\vec{x}_0, \vec{y}_i) \\
    &=\ && \vec{v}_3^T(\vec{y}_i - \vec{x}_0) \\
    &=\ && \vec{v}_3^T \left((\vec{y}_i - \vec{x}_i) + (\vec{x}_i -
    \vec{y}_{i-1}) + (\vec{y}_{i-1} - \vec{x}_0)\right) \\
    &=\ && \vec{v}_3^T(\vec{y}_i - \vec{x}_i) +
    \vec{v}_3^T(\vec{x}_i - \vec{y}_{i-1}) + \vec{v}_3^T(\vec{y}_{i-1} -
    \vec{x}_0)\!\!\! \\
    &=\ && f(\vec{x}_i, \vec{y}_i) + \vec{v}_3^T \mat{M} \parikh{\pi_{i-1}'}
    + f(\vec{x}_0, \vec{y}_{i-1}) \\
    &\geq\ && f(\vec{x}_0, \vec{y}_{i-1}) + f(\vec{x}_i, \vec{y}_i)
    && \text{(by \eqref{eq:farkas:3})} \\
    &\geq\ && f(\vec{x}_0, \vec{y}_{i-1}) + \lambda.
  \end{alignat*}
  By the above, we have $\lim_{i \to \infty} h(\vec{y}_i) = \infty$.
  
  Similarly, for every $i \geq 1$, we have \[f(\vec{x}_0, \vec{x}_i) =
  f(\vec{x}_0, \vec{y}_{i-1}) + \vec{v}_3^T(\vec{x}_i - \vec{y}_{i-1})
  \geq f(\vec{x}_0, \vec{y}_{i-1}),\] which implies $g(\vec{x}_i) \geq
  h(\vec{y}_{i-1})$. So, $\lim_{i \to \infty} g(\vec{x}_i) = \infty$.
\end{IEEEproof}

We now seek to show the following proposition.

\begin{proposition}\label{prop:from:points:to:sch}
  Let $X$, $Y$ and $Z$ be zones. Let $\vec{z}, \vec{z}' \in Z$,
  $\vec{x}_0, \vec{x}', \vec{x}_f \in X \cap Z$ and $\vec{y}_0,
  \vec{y}_f \in Y \cap Z$ be such that
  \begin{itemize}
  \item $\vec{z} \mreach{*}{Z} \vec{z}' \mreach{\pi''}{Z} \vec{x}_0
    \mreach{\pi}{Z} \vec{y}_0 \mreach{\pi'}{Z} \vec{x}'$,

  \item $\vec{x}' \mreach{\rho}{} \vec{x}_f \mreach{\rho'}{} \vec{y}_f
    \mreach{\rho''}{} \vec{x}_f$,

  \item $\supp{\rho} = \supp{\pi} = \supp{\pi'} = \supp{\pi''}$,

  \item $\supp{\rho'} \cup \supp{\rho''} \subseteq \supp{\rho}$ and
    $\norm{\parikh{\rho'} + \parikh{\rho''}} \geq 1$.
  \end{itemize}
  It is the case that $\vec{z} \models \G Z \land \G\F X \land \G\F
  Y$.
\end{proposition}

To prove the above proposition, we build a schedule from the
given schedules. By assumption, there exists a sufficiently
small $\epsilon \in \Rpos$ such that $\parikh{\rho} \geq \epsilon
\cdot (\parikh{\pi} + \parikh{\pi'})$. Let $\lambda \defeq 1 - (1 / (1
+ \epsilon))$. For every $n \in \N_{\geq 1}$, let
\begin{align*}
  \vec{x}_n
  &\defeq \vec{y}_{n-1} + \lambda^{n-1} \left(\effect{\pi'} +
  \frac{\effect{\rho} - \epsilon \effect{\pi\pi'}}{1+\epsilon}\right)
       {} \\
  &\hspace{5.25cm} {} + (1 - \lambda^{n-1})\effect{\rho''},\\
  \vec{y}_n &\defeq \vec{x}_n + \lambda^n \effect{\pi} + (1 -
  \lambda^n)\effect{\rho'}.
\end{align*}

\begin{restatable}{proposition}{propLassoSameLine}\label{prop:Lasso:SameLine}
  For every $n \in \N$, it is the case that $\vec{x}_n
  = \lambda^n \vec{x}_0 + (1 - \lambda^n)\vec{x}_f$ and $\vec{y}_n
  = \lambda^n \vec{y}_0 + (1-\lambda^n)\vec{y}_f$.
\end{restatable}

The following proposition proves \Cref{prop:from:points:to:sch}.

\begin{proposition}\label{prop:Lasso:ValidRun}
  It is the case that (1)~$\vec{x}_n \mreach{*}{Z} \vec{y}_n$ and
  (2)~$\vec{y}_n \mreach{*}{Z} \vec{x}_{n+1}$ for all $n \in \N$.
\end{proposition}

\begin{IEEEproof}
  (1)~Recall that $\vec{x}_0 \mreach{\pi}{Z} \vec{y}_0$ and $\vec{x}_f
  \in Z$. Therefore, by
  \Cref{prop:Lasso:SameLine,prop:Zsafe:execOnLine}, we have
  \begin{align}
    &\vec{x}_n = (\lambda^n \vec{x}_0 + (1-\lambda^n)\vec{x}_f) \notag \\
    &\hspace{3cm}
    \mreach{\lambda^n \pi}{Z} (\lambda^n \vec{y}_0 + (1 -
    \lambda^n)\vec{x}_f).\label{eq:req:1}
  \end{align}
  Similarly, by \Cref{prop:Lasso:SameLine,prop:Zsafe:execOnLine},
  we have    
  \begin{align}
    &(\lambda^n \vec{x}_0 + (1-\lambda^n)\vec{y}_f) \notag \\
    &\hspace*{2.5cm}
    \mreach{\lambda^n \pi}{Z} (\lambda^n \vec{y}_0 + (1 -
    \lambda^n)\vec{y}_f) = \vec{y}_n.\label{eq:req:2}
  \end{align}
  By definition, we have
  \begin{align}
    \vec{y}_n &= \vec{x}_n + \lambda^n \effect{\pi} + (1 -
    \lambda^n)\effect{\rho'}.\label{eq:req:3}
  \end{align}
  Altogether, \eqref{eq:req:1}--\eqref{eq:req:3} yield
  \[
  \vec{x}_n \mreach{\lambda^n \pi\, (1 - \lambda^n)\rho'}{} \vec{y}_n,\,
  \vec{x}_n \mreach{\lambda^n \pi}{Z} {}
  \text{ and }{}
  \mreach{\lambda^n \pi}{Z} \vec{y}_n.
  \]
  As $\supp{\rho'} \subseteq \supp{\pi}$, we have $\supp{\lambda^n
    \pi} = \supp{\lambda^n \pi (1 - \lambda^n)\rho'}$. So, by
  \Cref{prop:Zsafe:LineConstraints}, we conclude that $\vec{x}_n
  \mreach{*}{Z} \vec{y}_n$.

  \medskip\noindent(2)~By
  \Cref{prop:Lasso:SameLine,prop:Zsafe:execOnLine}, we have
  \begin{align}
    \vec{y}_n = \lambda^n \vec{y}_0 + (1-\lambda^n)\vec{y}_f
    \mreach{\lambda^n \pi'}{Z} (\lambda^n \vec{x}' + (1 -
    \lambda^n)\vec{y}_f).\label{eq:req:1:b}
  \end{align}
  Similarly, by \Cref{prop:Lasso:SameLine,prop:Zsafe:execOnLine}, we
  have
  \begin{align}
    (\lambda^{n+1} \vec{z}' + (1 - \lambda^{n+1})\vec{x}_f)
    \mreach{\lambda^{n+1} \pi''}{Z} \vec{x}_{n+1}.\label{eq:req:2:b}
  \end{align}
  Let $\pi_n$ be any finite schedule such that $\parikh{\pi}_n =
  \parikh{\rho} - \epsilon(\parikh{\pi} + \parikh{\pi'})$. We
  have
  \begin{align}
    \vec{y}_n \mreach{\lambda^n\pi'\, (\lambda^n / (1 +
      \epsilon))\pi_n\, (1 - \lambda^n)\rho''}{}
    \vec{x}_{n+1}.\label{eq:req:3:b}
  \end{align}
  By \eqref{eq:req:1:b}--\eqref{eq:req:3:b}, $\supp{\rho''} \subseteq
  \supp{\rho} = \supp{\pi} = \supp{\pi'} = \supp{\pi''}$ and
  \Cref{prop:Zsafe:LineConstraints}, we obtain $\vec{y}_n
  \mreach{*}{Z} \vec{x}_{n+1}$.
\end{IEEEproof}

We may now conclude this subsection by building a suitable first-order
formula. Let $\vec{x} \preach{\vec{\lambda}}{Z} \vec{y}$ be a
shorthand for formula $\varphi_Z(\vec{x}, \vec{\lambda}, \vec{y})$
from \Cref{ssec:transu}, and let $\vec{x} \preach{\vec{*}}{Z} \vec{y}$
stand for $\exists \vec{\lambda} \geq \vec{0} : \vec{x}
\preach{\vec{\lambda}}{Z} \vec{y}$. Let $\M = \{\vec{m}_1, \ldots,
\vec{m}_n\}$. We define $\varphi_{\G Z \land \G\F X \land \G\F
  Y}(\vec{z})$ by
\begin{align}
  &\hspace*{0px}
  \exists \vec{z}' \in Z; \vec{x}_0, \vec{x}', \vec{x}_f \in X \cap Z; \vec{y}_0, \vec{y}_f \in Y \cap Z; \notag \\
  & \parikh{\pi}, \parikh{\pi}', \parikh{\pi}'', \parikh{\rho}, \parikh{\rho}', \parikh{\rho}'' \geq \vec{0} : {} \notag\\
                                                                & \hspace*{0px} \vec{z} \preach{\vec{*}}{Z} \vec{z}' \preach{\parikh{\pi}''}{Z} \vec{x}_0 \preach{\parikh{\pi}}{Z} \vec{y}_0 \preach{\parikh{\pi}'}{Z} \vec{x}' \land {} \label{eq:RestrictedSchedule}\\
                                                                & \hspace*{0px} \vec{x'} \preach{\parikh{\rho}}{} \vec{x}_f \preach{\parikh{\rho'}}{} \vec{y}_f \preach{\parikh{\rho''}}{} \vec{x}_f \land {}  \label{eq:UnrestrictedSchedule}\\
                                                                & \hspace*{0px} \bigwedge_{j \in [1..n]} \theta_j \land \sum_{j \in [1..n]}(\parikh{\rho}'(j) + \parikh{\rho}''(j)) \geq 1, \notag
\end{align}
where
\begin{alignat*}{3}
  \theta_j &=\ && (\parikh{\pi}(j) > 0 \leftrightarrow \parikh{\pi}'(j) > 0 
  \leftrightarrow \parikh{\pi}''(j) > 0 \leftrightarrow \parikh{\rho}(j) > 0) \\
  &\land {} && (\parikh{\rho}'(j) > 0 \rightarrow \parikh{\rho}(j) > 0) \land (\parikh{\rho}''(j) > 0 \rightarrow \parikh{\rho}(j) > 0).
\end{alignat*}

\begin{proposition}
  It is the case that $\vec{z} \models_\M \G Z \land \G\F X \land \G\F
  Y$ iff $\varphi_{\G Z \land \G\F X \land \G\F Y}(\vec{z})$ holds.
\end{proposition}

\begin{IEEEproof}
  $\Leftarrow$) It follows directly from
  \Cref{prop:from:points:to:sch}, since $\varphi_Z$ is the same
  statement written in logic.

  $\Rightarrow$) Let $\pi$ be a non-Zeno infinite schedule such that
  $\sigma \defeq \induced{\pi}{\vec{z}} \models \G Z \land \G\F X
  \land \G\F Y$. Let $\M'$ be the set of modes used infinitely often
  in $\pi$. From $\vec{z}$, we can move along $\sigma$ to a point
  $\vec{z}'$ from which only modes of $\M'$ are used. We can further
  go to a point $\vec{x}_0 \in X \cap Z$ where all modes of $\M'$ are
  used from $\vec{z}'$ to $\vec{x}_0$. Similarly, we can use all modes
  of $\M'$ from $\vec{x}_0$ to some $\vec{y}_0 \in Y \cap Z$, and
  likewise from $\vec{y}_0$ to some $\vec{x}' \in X \cap Z$. The
  resulting sequence satisfies~\eqref{eq:RestrictedSchedule}.

  Note that $\G Z \land \G\F X \land \G\F Y$ is satisfied from
  $\vec{x}'$ in $\M'$. So, we can invoke \Cref{prop:Dual} to
  satisfy~\eqref{eq:UnrestrictedSchedule}. We are done since
  $\supp{\parikh{\pi''}} = \supp{\parikh{\pi}} = \supp{\parikh{\pi'}}
  = \supp{\parikh{\rho}} = \M'$, $\supp{\parikh{\rho'}} \subseteq \M'$
  and $\supp{\parikh{\rho''}} \subseteq M'$.
\end{IEEEproof}

\subsection{Expressing $\G Z$ in first-order logic}\label{ssec:GZ}

Let us now handle the special case $n = 0$ of the previous subsection. We
claim that $\vec{z} \models_\M \G Z$ iff $\varphi_{\G Z}(\vec{z})$
holds, where
\begin{align*}
  \varphi_{\G Z}(z) \defeq\
  &\exists \vec{z'} \in Z, \parikh{\pi}, \parikh{\pi}' \geq \vec{0} :
  \vec{z} \preach{\parikh{\pi}}{Z} \land \vec{z} \preach{\parikh{\pi}'}{} \vec{z'} \land {} \\
  & \mat{A}\vec{z'} \leq \mat{A}\vec{z} 
  \land \bigwedge_{\mathclap{j \in [1..n]}} (\parikh{\pi}'(j) > 0 \rightarrow \parikh{\pi}(j) > 0) \land{} \\
  &
  \sum_{\mathclap{j \in [1..n]}}\parikh{\pi}'(j) \geq 1.
\end{align*}
A simpler proof to the one of \Cref{prop:Dual} yields:

\begin{restatable}{proposition}{propfarkasGZ}\label{prop:farkas:GZ}
  If $\vec{z} \models_\M \G Z$, then there exist $\pi$ and $\vec{z}'$
  such that $\vec{z} \preach{\parikh{\pi}}{} \vec{z'}$,
  $\mat{A}\vec{z'} \leq \mat{A}\vec{z}$ and $\norm{\parikh{\pi}} \geq
  1$.
\end{restatable}

\begin{restatable}{proposition}{propStillSafe}\label{prop:StillSafe}
  Let $\vec{z}, \vec{z'} \in Z$ and $\rho$ be a finite schedule. If
  $\vec{z} \mreach{\rho}{Z}$ and $\mat{A}\vec{z}' \leq \mat{A}\vec{z}$
  then $\vec{z'} \mreach{\rho}{Z}$.
\end{restatable}

\begin{proposition}
  It is the case that $\vec{z} \models_\M \G Z$ iff $\varphi_{\G
    Z}(\vec{z})$.
\end{proposition}

\begin{IEEEproof}
  $\Leftarrow$) Let $\pi$ be the schedule such that $\vec{z}
  \mreach{\pi}{Z} {}$. By \Cref{prop:mode:Zsafe:exec}, we obtain some
  $\beta \in \N_{\geq 1}$ and $\vec{z} \mreach{\rho}{Z} \vec{z}_0$
  with $\vec{z}_0 \mreach{(1 /\beta) \vec{m}}{Z}$ for every $\vec{m} \in
  \supp{\pi}$. Let $\pi'$ be the schedule such that $\vec{z}
  \mreach{\pi'}{} \vec{z}'$.

  Let $\rho' \defeq (1 / \beta \cdot \tweight{\pi'}) \pi'$. For all $i
  \in \N$, let $\vec{z}_{i+1} \defeq \vec{z}_i +
  \effect{\rho'}$. Since $\supp{\pi'} \subseteq \supp{\pi}$, we have
  $\vec{z}_0 \mreach{\rho'}{Z} \vec{z}_1$. Moreover, as $\mat{A}
  \effect{\pi'} = \mat{A}(\vec{z}' - \vec{z}) \leq \vec{0}$, we have
  $\mat{A} \effect{\rho'} \leq \vec{0}$ and hence $\mat{A} \vec{z}_1
  \leq \mat{A} \vec{z}_0$. By \Cref{prop:StillSafe}, we obtain
  $\vec{z}_0 \mreach{\rho'}{Z} \vec{z}_1$. By the same
  reasoning, we conclude that
  \[
  \vec{z} \mreach{\rho}{Z} \vec{z}_0 \mreach{\rho'}{Z} \vec{z}_1
  \mreach{\rho'}{Z} \vec{z}_2 \mreach{\rho'}{Z} \cdots.
  \]

  $\Rightarrow$) Let $\pi''$ be a non-Zeno infinite schedule such that
  $\sigma \defeq \induced{\pi''}{\vec{z}} \models \G Z$. Let $\M'$ be
  the set of modes used in $\pi''$. From $\vec{z}$, we move along
  $\pi''$ to some point where all modes from $\M'$ have been used. We
  take $\pi$ as such a prefix. The other constraints hold by
  \Cref{prop:farkas:GZ}.
\end{IEEEproof}

\subsection{From $\G Z_0 \land \bigwedge_{i=1}^n \G\F Z_i$ to $\G Z \land \G\F X \land \G\F Y$}
\label{subsec:manyzones}

\newcommand{\x}{\vec{x}} \newcommand{\y}{\vec{y}}

\begin{lemma}\label{lem:n:to:two}
  Given a $d$-dimensional MMS $\M$, point $\x \in \R^d$ and zones $Z_0,
    \ldots, Z_n \subseteq \R^d$, it is possible to construct, in
  polynomial time, an $nd$-dimensional MMS $\M'$ and zones $X, Y, Z
    \subseteq \R^{nd}$ such that $\x \models_\M \G Z_0 \land \G\F Z_1
    \land \cdots \land \G\F Z_n$ iff $(\x, \ldots, \x) \models_{\M'} \G
    Z \land \G\F X \land \G\F Y$. Furthermore, zone $Z$ is bounded iff
  zone $Z_0$ is bounded, and zones $\{X, Y\}$ are all bounded iff
  zones $\{Z_1, \ldots, Z_n\}$ are all bounded.
\end{lemma}

\begin{IEEEproof}
  We consider each $\vec{s} \in \R^{nd}$ as a sequence of $n$ points
  from $\R^d$, \ie\ $\vec{s} = (\vec{s}[1], \ldots,
  \vec{s}[n])$. Formally, for all $\vec{s} \in \R^{nd}$ and $i \in
      [1..n]$, let $\vec{s}[i] \defeq (\vec{s}((i-1) \cdot d + 1),
      \ldots, \vec{s}(i \cdot d))$.

  For each $\vec{m} \in \M$, let $\vec{m}_i \in \R^{nd}$ be such that
  $\vec{m}_i[i] = \vec{m}$ and $\vec{m}_i[j] = \vec{0}$ for all $j
  \neq i$. Let,
  \begin{align*}
    \M' &\defeq \{\vec{m}_i : \vec{m} \in \M, i \in [1..n]\}, \\             
    Z   &\defeq Z_0 \times Z_0 \times \cdots \times Z_0, \\
    X   &\defeq Z_1 \times Z_2 \times \cdots \times Z_n, \text{ and} \\ 
    Y   &\defeq \{\vec{y} \in \R^{nd} : \vec{y}[1] = \cdots = \vec{y}[n]
    \in Z_1\}.
  \end{align*}

  Let $\varphi \defeq \G Z_0 \land \G\F Z_1 \land \cdots \land \G\F
  Z_n$ and $\varphi' \defeq \G Z \land \G\F X \land \G\F Y$. It is the
  case that $\x \models_\M \varphi$ iff $(\x, \ldots, \x)
  \models_{\M'} \varphi'$.
\end{IEEEproof}

\subsection{Model checking linear formulas}

We may now prove \Cref{thm:linear:P}, \ie\ show that the
model-checking problem for linear LTL formulas is in P.

\newcommand{\zone}[1]{\mathrm{zone}(#1)}

\begin{IEEEproof}[Proof of \Cref{thm:linear:P}]
  Let $\M$ be a $d$-dimensional MMS, let $\vec{x} \in \R^d$, and let
  $\psi$ be a semi-bounded linear LTL formula. We recursively build a
  formula $\varphi_\psi$ from the polynomial-time first-order logic
  of \Cref{ssec:poly:logic} such that $\vec{x} \models_\M \psi$ iff
  $\varphi_\psi(\vec{x})$.

  For every $A \subseteq \AP$, let $\zone{A}$ denote the zone obtained
  by taking the intersection of the zones from $A$.

  \medskip\noindent\emph{Case $\psi = A \land \psi'$}. We take
  $\varphi_\psi(\vec{x}) \defeq \vec{x} \in \zone{A} \land
  \varphi_{\psi'}(\vec{x})$, which can be expressed as $\zone{A}$
  is represented by a system of inequalities.

  \medskip\noindent\emph{Case $\psi = B \U (B' \land \psi')$}. Note
  that $\vec{x} \models_\M B \U B'$ almost amounts to
  $\vec{x} \mreach{*}{\zone{B}} \vec{y} \in \zone{B'}$, except that,
  contrary to the former, the latter requires $\vec{y}$ to be part of
  $\zone{B}$.

  In our case, we show that
  $\zone{B'} \subseteq \zone{B}$. Recall that $\upc{B} \supseteq \upc{B'}$ by definition of linear
  LTL formulas. Let
  $\vec{z} \in \zone{B'}$. We have $\indic{\vec{z}} \supseteq B'$ and
  $\indic{\vec{z}} \in \upc{B'} \subseteq \upc{B}$. Thus,
  $\indic{\vec{z}} \supseteq B$ and so $\vec{z} \in \zone{B}$.
  Thus, we take
  \begin{align*}
  \varphi_{\psi}(\vec{x}) \defeq
  \exists \vec{\lambda} \geq \vec{0}, \vec{y} \in \zone{B'} :
  \varphi_{\zone{B}}(\vec{x}, \vec{\lambda}, \vec{y}) \land
  \varphi_{\psi'}(\vec{y}),
  \end{align*}
  where $\varphi_Z$ is the formula of \Cref{ssec:transu} with $Z
  \defeq \zone{B}$.

  \medskip\noindent\emph{Case~$\psi = (\G C_0) \land \bigwedge_{i=1}^n
    \G\F C_i$}. Let
  $Z_i \defeq \zone{C_i}$ for all $i \in [0..n]$. If $n = 0$, then we
  use formula $\varphi_{\G Z_0}$ from \Cref{ssec:GZ}. If $n = 1$, then
  we artificially define $Z_2 \defeq Z_1$. So, assume that $n \geq
  2$. By \Cref{lem:n:to:two}, we can construct, in polynomial time, an
  $nd$-dimensional MMS $\M'$ and zones $X, Y, Z \subseteq \R^{nd}$
  such that $\x \models_\M \G Z_0 \land \G\F Z_1 \land \cdots \land
  \G\F Z_n$ iff $(\x, \ldots, \x) \models_{\M'} \G Z \land \G\F X
  \land \G\F Y$. Furthermore, zones $X$ and $Y$ are bounded since
  $\{Z_1, \ldots, Z_n\}$ are all bounded. Thus, we take
  $\varphi_{\psi}(\vec{x}) \defeq \varphi_{\G Z \land \G\F X \land
    \G\F Y}((\vec{x}, \ldots, \vec{x}))$ from \Cref{ssec:lasso} for
  $\M'$.
\end{IEEEproof}

\section{NP-complete fragments}
\label{sec:np}
In this section, we establish the NP-completeness of fragments
\fragmentB{\F, \G, \land}, \fragmentB{\F, \land, \lor} and
\fragmentB{\F, \land}.

\subsection{Membership} \label{subsection:membership}

\begin{theorem}\label{thm:fga:NP}
  The fragment $\text{\fragmentB{\F, \G, \land}}$ belongs to NP.
\end{theorem}

\newcommand{\lin}[1]{\mathrm{lin}(#1)}
\newcommand{\lps}[1]{\mathrm{lps}(#1)}
\newcommand{\sch}[1]{\mathrm{sch}(#1)}

\begin{IEEEproof}
  Let $\M$ be a $d$-dimensional MMS, $\vec{x} \in \R^d$ and
  $\varphi \in \text{\fragmentB{\F, \G, \land}}$. Let $\sch{\M}$
  denote the set of non-Zeno infinite schedules of $\M$. Let
  $\lps{\A_\varphi}$ denote the (finite) set of LPS of $\A_\varphi$
  starting from $q_0$. Let $\lin{S}$ be the linear LTL formula
  obtained from LPS $S$ by \Cref{prop:lps:to:formula}. We have
  \begin{alignat}{3}
     &        &  &
    \vec{x} \models \varphi \notag                          \\
     & \iff\  &  &
    \exists \pi \in \sch{\M} :
    \induced{\pi}{\vec{x}} \models \varphi \notag           \\
     & \iff\  &  &
    \exists \pi \in \sch{\M},
    w \in \trace{\induced{\pi}{\vec{x}}} :
    w \models \varphi \label{eq:lin:1}                      \\
     & \iff\  &  &
    \exists \pi \in \sch{\M},
    w \in \trace{\induced{\pi}{\vec{x}}} : {} \notag \\
    &        &  &
    \hspace{4.5cm}
    w \in \lang{\A_\varphi} \label{eq:lin:2}                \\
     & \iff\  &  &
    \exists \pi \in \sch{\M},
    w \in \trace{\induced{\pi}{\vec{x}}}, \notag            \\
     &        &  &
    \hspace{2cm}
    S \in \lps{\A_\varphi} :
    w \models \lin{S} \label{eq:lin:3}                      \\
    & \iff\  &  &
    \exists \pi \in \sch{\M},
    S \in \lps{\A_\varphi} : {} \notag \\
    &&& \hspace{2.825cm}
    \induced{\pi}{\vec{x}} \models \lin{S} \label{eq:lin:4} \\
     & \iff\  &  &
    \exists S \in \lps{\A_\varphi} :
    \vec{x} \models \lin{S}, \notag
  \end{alignat}
  where
  \begin{itemize}
  \item \eqref{eq:lin:1} follows from
    \Cref{prop:has:trace,prop:ltl:real:discrete},

  \item \eqref{eq:lin:2} follows from
    \Cref{prop:equiv:automata:formula},
          
  \item \eqref{eq:lin:3} follows from \Cref{prop:lps:to:formula} and
    from the fact that $\lang{\A_\varphi} = \bigcup_{S \in
      \lps{\A_\varphi}} \lang{S}$,

  \item \eqref{eq:lin:4} follows from
    \Cref{prop:has:trace,prop:ltl:real:discrete}.
  \end{itemize}
  Thus, to check $\vec{x} \models \varphi$, we can nondeterministically
  construct a linear path scheme $S$ of $\A_\varphi$, which is of
  linear size by \Cref{prop:automata:nocycle}, and the linear formula
  $\psi = \lin{S}$, and check whether $\vec{x} \models \psi$ in
  polynomial time by \Cref{thm:linear:P}.
\end{IEEEproof}

\begin{theorem}\label{thm:fao:NP}
  The fragment $\text{\fragmentB{\F, \land, \lor}}$ belongs to NP.
\end{theorem}

\begin{IEEEproof}
  We give a nondeterministic polynomial-time procedure which, given an
  MMS $\M$, $\vec{x}$ and $\varphi \in \text{\fragmentB{\F, \land,
      \lor}}$, decides $\vec{x} \models_\M \varphi$. For each
  disjunction $\varphi_1 \lor \varphi_2$ of $\varphi$, we
  nondeterministically replace $\varphi_1 \lor \varphi_2$ with either
  $\varphi_1$ or $\varphi_2$. Let $\varphi'$ denote the resulting
  formula. A simple induction shows that $\vec{x} \models_\M \varphi$
  iff $\vec{x} \models_\M \varphi'$ for some such formula $\varphi'$. Moreover, $\varphi'$ has no
  disjunctions and so $\varphi' \in \text{\fragmentB{\F, \land}}
  \subseteq \text{\fragmentB{\F, \G, \land}}$. Therefore, deciding
  whether $\vec{x} \models_\M \varphi'$ can be done in NP
  by~\Cref{thm:fga:NP}.
\end{IEEEproof}

\subsection{NP-hardness}

\begin{theorem}\label{thm:fa:NPhard}
  Fragment \fragmentB{\F, \land} is strongly NP-hard.
\end{theorem}

\begin{IEEEproof}
  We reduce from the rational variant of SUBSET-SUM~\cite{Woj18},
  which asks, given $S \subseteq \Q$ and $t \in \Q$, whether some
  subset $V \subseteq S$ satisfies $\sum_{v \in V} v = t$. Given an
  instance where $S = \{s_1, \ldots, s_n\}$, we give a $(4n +
  1)$-dimensional MMS $\M$ and a formula $\varphi \in
  \text{\fragmentB{F, \land}}$ such that $\vec{0} \models_\M \varphi$
  holds iff there is a solution for $(S, t)$.

  \newcommand{\cstar}{c^*}

  A simple faulty approach is as follows. For each $s_i \in S$, we
  could associate the modes $\vec{y}_i = (0, \ldots 0, 1, 0, \ldots,
  0, s_i)$ and $\vec{n}_i = (0, \ldots 0, 1, 0, \ldots, 0, 0)$,
  where ``$1$'' appears in dimension $i$. The goal would be to sum the
  modes in order to obtain $(1, \ldots, 1, t)$. However, this is too
  naive. For example, consider $S = \{8, 9\}$ and $t = 4$. By taking
  $(0.5, \vec{y}_1)\, (0.5, \vec{n}_1)\, (1, \vec{n}_2)$, we obtain
  $(1, 1, 4)$ even though $4$ cannot be obtained from $S$. We need a
  mechanism to ensure that, for each $i$, either $\vec{y}_i$ or
  $\vec{n}_i$ is used by exactly one unit. For this reason, we will
  introduce the additional modes $\overline{\vec{y}}_i$ and
  $\overline{\vec{n}}_i$, and zones $Y_i$, $N_i$ and $C_i$. We will
  require that $Y_i$, $N_i$ and $C_i$ are all reached. As (partially)
  depicted in \Cref{fig:np:hard}, the only way to do so will be to
  either use schedule $(1, \vec{y}_i)\, (1, \overline{\vec{y}}_i)$ or
  schedule $(1, \vec{n}_i)\, (1, \overline{\vec{n}}_i)$. Moreover, the
  first zone reached, among $Y_i$ and $N_i$, will determine whether
  $s_i \in S$ has been used.

  \paragraph*{Definition of $\M$ and $\varphi$}

  Let us now proceed. We will refer to the first $4n$ dimensions as
  $c_{i,1}, c_{i,2}, c_{i,3}, c_{i,4}$ for every $i \in [1..n]$, and
  to the remaining dimension as $\cstar$. Intuitively, at the end of a
  satisfying execution, $\cstar$ will store the number $t$, which was
  derived by summing up elements from $S$. The other dimensions ensure
  that each element of $S$ is added to $\cstar$ by a factor of $1$ or
  $0$, \ie\ neither partially nor more than once.

  For all $i \in [1..n]$, modes $\{\vec{y}_i, \vec{n}_i,
    \overline{\vec{y}}_i, \overline{\vec{n}}_i\}$ are defined by:
  \[
    \begin{array}{lp{2pt}cccc}
      \toprule
      j                       &                         & \vec{y}_i(j) & \vec{n}_i(j) &
      \overline{\vec{y}}_i(j) & \overline{\vec{n}}_i(j)                                        \\
      \midrule
      c_{i, 1}                &                         & 0.5          & -0.5         & -1 & 1 \\
      c_{i, 2}                &                         & 1            & 1            & 1  & 1 \\
      c_{i, 3}                &                         & 1            & 1            & 0  & 0 \\
      c_{i, 4}                &                         & 0            & 0            & 1  & 1 \\
      \cstar                  &                         & s_i          & 0            & 0  & 0 \\
      \text{else}             &                         & 0            & 0            & 0  & 0 \\
      \bottomrule
    \end{array}
  \]

  \newcommand{\xc}[1]{c_{i,#1}}

  Let $\gamma \defeq \max(2, |t|, n \cdot \max(|s_1|, \ldots,
      |s_n|))$. For every $i \in [1..n]$, we define zones $Y_i$, $N_i$ and $C_i$ by the constraints:
  \[
    \begin{array}{lp{2pt}llll}
      \toprule
                  &                       & Y_i                   & N_i                   & C_i             \\
      \midrule
      c_{i, 1}    &                       & = 0.5                 & = -0.5                & \in [-0.5, 0.5] \\
      c_{i, 2}    &                       & \in [1, 2]            & \in [1, 2]            & = 2             \\
      c_{i, 3}    &                       & = 1                   & = 1                   & = 1             \\
      c_{i, 4}    &                       & \in [0, 1]            & \in [0, 1]            & = 1             \\
      \text{else} &                       & \in [-\gamma, \gamma] & \in [-\gamma, \gamma]
                  & \in [-\gamma, \gamma]                                                                   \\
      \bottomrule
    \end{array}
  \]

  Let $T$ be the zone $T \defeq \{\vec{x} \in C_1 \cap \cdots \cap C_n
    : \vec{x}(\cstar) = t\}$. We define $\varphi \defeq \F T \wedge
    \bigwedge_{i=1}^n (\F Y_i \wedge \F N_i)$.

  Intuitively, the first zone that is reached among $Y_i$ and $N_i$
  indicates whether number $s_i$ is used in the solution to the
  SUBSET-SUM instance. Mode $\vec{y}_i$ can be used to reach $Y_i$
  first, and likewise with $\vec{n}_i$ for $N_i$. Mode
  $\overline{\vec{y}}_i$ can be used to go from $Y_i$ to $N_i$; and
  likewise for $\overline{\vec{n}}_i$ for $N_i$ to $Y_i$. See
  \Cref{fig:np:hard}.

  \begin{figure}
    \begin{center}
      \begin{tikzpicture}[scale=1.1, transform shape]
  \draw[step=0.5, black, opacity=0.25, thin]
  (-0.75, -0.25) grid (2.25, 2.25);
  
  \node[inner sep=0] (zero) at (0,0) {};
  \node[inner sep=0] (y) at (0.5, 1) {};
  \node[inner sep=0] (n) at (-0.5, 1) {};
  
  \foreach \i in {0,...,2} {
    \node[font=\small, opacity=0.5] at (\i, -0.5) {\i};
    \node[font=\small, opacity=0.5] at (-1.15, \i) {\i};
  }
  
  \draw[->, very thick] (zero) edge node[right] {$\vec{y}_i$} (y);
  \draw[->, very thick, densely dotted] (zero) edge node[left]  {$\vec{n}_i$} (n);
  
  \node[inner sep=0] (yy) at (-0.5, 2) {};
  \node[inner sep=0] (nn) at (0.5, 2) {};
  
  \draw[-, colA, opacity=0.5, line width=3pt]   (0.5, 1) --   (0.5, 2);
  \draw[-, colB, opacity=0.5, line width=3pt] (-0.55, 2) --  (0.55, 2);
  \draw[-, colC, opacity=0.5, line width=3pt]  (-0.5, 1) --  (-0.5, 2);
  
  \node[colA!70!black, xshift=2pt,  yshift=6pt] at  (0.5, 2) {$Y_i$};
  \node[colB!60!black, xshift=0pt,  yshift=6pt] at    (0, 2) {$C_i$};
  \node[colC!70!black, xshift=-2pt, yshift=6pt] at (-0.5, 2) {$N_i$};
  
  \draw[->, very thick] (y) edge node[near end, left, xshift=-6pt] {
    $\overline{\vec{y}}_i$
  } (yy);

  \draw[->, very thick, densely dotted] (n) edge node[near end, right, xshift=7pt] {
    $\overline{\vec{n}}_i$
  } (nn);
\end{tikzpicture}
    \end{center}
    \caption{Schedules $(1, \vec{y}_i)\, (1, \overline{\vec{y}}_i)$
      and $(1, \vec{n}_i)\, (1, \overline{\vec{n}}_i)$, where the $x$
      and $y$ axes respectively correspond to $c_{i,1}$ and
      $c_{i,2}$.}
    \label{fig:np:hard}
  \end{figure}
  
  \paragraph*{Correctness} The proof appears in the full version.
\end{IEEEproof}

\section{Undecidable fragments}
\label{sec:undecidable}
\newcommand{\comp}[1]{\vec{\sim}_{#1}}
\newcommand{\guard}[1]{\vec{g}_{#1}}

In this section, we show that \fragmentB{\U} and \fragmentB{\G, \lor}
are undecidable, by reducing from the reachability problem for Petri
nets with inhibitor arcs (\ie\ zero-tests). A \emph{Petri net with
  inhibitor arcs} is a tuple $\pn = (P, T, {\comp{}}, {\effect{}})$
where
\begin{itemize}
\item $P$ is a finite set of elements called \emph{places},

\item $T$ is a disjoint finite set of elements called
  \emph{transitions},

\item $\comp{} \colon T \to \{\geq, =\}^P$, and

\item $\effect{} \colon T \to \Z^P$.
\end{itemize}

A transition $t$ is \emph{enabled} in $\vec{x} \in \N^P$ if $\vec{x}
\comp{t} \vec{0}$ and $\vec{x} + \effect{t} \geq \vec{0}$. If it is
enabled, then its \emph{firing} leads to $\vec{x}' \defeq \vec{x} +
\effect{t}$, denoted $\vec{x} \trans{t} \vec{x}'$. We write $\vec{x}
\trans{} \vec{x}'$ if $\vec{x} \trans{t} \vec{x}'$ for some $t$. We
define ${\trans{+}}$ as the transitive closure of ${\trans{}}$, and
${\trans{*}}$ as the reflexive closure of ${\trans{+}}$. The
\emph{reachability problem} asks, given a Petri nets with inhibitor
arcs $\pn$, and $\xsrc, \xtgt$, whether $\xsrc \trans{+} \xtgt$ in
$\pn$. This problem is undecidable, \eg see \cite{Rei08}.

\subsection{From Petri nets with inhibitor arcs to MMS}

In this subsection, we will prove the following proposition through a
series of intermediate propositions:

\begin{proposition}\label{prop:pni:mms}
  Given a Petri net with inhibitor arcs $\pn$ and $\xsrc, \xtgt$, it
  is possible to compute an MMS $\M$, two points $\vec{x}, \vec{x}'$,
  and a finite set of bounded zones $\AP$ such that
  \begin{enumerate}
  \item $\xsrc \trans{+} \xtgt$ in $\pn$ iff $\vec{x} \mreach{*}{\AP}
    \vec{x}'$ in $\M$, and

  \item no infinite non-Zeno schedule $\pi$ satisfies $\vec{x}
    \mreach{\pi}{\AP}$ in $\M$.\label{itm:no:inf:schedule}
  \end{enumerate}
\end{proposition}

Let $\pn = (P, T, {\comp{}}, {\effect{}})$ be a Petri net with
inhibitor arcs. We define a $(|P| + 3|T|)$-dimensional MMS $\M$
together with zones $\AP \defeq \bigcup_{t \in T} \{A_t, A_t', B_t,
B_t', C_t, C_t'\}$. We associate $|P|$ dimensions to $P$, which we
collectively denote $\vec{p}$. Each transition $t \in T$ is associated
to dimensions $\{t_A, t_B, t_C\}$.

Each transition $t \in T$ gives rise to modes $\{\vec{a}_t, \vec{b}_t,
\vec{c}_t\}$. Informally, these three modes are respectively used to
``request the firing of $t$'', ``fire $t$'' and ``release the control
on $t$''. For every $s \neq t$ and $I \in \{A, B, C\}$, we have
$\vec{a}_t(s_I) = \vec{b}_t(s_I) = \vec{c}_t(s_I) \defeq 0$. The rest
of the values are defined as follows:
\[
\begin{array}{lp{2pt}rrr}
  \toprule
  j        && \vec{a}_t(j) & \vec{b}_t(j) & \vec{c}_t(j) \\
  \midrule
  \vec{p}  && \vec{0}      & \effect{t}   & \vec{0}   \\[0pt]
  \midrule \\[-12pt]
  t_A      && -1           &  0           &  1        \\
  t_B      &&  1           & -1           &  0        \\
  t_C      &&  0           &  1           & -1        \\
  \bottomrule
\end{array}
\]

The six zones associated to $t \in T$ are defined by these
constraints, where $s$ stands for ``any transition $s \neq t$'':
\[
\begin{array}{lp{2pt}rrp{2pt}rrp{2pt}rr}
  \toprule
  && A_t & A_t' && B_t & B_t' && C_t & C_t' \\
  \midrule
  \vec{p} && \geq \vec{0} &  \geq \vec{0} && \comp{t} \vec{0}
           & \geq \vec{0} && \geq \vec{0} &      \geq \vec{0}  \\[0pt]
  \midrule                                                     \\[-12pt]
  t_A && \geq 0 & \geq 0 &&    = 0 &    = 0 &&    = 0 & \geq 0 \\
  t_B &&    = 0 & \geq 0 && \geq 0 & \geq 0 &&    = 0 &    = 0 \\
  t_C &&    = 0 &    = 0 &&    = 0 & \geq 0 && \geq 0 & \geq 0 \\[0pt]
  \midrule                                                     \\[-12pt]
  s_A && \geq 0 & \geq 0 && \geq 0 & \geq 0 && \geq 0 & \geq 0 \\
  s_B &&    = 0 &    = 0 &&    = 0 &    = 0 &&    = 0 &    = 0 \\
  s_C &&    = 0 &    = 0 &&    = 0 &    = 0 &&    = 0 &    = 0 \\
  \bottomrule
\end{array}
\]

Since $A_t = A_s$ for all $s, t \in T$, we simply call this zone
$A$. Informally, the MMS operates as follows:
\begin{itemize}
\item from $A$, we can take a mode $\vec{a}_t$, must go through
  $A_t'$, and end up in $B_t$ after maximally taking $\vec{a}_t$;

\item in $B_t$, we test whether $\vec{p} \comp{t} \vec{0}$ through the
  constraints;

\item from $B_t$, we must take mode $\vec{b}_t$, go through
  $B_t'$ and end up in $C_t$ after maximally taking $\vec{b}_t$ (adding $\effect{t}$ to $\vec{p}$);

\item from $C_t$, we must take mode $\vec{c}_t$, go through
  $C_t'$, and end up in $A$ after maximally taking $\vec{c}_t$.
\end{itemize}

\noindent More formally, the following holds.

\begin{restatable}{lemma}{lemMmsAbc}\label{lem:mms:ABC}
  Let $\vec{x}_A, \vec{x}_A' \in A$ and let $\pi$ be a finite schedule
  such that $\vec{x}_A(t_A) = 1$, $|\pi| > 0$, and $\vec{x}_A
  \mreach{\pi}{\AP} \vec{x}_A'$ holds with no intermediate points in
  $A$, \ie\ $\induced{\pi}{\vec{x}_A}(\tau) \in A$ iff $\tau \in \{0,
  \tweight{\pi}\}$. It is the case that $\pi \equiv \vec{a}_t
  \vec{b}_t \vec{c}_t$ and there exist $\vec{x}_B \in B_t$, $\vec{x}_C
  \in C_t$ such that
  \[
  \vec{x}_A \mreach{\vec{a}_t}{A_t'}
  \vec{x}_B \mreach{\vec{b}_t}{B_t'}
  \vec{x}_C \mreach{\vec{c}_t}{C_t'}
  \vec{x}_A'.
  \]
\end{restatable}

A zone $Z$ is \emph{closed under scaling} if $\lambda \vec{z} \in Z$
holds for all $\lambda \in \Rpos$ and $\vec{z} \in Z$. A set of zones
$\AP$ is closed under scaling if each $Z \in \AP$ is closed under
scaling. From \Cref{lem:mms:ABC}, the following can be shown:

\begin{restatable}{proposition}{propPniMmsUnbounded}\label{prop:pni:mms:unbounded}
  Given a Petri net with inhibitor arcs $\pn$ and $\xsrc, \xtgt$, it
  is possible to compute an MMS $\M$, points $\vec{x}, \vec{x}'$,
  and a finite set of zones $\AP$ closed under scaling, such that
  \begin{enumerate}
  \item $\xsrc \trans{+} \xtgt$ in $\pn$ iff $\vec{x}
    \mreach{\pi}{\AP} \vec{x}'$ in $\M$ for some finite schedule $\pi$
    with $\tweight{\pi} \geq 1$,

  \item $\vec{0} \not\mreach{+}{\AP} \vec{0}$ in
    $\M$.\label{itm:non:zero:cycle}
  \end{enumerate}
\end{restatable}

\begin{proposition}\label{prop:pni:mms:existential}
  Given a Petri net with inhibitor arcs $\pn$ and $\xsrc, \xtgt$, one
  can compute an MMS $\M$, points $\vec{x}, \vec{x}'$, and a finite
  set of bounded zones $\AP$ such that $\xsrc \trans{+} \xtgt$ in
  $\pn$ iff $\lambda \vec{x} \mreach{\pi}{\AP} \lambda \vec{x}'$ in
  $\M$ for some $\lambda \in [0, 1]$ and $\pi$ with $\tweight{\pi} =
  1$.
\end{proposition}

\begin{IEEEproof}
  Let $\M$, $\vec{x}$, $\vec{x}'$ and $\AP$ be given by
  \Cref{prop:pni:mms:unbounded}. Let $\gamma \defeq \norm{\vec{x}} +
  \norm{\M}$ and let $d$ denote the dimension of $\M$. We show the
  proposition with $\AP' \defeq \{Z \cap [-\gamma, \gamma]^d : Z \in
  \AP\}$.
  
  $\Rightarrow$) As $\xsrc \trans{+} \xtgt$ in $\pn$,
  \Cref{prop:pni:mms:unbounded} yields $\vec{x} \mreach{\pi}{\AP}
  \vec{x}'$ in $\M$ and $\tweight{\pi} \geq 1$ for some finite schedule
  $\pi$. As $\AP$ is closed under scaling, we have $\lambda \vec{x}
  \mreach{\lambda \pi}{\AP} \lambda \vec{x}'$ in $\M$ for any $\lambda
  \in \Rpos$.  By picking $\lambda \defeq 1 / \tweight{\pi}$, each
  point $\vec{y}$ along the resulting execution satisfies
  $\norm{\vec{y}} \leq \gamma$, and hence $\vec{y} \in Z'$ for some
  $Z' \in \AP'$. Further, $\tweight{\lambda \pi} = 1$.

  $\Leftarrow$) Let $\lambda \vec{x} \mreach{\pi}{\AP'} \lambda
  \vec{x}'$ with $\tweight{\pi} = 1$. As $\pi$ is nonempty, and
  $\vec{0} \not\mreach{+}{\AP} \vec{0}$ by
  \Cref{prop:pni:mms:unbounded}, we have $\lambda > 0$.

  Let $Z \in \AP$. If $\vec{y} \in Z \cap [-\gamma, \gamma]^d$, then
  in particular $\vec{y} \in Z$. Since $Z$ is closed under scaling, we
  have $\lambda \vec{y} \in Z$. Thus,
  \[
  \vec{x} =
  (1 / \lambda) \cdot \lambda \vec{x} \mreach{\frac{1}{\lambda} \pi}{\AP}
  (1 / \lambda) \cdot \lambda \vec{x}'
  = \vec{x}' \text{ in } \M.
  \]
  By \Cref{prop:pni:mms:unbounded}, this implies that $\xsrc \trans{+}
  \xtgt$ in $\pn$.
\end{IEEEproof}

We are now ready to prove \Cref{prop:pni:mms}.

\begin{IEEEproof}[Proof of \Cref{prop:pni:mms}]
  Let $\M$, $\vec{x}$, $\vec{x}'$ and $\AP$ be given by
  \Cref{prop:pni:mms:existential}. Let us define an MMS $\M'$ and a
  set of zones $\AP'$. MMS $\M'$ has the same dimensions as $\M$,
  plus four more: $\{\top, \vdash, \bot, \star\}$. The modes of
  $\M'$ are $\{\vec{a}_\top, \overline{\vec{a}}_\top\} \cup
  \{\vec{m}_\vdash : \vec{m} \in \M\} \cup \{\vec{a}_\bot,
  \overline{\vec{a}}_\bot\}$. They are defined as follows:
  \[
  \begin{array}{lp{1pt}rrp{1pt}rp{1pt}rr}
    \toprule
    j              && \vec{a}_\top(j)   & \overline{\vec{a}}_\top(j)
                   && \vec{m}_\vdash(j)
                   && \vec{a}_\bot(j)   & \overline{\vec{a}}_\bot(j) \\
    \midrule
    \top           && -1 &  -1 && 0 &&  0 & 0 \\
    \vdash         &&  0 &   0 && 1 &&  0 & 0 \\
    \bot           &&  0 &   0 && 0 &&  1 & 1 \\
    \star          &&  1 &   0 && 0 && -1 & 0 \\
    \textbf{rest}  &&  \vec{x}  & \vec{0} && \vec{m}
                   && -\vec{x}' & \vec{0} \\
    \bottomrule
  \end{array}
  \]

  The set $\AP'$ contains two new zones, plus each zone from $\AP$
  extended with the constraint $\top = \bot = 0$ and $\vdash, \star
  \in [0, 1]$. Since $\AP$ consists of bounded zones, we can extract
  $\gamma \in \N$ such that each dimension must remain within
  $[-\gamma, \gamma]$. We add zones $\{A_\top, A_\bot\}$ defined by
  these constraints:
  \[
  \begin{array}{lp{2pt}lll}
    \toprule
                && A_\top                & A_\bot     \\
    \midrule
    \top        && \in [0, 1]            & = 0        \\
    \vdash      && = 0                   & = 1        \\
    \bot        && = 0                   & \in [0, 1] \\
    \star       && \in [0, 1]            & \in [0, 1] \\
    \text{else} && \in [-\gamma, \gamma] & \in [-\gamma, \gamma] \\
    \bottomrule
  \end{array}
  \]

  Informally, the MMS operates as follows:
  \begin{itemize}
  \item from $A_\top$, modes $\vec{a}_\top$ and
    $\overline{\vec{a}_\top}$ empty $\top$ to generate $\lambda
    \vec{x}$, and keep a copy of $\lambda \in [0, 1]$ in $\star$;

  \item modes of $\M$ are used until the time reaches ${\vdash} = 1$;

  \item from $A_\bot$, modes $\vec{a}_\bot$ and
    $\overline{\vec{a}_\bot}$ increase $\bot$ to $1$, in order to
    consume $\lambda \vec{x}'$, using $\star$ to infer $\lambda$.
  \end{itemize}

  Formally, by definition of modes and zones, there exist $\lambda \in
  [0, 1]$ and $\lambda \vec{x} \mreach{\pi}{\AP} \lambda \vec{x}'$ in
  $\M$ such that $\tweight{\pi} = 1$ iff there exist finite schedules
  $\rho_\top$ and $\rho\bot$, respectively using only modes
  $\{\vec{a}_\top, \overline{\vec{a}_\top}\}$ and $\{\vec{a}_\bot,
  \overline{\vec{a}_\bot}\}$, such that
  \[
  \begin{pmatrix}
    1 \\
    0 \\
    0 \\
    0 \\
    \vec{0}
  \end{pmatrix}
  \mreach{\rho_\top}{A_\top}
  \begin{pmatrix}
    0 \\
    0 \\
    0 \\
    \lambda \\
    \lambda \vec{x}
  \end{pmatrix}
  \mreach{\pi_\vdash}{\AP' \setminus \{A_\top, A_\bot\}}
  \begin{pmatrix}
    0 \\
    1 \\
    0 \\
    \lambda \\
    \lambda \vec{x}'
  \end{pmatrix}
  \mreach{\rho_\bot}{A_\bot}
  \begin{pmatrix}
    0 \\
    1 \\
    1 \\
    0 \\
    \vec{0}
  \end{pmatrix}
  \]
  in $\M'$.
  The above holds iff $(1, 0, 0, 0, \vec{0}) \mreach{*}{\AP'} (0, 1,
  1, 0, \vec{0})$ in $\M'$ since zones enforce this ordering.

  It remains to show \Cref{itm:no:inf:schedule}. For the sake of
  contradiction, suppose there exists an infinite non-Zeno schedule
  $\pi$ such that $\vec{x} \mreach{\pi}{\AP}$. All zones of $\AP$
  enforce $\top, \bot, \vdash, \star \in [0, 1]$. Thus, we obtain a
  contradiction since:
  \begin{itemize}
  \item If $\weight{\pi}{\vec{a}_\top} +
    \weight{\pi}{\overline{\vec{a}}_\top} = \infty$, then $\top$
    drops below $0$;
    
  \item If $\sum_{\vec{m} \in \M} \weight{\pi}{\vec{m}_\vdash} =
    \infty$, then $\vdash$ exceeds $1$;

  \item If $\weight{\pi}{\vec{a}_\bot} +
    \weight{\pi}{\overline{\vec{a}}_\bot} = \infty$, then $\bot$
    exceeds $1$. \hfill\qedhere
  \end{itemize}
\end{IEEEproof}

\subsection{Undecidability}

We prove the undecidability of the fragments \fragmentB{\U} and
\fragmentB{\G, \lor} using \Cref{prop:pni:mms}.

\begin{restatable}{lemma}{lemOrU}\label{lem:or:U}
  Given $\psi_1, \ldots, \psi_n, \varphi \in
  \text{\fragmentB{\U}}$, it is possible to compute a formula from
  \fragmentB{\U}\ that is equivalent to formula $(\psi_1 \lor \cdots
  \lor \psi_n) \U \varphi$.
\end{restatable}

\begin{restatable}{theorem}{thmUGoUndec}\label{thm:u:go:undec}
  \fragmentB{\U} and \fragmentB{\G, \lor} are undecidable.
\end{restatable}

\begin{IEEEproof}
  Let $\pn$ be a Petri net with inhibitor arcs and let $\xsrc,
  \xtgt$. Let $\M$, $\vec{x}$, $\vec{x}'$ and $\AP$ be given by
  \Cref{prop:pni:mms}. Let $X' \defeq \{\vec{x}'\}$ and $\psi \defeq
  (\bigvee_{Z \in \AP} Z) \U X'$. By \Cref{lem:or:U}, we can compute a
  formula $\varphi \in \text{\fragmentB{\U}}$ with $\varphi \equiv
  \psi$. By \Cref{prop:pni:mms}, we have $\xsrc \trans{+} \xtgt$ in
  $\pn$ iff $\vec{x} \mreach{*}{\AP} \vec{x}'$ in $\M$ iff $\vec{x}
  \models_\M \psi$ iff $\vec{x} \models_\M \varphi$.

  The proof for \fragmentB{\G, \lor} is essentially the same, but
  requires an extra ``dummy dimension'' that can be increased and
  decreased once (and only once) $\vec{x}'$ is reached.
\end{IEEEproof}

\section{Conclusion}
\label{sec:conlusion}
We have introduced a linear temporal logic for MMS and established the
complexity of model checking for each syntactic fragments: Each one is
either P-complete, NP-complete or undecidable. This generalizes and
unifies existing work on MMS and continuous vector addition
systems/Petri nets.

Future work includes fully dealing with unbounded zones; allowing for
time constraints on temporal operators; and algorithmically optimizing
objective functions on schedules satisfying a given LTL
specification. It would also be interesting to go from theory to
practice by providing a solver for linear LTL formulas, and more
generally \fragmentB{\F, \G, \land}.

\balance
\bibliographystyle{IEEEtran}
\bibliography{references}

\clearpage
\appendix
\label{sec:appendix}
\section{Missing proofs of \Cref{ssec:ltl}}

Before proving \Cref{prop:has:trace}, which we recall shortly, let us
prove the following technical lemma.

\begin{lemma}\label{lem:between:convex}
  Let $\sigma$ be an execution and let $\tau, \tau' \in \dom \sigma$
  belong to a common interval $I_j$ of $\sigma$. Let $X \subseteq
  \R^d$ be a convex set. If $\tau'' \in [\tau, \tau']$ and
  $\sigma(\tau), \sigma(\tau') \in X$, then $\sigma(\tau'') \in X$.
\end{lemma}

\begin{IEEEproof}
  Let $\lambda \in [0, 1]$ be such that $\tau'' = \lambda \cdot \tau +
  (1 - \lambda) \cdot \tau'$. Let $b \defeq \min I_j$, $c \defeq \max
  I_j - \min I_j$ and $\vec{y} \defeq \vec{x}_{j+1} - \vec{x}_j$. We
  have:
  \begin{alignat*}{4}
    &&& \sigma(\tau'') \\
    &=\ &&
    \vec{x}_j + (\tau'' - b) / c \cdot \vec{y} \\
    &=\ &&
    \vec{x}_j + (\lambda\tau + (1 - \lambda) \tau' - b) /
    c \cdot \vec{y} \\
    &=\ &&
    \lambda \left(\vec{x}_j + (\tau - b) / c \cdot
    \vec{y}\right) + 
    (1 - \lambda)\left(\vec{x}_j + (\tau' - b) /
    c \cdot \vec{y}\right) \\
    &=\ &&
    \lambda \sigma(\tau) + (1 - \lambda) \sigma(\tau') \\
    &\in\ && X. \tag*{\qedhere}
  \end{alignat*}
\end{IEEEproof}

\propHasTrace*

\begin{IEEEproof}
  Let $\sigma = \vec{x}_0 I_0 \vec{x}_1 \cdots$. Let $\tau_0 \defeq
  0$. We construct the rest of the sequence inductively. Let $X_i
  \defeq \{\tau' \in \dom \sigma : \tau' > \tau_{i-1} \text{ and }
  \indic{\tau'} \neq \indic{\tau_{i-1}}\}$. Let us make a case
  distinction.

  \medskip\noindent\emph{$X_i$ is empty}. If $\tau_{i-1} = \sup \dom
  \sigma$, then the process ends. Otherwise, we add $\tau_i \defeq
  \max I_j$, where $j \in \N$ is the maximal index such that
  $\tau_{i-1} \in I_j$.

  \medskip\noindent\emph{$\inf X_i > \tau_{i-1}$}. We add $\tau_i
  \defeq \inf X_i$, as it satisfies $\indic{\sigma(\tau')} =
  \indic{\sigma(\tau_{i-1})}$ for all $\tau' \in [\tau_{i-1}, \tau_i)$.

  \medskip\noindent\emph{$\inf X_i = \tau_{i-1}$}. Let $j$ be the
  maximal index such that $\tau_{i-1} \in I_j$. There is a sequence
  $\alpha_0 > \alpha_1 > \cdots \in X_i \cap I_j$ that converges to
  $\tau_{i-1}$. By \Cref{lem:between:convex} and convexity of zones,
  there exists $k \in \N$ such that $\indic{\sigma(\tau')} =
  \indic{\sigma(\alpha_k)}$ for all $\tau' \in (\tau_{i-1},
  \alpha_k]$. Thus, we add $\tau_i \defeq \alpha_k$ to the sequence.

  \medskip
  It remains to show that $\dom \sigma = [\tau_0, \tau_1] \cup
  [\tau_1, \tau_2] \cup \cdots$. For the sake of contradiction,
  suppose that $\tau_0, \tau_1, \ldots$ is infinite and converges to
  some $\alpha \leq \sup \dom \sigma$. Let $j \in \N$ be the minimal
  index such that $\alpha \in I_j$. There exists $k \in \N$ such that
  $\tau_k, \tau_{k+1}, \ldots \in I_j$. By \Cref{lem:between:convex}
  and convexity of zones, there exists $\ell \geq k$ such that
  $\indic{\sigma(\tau')} = \indic{\sigma(\alpha)}$ holds for all
  $\tau' \in [\tau_\ell, \alpha)$. This means that either $X_{\ell+1}
    = \emptyset$ or $\inf X_{\ell+1} \geq \alpha$. In both cases, this
    means that $\tau_{\ell+1} \geq \alpha$. If $\tau_{\ell+1} >
    \alpha$, then this contradicts $\lim_{i \to \infty} \tau_i =
    \alpha$. If $\tau_{\ell+1} = \alpha$, then we must have
    $\tau_{\ell+1} = \sup \dom \sigma$, which contradicts the
    fact that the sequence is infinite.
\end{IEEEproof}

Before proving \Cref{prop:ltl:real:discrete}, which we recall shortly,
let us prove the following technical lemma.

\begin{lemma}\label{lem:has:min}
  Let $\varphi$ be a negation-free LTL formula, let $\sigma$ be an
  execution with $\dom \sigma = \Rnon$, let $\tau, \tau' \in \Rnon$
  and let $T_\varphi \defeq \{\tau'' \in [\tau, \tau') : \sigma,
    \tau'' \models \varphi\}$. If the set $T_\varphi$ is nonempty,
    then it has a minimum.
\end{lemma}

\begin{IEEEproof}
  We proceed by induction on the structure of $\varphi$.

  \medskip\noindent\emph{Case $\varphi = \true$.} We trivially have
  $\min T_\varphi = \tau$.
  
  \medskip\noindent\emph{Case $\varphi = Z \in \AP$.} Informally, the
  claim follows from the fact that $Z$ is defined as the intersection
  of \emph{closed} half-spaces, and hence $\sigma$ must intersect with
  a face of $Z$.

  Formally, let $\sigma = \vec{x}_0 I_0 \vec{x}_1 \cdots$ and let $Z$
  be defined by the system $\mat{A} \vec{x} \leq \vec{b}$. Let $\alpha
  \defeq \inf T_Z$. If $\sigma(\alpha) \in Z$, then we are done. For
  the sake of contradiction, assume that this is not the case. We have
  $(\mat{A} \cdot \sigma(\alpha))(\ell) > \vec{b}(\ell)$ for some
  $\ell$.

  Let $\alpha_0 > \alpha_1 > \cdots$ be a sequence from $T_Z$ that
  converges to $\alpha$. For every $i \in \N$, let $j_i \in \N$ be the
  last index such that $\alpha_i \in I_{j_i}$. Let $j \in \N$ be the
  last index such that $\alpha \in I_j$. There exists $k \in \N$ such
  that $j_k = j_{k+1} = \cdots = j$. Let $b \defeq \min I_j$, $c
  \defeq \max I_j - \min I_j$ and $\vec{y} \defeq \vec{x}_{j+1} -
  \vec{x}_j$. By definition of an execution, we have
  \[
  \mat{A} \cdot \sigma(\alpha) = \mat{A} \vec{x}_j + \frac{\alpha -
    b}{c} \cdot \mat{A} \vec{y}.
  \]
  Moreover, for every $k' \geq k$, we have
  \[
  \mat{A} \cdot \sigma(\alpha_{k'}) = \mat{A} \cdot \sigma(\alpha) +
  \frac{\alpha_{k'} - \alpha}{c} \cdot \mat{A} \vec{y}.
  \]
  For every $k' \geq k$, we have $\alpha_{k'} > \alpha$ and $(\mat{A}
  \cdot \sigma(\alpha_{k'}))(\ell) \leq \vec{b}(\ell)$. Hence,
  $(\mat{A} \vec{y})(\ell) < 0$. Let $k' \geq k$ be sufficiently large
  so that $\alpha_{k'} - \alpha$ is small enough for the following to
  hold:
  \[
  (\mat{A} \cdot \sigma(\alpha))(\ell) + \frac{\alpha_{k'} -
    \alpha}{c} \cdot (\mat{A} \vec{y})(\ell) > \vec{b}(\ell).
  \]
  We obtain $\sigma(\alpha_{k'})(\ell) > \vec{b}(\ell)$, which is a
  contradiction.

  \medskip\noindent \emph{Case $\varphi = \psi \land \psi'$.} Let
  $\alpha \defeq \inf T_\varphi$, and let
  \begin{align*}
    B &\defeq \{\alpha' \in [\alpha, \tau') : \sigma, \alpha' \models
      \psi\}, \\
    B' &\defeq \{\alpha'  \in [\alpha, \tau') : \sigma, \alpha' \models
      \psi'\}.
  \end{align*}
  As $T_\varphi \neq \emptyset$, both $B$ and $B'$ are nonempty. Thus,
  by induction, $\beta \defeq \min B$ and $\beta' \defeq \min B'$ are
  well-defined. We must have $\inf B = \inf B' = \inf T_\varphi$,
  since $\varphi = \psi \land \psi'$. Thus, $\min B = \min B' =
  \alpha$, which means that $\min T_\varphi = \alpha$.

  \medskip\noindent \emph{Case $\varphi = \psi \lor \psi'$.} Follows
  from $T_\varphi = T_\psi \cup T_{\psi'}$ and induction.

  \medskip\noindent\emph{Case $\varphi = \psi \U \psi'$.} Let $\alpha
  \defeq \inf T_\varphi$ and let
  \[
  T' \defeq \{\alpha' \in [\alpha, \tau') : \sigma, \alpha' \models
    \psi'\},
  \]
  Since $T_\varphi \neq \emptyset$, we must have $T' \neq
  \emptyset$. By induction hypothesis, $\beta \defeq \min T'$ is
  well-defined. If $\beta = \alpha$, then we are done as $\sigma,
  \alpha \models \psi \U \psi'$ and hence $\min T_\varphi = \alpha$.
    
  Otherwise, $\alpha < \beta$. We claim that $\sigma, \gamma \models
  \psi$ for all $\gamma \in (\alpha, \beta)$. By induction on
  $\{\alpha' \in [\alpha, \beta) : \sigma, \alpha' \models \psi\}$,
    the claim implies $\min T_\varphi = \alpha$. It remains to show
    the claim. Let $\gamma \in (\alpha, \beta)$. As $\alpha = \inf
    T_\varphi$, there is $\gamma' \in (\alpha, \gamma)$ such that
    $\sigma, \gamma' \models \psi \U \psi'$. By minimality of $\beta$,
    we must have $\sigma, \delta \models \psi$ for all $\delta \in
    [\gamma', \beta)$. Thus, $\sigma, \gamma' \models \psi \U \psi'$,
      and so $\sigma, \gamma \models \psi \U \psi'$.

  \medskip\noindent \emph{Case $\varphi = \F \psi$.} Follows from $\F
  \psi \equiv \true \U \psi$.

  \medskip\noindent \emph{Case $\varphi = \G \psi$.} Let $\alpha
  \defeq \inf T_\varphi$. It is the case that $\sigma, \beta \models
  \G \psi$ for infinitely many $\beta \in (\alpha, \tau')$ arbitrarily
  closer to $\alpha$. Thus, we have $\sigma, \beta \models \psi$ for
  all $\beta > \alpha$. By induction hypothesis, the set $ \{\beta \in
  [\alpha, \tau') : \sigma, \beta \models \psi\}$ has a minimum, which
    must be $\alpha$. Hence, $\sigma, \alpha \models \G \psi$, which
    means that $\min T_\varphi = \alpha$.
\end{IEEEproof}

\propLtlRealDiscrete*

\begin{IEEEproof}
  Let $\tau_0, \tau_1, \ldots \in \Rnon$ yield $w$. Let $f \colon
  \Rnon \to \N$ be the function that satisfies the following, for all
  $\tau \in [\tau_i, \tau_{i+1})$:
  \[
  f(\tau) \defeq
  \begin{cases}
    i
    & \text{if } \indic{\sigma(\tau)} = \indic{\sigma(\tau_i)}, \\
    i + 1
    & \text{otherwise}.
  \end{cases}
  \]
  By definition of traces, $f$ is non-decreasing. We show that
  $\sigma, \tau \models \varphi$ iff $w, f(\tau) \models \varphi$ by
  induction on the structure of $\varphi$.

  \medskip\noindent\emph{Case $\varphi = Z \in \AP$.} We have $w(i) =
  \indic{\sigma(\tau_i)}$ for all $i \in \N$. Thus, $\sigma, \tau
  \models \varphi$ iff $\sigma(\tau) \in Z$ iff $Z \in w(f(\tau))$ iff
  $w, f(\tau) \models \varphi$.

  \medskip\noindent \emph{Case $\varphi = \psi \land \psi'$.} We have
  $\sigma, \tau \models \varphi$ iff $\sigma, \tau \models \psi \land
  \sigma, \tau \models \psi'$ iff $w, f(\tau) \models
  \psi \land w, f(\tau) \models \psi'$ iff
  $w, \allowbreak f(\tau) \models \varphi$.

  \medskip\noindent \emph{Case $\varphi = \psi \lor \psi'$.} Symmetric
  to $\land$.

  \medskip\noindent\emph{Case $\varphi = \psi \U \psi'$.}
  $\Rightarrow$) Since $\sigma, \tau \models \varphi$, there exists
  $\tau' \geq \tau$ such that $\sigma, \tau' \models \psi'$, and
  $\sigma, \tau'' \models \psi$ for all $\tau'' \in [\tau, \tau')$.
  By induction hypothesis, we have $w, f(\tau') \models \psi'$,
  and $w, f(\tau'') \models \psi$ for all $\tau'' \in [\tau, \tau')$.
  By definition of $f$, we have
  \[
    [f(\tau)..f(\tau') - 1] \subseteq \{f(\tau'') : \tau'' \in
    [\tau, \tau')\}.
  \]
  Thus, $w, j \models \psi$ holds for all $j \in [f(\tau)..f(\tau') -
    1]$. This means that $w, f(\tau) \models \psi \U \psi'$.

  $\Leftarrow$) Since $w, f(\tau) \models \varphi$, there is a minimal
  $i \geq f(\tau)$ such that $w, i \models \psi'$, and $w, j \models
  \psi$ for all $j \in [f(\tau)..i-1]$.

  If $f(\tau) = i$, then we are done by the induction
  hypothesis. Thus, we assume $f(\tau) \neq i$. We claim that $\tau <
  \tau_i$. Indeed, if $\tau_i \leq \tau$ would hold, then, as $f$ is
  non-decreasing, we would have $i = f(\tau_i) \leq f(\tau) \leq i$,
  which contradicts $f(\tau) \neq i$.

  By induction hypothesis, it is the case that $\sigma, \tau_i \models
  \psi'$. Let $\tau' \in [\tau, \tau_i)$. It remains to show that
  $\sigma, \tau' \models \psi$. We have $f(\tau) \leq f(\tau') \leq
  f(\tau_i) = i$ as $f$ is non-decreasing. So, $f(\tau') \in
  [f(\tau)..i]$. If $f(\tau') \leq i - 1$, then we are done by the
  induction hypothesis. Therefore, suppose that $f(\tau') = i$. Let
  \[
  A \defeq \{\alpha \in [\tau_{i-1}, \tau_i) : \sigma, \alpha \models
    \psi'\}.
  \]
  As $\tau' < \tau_i$, the definition of $f$ yields $f(\alpha) = i$
  for all $\alpha \in (\tau_{i-1}, \tau_i)$. So, by induction
  hypothesis, we have $\inf A = \tau_{i-1}$, and hence $\min A =
  \tau_{i-1}$ by \Cref{lem:has:min}. Thus, $\sigma, \tau_{i-1} \models
  \psi'$, and so $w, i - 1 \models \psi'$, which contradicts the
  minimality of $i$.

  \medskip\noindent \emph{Case $\varphi = \F \psi$.} Follows from $\F
  \psi \equiv \true \U \psi$.

  \medskip\noindent \emph{Case $\varphi = \G \psi$.} We have
  \begin{alignat*}{4}
    &&& \sigma, \tau \models \G \psi \\
    &\iff\ && \sigma, \tau' \models \psi \text{ for all } \tau' \geq \tau \\
    &\iff\ && w, f(\tau') \models \psi \text{ for all } \tau' \geq \tau \quad
    && \text{(by ind.\ hyp.)} \\
    &\iff\ && w, i \models \psi \text{ for all } i \geq f(\tau) \quad
    && \text{(by def.\ of $f$)} \\
    &\iff\ && w, f(\tau) \models \G \psi. && \tag*{\qedhere}
  \end{alignat*}  
\end{IEEEproof}

\section{Missing proofs of \Cref{sec:buechi}}

Recall that in the forthcoming proposition, we have
\[
\varphi = \psi \land \bigwedge_{i \in I} \G \varphi_i \land
\bigwedge_{j \in J} \F \varphi_j.
\]

\propEquivFlatFormula*

\begin{IEEEproof}
  We claim that the following equivalences hold:
  \begin{enumerate}
  \item $\G \varphi \equiv \G \psi \land \bigwedge_{i \in I} \G
    \varphi_i \land \bigwedge_{j \in J} \G\F \varphi_j$,
    
  \item $\G\F \varphi \equiv \G\F \psi \land \bigwedge_{i \in I}
    \F\G \varphi_i \land \bigwedge_{j \in J} \G\F \varphi_j$,
    
  \item $\F\G \varphi \equiv \F\G \psi \land \bigwedge_{i \in I}
    \F\G \varphi_i \land \bigwedge_{j \in J} \G\F\varphi_j$.
  \end{enumerate}
   By a routine induction, these equivalences yield $\flat_\G(\varphi)
  \equiv \G \varphi$, $\flat_{\G\F}(\varphi) \equiv \G\F \varphi$,
  $\flat_{\F\G}(\varphi) \equiv \F\G \varphi$ and $\flat(\varphi) \equiv
  \varphi$. It remains to prove the equivalences.
  \begin{enumerate}
  \item It follows from distributivity of $\G$ over $\land$, and
    idempotence of $\G$.
  
  \medskip
  \item
    $\Rightarrow$) Let $w \models \G\F\varphi$. Let $i_0 < i_1 <
    \cdots \in \N$ be such that $w, i_k \models \varphi$ for every $k
    \in \N$. Recall that $\varphi = \psi \land \bigwedge_{i \in I} \G
    \varphi_i \land \bigwedge_{j \in J} \F \varphi_j$. Clearly, we
    have $w \models \G\F \psi$ and $w \models \bigwedge_{j \in J} \G
    \F \varphi_j$. Moreover, since $w, i_0 \models \bigwedge_{i \in I}
    \G \varphi_i$, we have $w \models \bigwedge_{i \in I} \F \G
    \varphi_i$.\medskip

    $\Leftarrow$) Let $w \models \G\F \psi \land \bigwedge_{i \in I}
    \F\G\varphi_i \land \bigwedge_{j \in J} \G\F\varphi_j$. Let $i_0
    \in \N$ satisfy $w, i_0 \models \G\F \psi \land \bigwedge_{i \in
      I} \G\varphi_i \land \bigwedge_{j \in J} \G\F\varphi_j$. There
    exist $i_0 < i_1 < \cdots \in \N$ such that $w, i_k \models \psi
    \land \bigwedge_{i \in I} \G\varphi_i \land \bigwedge_{j \in J} \F
    \varphi_j$ for every $k \geq 1$. Thus, $w \models \G\F \varphi$.

  \medskip
  \item It follows from distributivity of $\G$ and $\F\G$ over
    $\land$, and idempotence of $\G$. \hfill\qedhere
  \end{enumerate}
\end{IEEEproof}

\propFCount*

We prove
\Cref{itm:greater:Foperator,itm:Formulasize:Foperator}. \Cref{itm:bracketoperator:Fcount}
follows easily by definition.

\begin{IEEEproof}[Proof of \Cref{prop:F:count}\eqref{itm:greater:Foperator}]
  We proceed by induction on the structure of the flat formula.

  \medskip\noindent\emph{True, atomic propositions and operator
    $\G$}. It follows trivially from $\mathfrak{U}(\varphi) \setminus
  \{\varphi\} = \emptyset$.

  \medskip\noindent\emph{Conjunction}. Let $\varphi' \in
  \mathfrak{U}(\varphi_1 \land \varphi_2) \setminus \{\varphi_1 \land
  \varphi_2\}$. By definition, \[\varphi' \in \{\psi_1 \land \psi_2:
  \psi_1 \in \mathfrak{U}(\varphi_1), \psi_2 \in
  \mathfrak{U}(\varphi_2)\} \setminus \{\varphi_1 \land \varphi_2\}.\]
  Thus, we have $\varphi' = \psi_1 \land \psi_2$ where $\psi_1 \in
  \mathfrak{U}(\varphi_1)$, $\psi_2 \in \mathfrak{U}(\varphi_2)$, and
  either $\psi_1 \neq \varphi_1 \text{ or } \psi_2 \neq \varphi_2 $.
  Let us consider the first case. The second one is symmetric. We
  have:
  \begin{align*}
    \Fcount{\varphi'}
    &= \Fcount{\psi_1 \land \psi_2} \\
    &= \Fcount{\psi_1} + \Fcount{\psi_2} \\
    &< \Fcount{\varphi_1} + \Fcount{\psi_2}
    && \text{(by ind.\ hyp.\ as $\psi_1 \neq \varphi_1$)} \\
    &\leq \Fcount{\varphi_1} + \Fcount{\varphi_2}
    && \text{(by ind.\ hyp.\ or $\psi_2 = \varphi_2$)} \\
    &= \Fcount{\varphi_1 \land \varphi_2}.
  \end{align*}
  
  \medskip\noindent\emph{Operator $\F$}. Let $\varphi' \in \unfold{\F
    \varphi} \setminus \{\F \varphi\}$. By definition, it is the case
  that $\varphi' \in \unfold{\varphi}$. We have $\Fcount{\F \varphi} >
  \Fcount{\varphi'}$ since
  \begin{align*}
    \Fcount{\F \varphi}
    &= 1 + \Fcount{\varphi} \\
    &\geq 1 + \Fcount{\varphi'}
    && \text{(by ind.\ hyp.\ or $\varphi' = \varphi$)}. \tag*{\qedhere}
  \end{align*}
\end{IEEEproof}

\begin{IEEEproof}[Proof of \Cref{prop:F:count}\eqref{itm:Formulasize:Foperator}]
  We prove this by structural induction: $\Fcount{\flat(\varphi)} \leq
  |\varphi|$, $\Fcount{\flat_{\G}(\varphi)} \leq |\varphi|$,
  $\Fcount{\flat_{\G\F}(\varphi)} \leq |\varphi|$ and
  $\Fcount{\flat_{\F\G}(\varphi)} \leq |\varphi|$.

  Let $\varphi$ be of the form $\psi \land \bigwedge_{i \in I} \G
  \varphi_i \land \bigwedge_{j \in J} \F \varphi_j$.

  \medskip\noindent\emph{Case $\G$}. We have
  \begin{align*}
    \Fcount{\flat_{\G}(\varphi)}
    &= \Sigma_{i \in I} \Fcount{\flat_{\G}(\varphi_i)} + \Sigma_{j \in
      J} \Fcount{\flat_{\G\F}(\varphi_j)} \\    
    &\leq \Sigma_{i \in I} |\varphi_i| + \Sigma_{j \in J} |\varphi_j|
    && \hspace{-21pt}(\text{by ind.}) \\
    &\leq |\varphi|.
  \end{align*}
  
  \medskip\noindent\emph{Case $\G\F$}. We have
  \begin{align*}
    \Fcount{\flat_{\G\F}(\varphi)}
    &= \Sigma_{i \in I}
    \Fcount{\flat_{\F\G}(\varphi_i)} + \Sigma_{j \in J}
    \Fcount{\flat_{\G\F}(\varphi_j)} \\   
    &\leq \Sigma_{i \in I} |\varphi_i| + \Sigma_{j \in J} |\varphi_j|   
    && \hspace{-29pt}(\text{by ind.}) \\
    &\leq |\varphi|.
  \end{align*}

  \medskip\noindent\emph{Case $\F\G$}. We have
  \begin{align*}
    \Fcount{\flat_{\F\G}(\varphi)}
    &= 1 + \Sigma_{i \in I}
    \Fcount{\flat_{\F\G}(\varphi_i)} + \Sigma_{j \in J}
    \Fcount{\flat_{\G\F}(\varphi_j)} \\
    &\leq |\psi| + \Sigma_{i \in I} |\varphi_i| + \Sigma_{j \in J}
    |\varphi_j| \\
    %
    &\leq |\varphi|.
  \end{align*}

  \medskip\noindent\emph{General case}. We have
  \begin{align*}
    \Fcount{\flat(\varphi)}
    &= \Fcount{\psi} + \Sigma_{i \in I} \Fcount{\flat_{\G}(\varphi_i)} +
    \Sigma_{j \in J} \Fcount{\flat(\varphi_j)} \\
    &\leq |\psi| + \Sigma_{i \in I} |\varphi_i| + \Sigma_{j \in J} |\varphi_j|
    &&\hspace{-35pt}(\text{by ind.}) \\
    &\leq |\varphi|. && \tag*{\qedhere}
  \end{align*}
\end{IEEEproof}

\begin{proposition}\label{prop:equivalence:Bracketoperator}
  Let $\varphi \in \text{\fragment{\F, \G, \land}}$ be flat, and let
  $A, A' \subseteq \AP$. It is the case that $\varphi[A][A'] \in
  \{\varphi[A], \false\}$.
\end{proposition}

\begin{IEEEproof}
  If $\varphi[A] = \false$, then $\varphi[A][A'] = \false$. Otherwise,
  by definition, $\varphi[A][A']$ is either $\varphi[A]$ or $\false$.
\end{IEEEproof}

\propAutomataFcountLowering*

\begin{IEEEproof}
  Let $i \in [1..n-1]$. By definition of $\A_\varphi$, there exist
  $\psi_i \in \unfold{r_{i-1}}$ and $\psi_{i+1} \in \unfold{r_i}$ such
  that $r_i = \psi_i[A_i] \neq \false$ and $r_{i+1} =
  \psi_{i+1}[A_{i+1}] \neq \false$. There are two cases.

  \medskip\noindent\emph{Case $\psi_{i+1} = r_i$}. We have
  \[
  r_{i+1} = \psi_{i+1}[A_{i+1}] = r_i[A_{i+1}] =
  \psi_i[A_i][A_{i+1}].
  \]
  As $r_{i+1} \neq \false$, it is the case that $\psi_i[A_i][A_{i+1}]
  = \psi_i[A_i] = r_i$ by
  \Cref{prop:equivalence:Bracketoperator}. Therefore, $r_i = r_{i+1}$,
  which contradicts the fact that the path is simple.

  \medskip\noindent\emph{Case $\psi_{i+1} \neq r_i$}. We have
  $\Fcount{r_{i+1}} = \Fcount{\psi_{i+1}[A_{i+1}]} \leq
  \Fcount{\psi_{i+1}}$ by \Cref{itm:bracketoperator:Fcount} of
  \Cref{prop:F:count}. Moreover, we have $\Fcount{\psi_{i+1}} <
  \Fcount{r_i}$ by \Cref{itm:greater:Foperator} of
  \Cref{prop:F:count}, since $\psi_{i+1} \in \unfold{r_i} \setminus
  \{r_i\}$.
\end{IEEEproof}

\lemConsume*


\begin{IEEEproof}[Proof of Item~1]
  We proceed by induction on $\Fcount{\varphi}$.

  If $\Fcount{\varphi} = 0$, then $\varphi$ is of the form $\psi \land
  \G \psi' \land \bigwedge_{i \in I} \G\F \psi_i''$. Let $\varphi'
  \defeq \varphi[w(0)]$. Note that $\unfold{\varphi} =
  \{\varphi\}$. We have
  \begin{alignat*}{3}
    &&& w \models \varphi \\
    &\iff\ && \prop{\psi \land \psi'} \subseteq w(0) \land
    \FromTo{w}{1}{} \models
    (\G \psi' \land \bigwedge_{i \in I} \G\F \psi_i'') \\
    &\iff\ && \varphi[w(0)] \neq \false \land \FromTo{w}{1}{} \models
    \varphi[w(0)] \\
    &\iff\ && \varphi \trans{w(0)} \varphi' \land \FromTo{w}{1}{}
    \models \varphi'.
  \end{alignat*}

  Now, assume that $\varphi = \theta \land \F \psi$ where $\theta$ and
  $\psi$ are flat.

  $\Rightarrow$) Let $w \models \varphi$. We have $w \models \theta$
  and $w \models \F \psi$. By induction hypothesis, there exists
  $\theta'$ such that $\theta \trans{w(0)} \theta'$ and
  $\FromTo{w}{1}{} \models \theta'$. By definition of
  $\unfold{\cdot}$, we have $\theta \land \F \psi \trans{w(0)} \theta'
  \land \F \psi$. So, if $\FromTo{w}{1}{} \models \F \psi$, then we
  are done by taking $\varphi' \defeq \theta' \land \F
  \psi$. Otherwise, we must have $w \models \psi$. By induction
  hypothesis, there exists $\psi'$ such that $\psi \trans{w(0)} \psi'$
  and $\FromTo{w}{1}{} \models \psi'$. Since $\unfold{\psi} \subseteq
  \unfold{\F \psi}$, we have $\F \psi \trans{w(0)} \psi'$, and hence
  $\theta \land \F \psi \trans{w(0)} \theta' \land \psi'$. So, we are
  done by taking $\varphi' \defeq \theta' \land \psi'$.

  $\Leftarrow$) Let $\varphi'$ satisfy $\varphi \trans{w(0)} \varphi'$
  and $\FromTo{w}{1}{} \models \varphi'$. By definition of
  ${\trans{}}$, there exist $\theta', \psi'$ such that $\theta
  \trans{w(0)} \theta'$, $\F \psi \trans{w(0)} \psi'$ and $\varphi' =
  \theta' \land \psi'$. Thus, $\FromTo{w}{1}{} \models \theta'$ and
  $\FromTo{w}{1}{} \models \psi'$. By induction hypothesis, we have $w
  \models \theta$. It remains to show that $w \models \F \psi$.

  If $\psi' \neq \F \psi$, then $\Fcount{\psi'} < \Fcount{\F \psi}$ by
  \Cref{prop:F:count}, and hence $w \models \F \psi$ by induction
  hypothesis. Otherwise, we have $\psi' = \F \psi$ and so
  $\FromTo{w}{1}{} \models \F \psi$, and in particular $w \models \F
  \psi$.
\end{IEEEproof}


\begin{IEEEproof}[Proof of Item~2]
  We proceed by induction on $\Fcount{\varphi}$. If $\Fcount{\varphi}
  = 0$, then the claim trivially holds. Let $\varphi = \theta \land \F
  \psi$ where $\theta$ and $\psi$ are flat.

  $\Leftarrow)$ Since $\varphi \trans{w(0) \cdots w(i-1)} \varphi'$
  and $\FromTo{w}{i}{} \models \varphi'$, repeated applications of
  Item~1 yields $w \models \varphi$.

  $\Rightarrow$) Let $w \models \varphi$. It is the case that $w \models \theta$
  and $w \models \F \psi$. Let $j \in \N$ be such that
  $\FromTo{w}{j}{} \models \psi$. By Item~1, there exists $\psi'$ such
  that $\psi \trans{w(j)} \psi'$ and $\FromTo{w}{j+1}{} \models
  \psi'$. By definition of ${\trans{}}$, there exists $\psi'' \in
  \unfold{\psi}$ such that $\psi' = \psi''[w(j)]$. As $\unfold{\psi}
  \subseteq \unfold{\F \psi}$, we have $\psi'' \in \unfold{\F \psi}$,
  and so $\F \psi \trans{w(j)} \psi'$. Moreover, we have $\F \psi
  \trans{w(0) \cdots w(j-1)} \F \psi$. Thus, $\F \psi \trans{w(0)
    \cdots w(j)} \psi'$.

  By repeated applications of Item~1, there exists $\theta'$ such that
  $\theta \trans{w(0) \cdots w(j)} \theta'$ and $\FromTo{w}{j+1}{}
  \models \theta'$. Hence, \[(\theta \land \F \psi) \trans{w(0) \cdots
    w(j)} (\theta' \land \psi') \text{ and } \FromTo{w}{j+1}{} \models
  \theta \land \psi'.\] By \Cref{prop:F:count}, we have
  $\Fcount{\theta} \geq \Fcount{\theta'}$ and $\Fcount{\F \psi} >
  \Fcount{\psi'}$. Thus, $\Fcount{\theta \land \F \psi} >
  \Fcount{\theta' \land \psi'}$. So, by induction hypothesis, there
  exist $k \geq j+1$ and $\varphi'$ such that $(\theta' \land \psi')
  \trans{w(j+1) \cdots w(k-1)} \varphi'$, $\Fcount{\varphi'} = 0$ and
  $\FromTo{w}{k}{} \models \varphi'$. We are done since
  \[
  \varphi =
  (\theta \land \F \psi)
  \trans{w(0) \cdots w(j)}
  (\theta' \land \psi')
  \trans{w(j+1) \cdots w(k-1)}
  \varphi'. \tag*{\qedhere}
  \]

\end{IEEEproof}

\propAutTransProp*

\begin{IEEEproof}
  \leavevmode
  \begin{enumerate}
  \item We assume that $X_{q, r} \neq \emptyset$, as we are otherwise
    trivially done. Let us first show that $X_{q, r}$ is closed under
    intersection. Let $A, B \in X_{q, r}$. By definition of
    ${\trans{}}$, there must exist $\theta \in \unfold{q}$ such that
    $r = \theta[A] = \theta[B]$. Moreover, $\theta$ is of the form
    \[
    \theta = \psi \land \G \psi' \land \bigwedge_{i \in I} \G\F
    \psi_i'' \land \bigwedge_{j \in J} \F \varphi_j,
    \]
    and it is the case that $\prop{\psi \land \psi'} \subseteq A$ and
    $\prop{\psi \land \psi'} \subseteq B$. This means that $\prop{\psi
      \land \psi'} \subseteq A \cap B$, and hence that $r = \theta[A
      \cap B]$, which implies $q \trans{A \cap B} r$.

    \medskip

    Since $X_{q, r}$ is closed under intersection, it has a minimal
    element $A$, which is in fact $A \defeq \prop{\psi \land \psi'}$
    and can thus be obtained in polynomial time from $\theta$. We have
    $X_{q, r} = \upc{A}$, as for any $A' \supseteq A$, we have
    $\prop{\psi \land \psi'} = A \subseteq A'$, and so $q \trans{A'}
    \theta[A'] = r$.

    \medskip
  \item Since $X_{q, r} \neq \emptyset$, there exist $\theta \in
    \unfold{q}$ and $A \subseteq \AP$ such that $r = \theta[A]$. Thus,
    $r$ is of the form $\G \psi' \land \bigwedge_{i \in I} \G\F
    \psi_i'' \land \bigwedge_{j \in J} \F \varphi_j$. Note that $r \in
    \unfold{r}$ and $r[B] = r$ where $B \defeq \prop{\psi'}$. Thus, $r
    \trans{B} r$, and so $B \in X_{r, r}$.

    \medskip
  \item Since $X_{q, q} \neq \emptyset$, $q$ is of the form $q = \G
    \psi' \land \bigwedge_{i \in I} \G\F \psi_i'' \land \bigwedge_{j
      \in J} \F \varphi_j$. Let $\theta \in \unfold{q}$. The latter is
    of the form
    \begin{align*}
      &&&&
      \theta &= \psi_{\mathfrak{U}} \land \G (\psi' \land
      \psi_{\mathfrak{U}}') \land \bigwedge_{i \in I_{\mathfrak{U}}}
      \G\F \psi_i'' \land \bigwedge_{j \in J_{\mathfrak{U}}} \F
      \varphi_j.
    \end{align*}
    Testing $\prop{\psi'} \subseteq A$ is less restrictive than
    $\prop{\psi_{\mathfrak{U}} \land \psi' \land \psi_{\mathfrak{U}}'}
    \subseteq A$. So, if $\theta[A] \neq \false$, then $q[A] \neq
    \false$. As $q \in \unfold{q}$, this means that $q \trans{A} r$
    implies $q \trans{A} q$. \hfill\qedhere
  \end{enumerate}
\end{IEEEproof}

\section{Missing proofs of \Cref{sec:ptime}}

\thmFGNforP*

\begin{IEEEproof}
  Let us give the details missing from the main text.  
  \begin{enumerate}
  \item We show that $\vec{x} \models_\M \F \neg Z$ iff $\vec{x} \not
    \in Z$ or there exists a mode $\vec{m} \in \M$ such that $\vec{m}
    \neq \vec{0}$.\label{itm:1}

    \smallskip
    $\Rightarrow)$ For the sake of contradiction, suppose that
    $\vec{x} \in Z$ and that there is no $\vec{m} \in M$ with $\vec{m}
    \neq \vec{0}$. Clearly, we can never reach a point other than
    $\vec{x}$.

    \smallskip
    $\Leftarrow)$ If $\vec{x} \not\in Z$, then clearly $\vec{x}
    \models_\M \F \neg Z$.  Otherwise, we have $\vec{m} \in Z$ with
    $\vec{m} \neq \vec{0}$. Since we assume $Z$ to be a bounded zone,
    it must hold that by simply scheduling $\vec{m}$ forever, we eventually leave $Z$.

  \medskip
  \item We show that $\vec{x} \models_\M \G \neg Z$ iff $\vec{x} \not
    \in Z$ and there exists a mode $\vec{m} \in \M$ such that for all
    $\alpha \in \Rpos$ it is the case that $\vec{x} + \alpha \vec{m}
    \not \in Z$.

    \smallskip
    $\Leftarrow)$ It is easy to see that the execution obtained by
    scheduling $\vec{m}$ for an infinite duration satisfies the
    formula since $\vec{x} + \alpha \vec{m} \not \in Z$ for all
    $\alpha \in \Rpos$.

    \smallskip
    $\Rightarrow)$ Clearly, it must hold that $\vec{x} \not \in Z$.
    Let the modes of $\M$ be $\{\vec{m}_1, \dots, \vec{m}_n\}$.
    To prove the other half, assume for contradiction that for all
    $i \in [1..n]$, it holds that $\vec{x} + \alpha_{i}
    \vec{m}_i \in Z$ for some $\alpha_i \in \Rpos$.
    Let us
    demonstrate that for any non-Zeno infinite schedule $\rho$, it is the case that $\induced{\rho}{\vec{x}} \not\models \G \neg Z$. We do so by
    showing the existence of $\tau \in \Rnon$ and $\vec{z} \in Z$ such
    that $\vec{x} \trans{\FromTo{\rho}{}{\tau}} \vec{z}$.

    Let $\vec{m}'_i \defeq \alpha_i \vec{m}_i$, and
    let us consider the MMS $\M' \defeq \{\alpha_i \vec{m}_i :
    i \in [1..n]\}$. Clearly, schedules of $\M$ and $\M'$ can be
    related: A schedule $\pi$ of $\M$ amounts to the schedule $\pi'$
    of $\M'$ where we replace we replace each
    occurrence $(\beta, \vec{m}_i)$ with $(\beta / \alpha_{i},
    \vec{m}'_i)$. Thus, for every $\vec{y}, \vec{y}'$, we have $\vec{y}
    \trans{\pi} \vec{y}'$ iff $\vec{y} \trans{\pi'} \vec{y}'$.
      
    Let us consider a schedule $\pi$ of $\M'$, and let $\rho =
    \pi[..1]$. It is the case that $\vec{x} \trans{\rho} \vec{z}$ with 
    \[
    \vec{z} = \vec{x} + \sum_{i=1}^n \lambda_i \vec{m}'_i,
    \]
    for some $\lambda_1, \dots, \lambda_n$ such that
    $\sum_{i=1}^n \lambda_i = 1$. Thus, $\vec{z}$ is a
    convex combination of the points $\vec{x} + \vec{m}_i'$ for any
    $i \in [1..n]$. But note that $\vec{x} + \vec{m}_i' = \vec{x} +
    \alpha_{i} \vec{m}_i$ by definition of $\M'$, and $\vec{x} +
    \alpha_{i} \vec{m}_i \in Z$ by definition of
    $\alpha_{i}$. So, $\vec{z}$ is a convex combination of
    points from $Z$, and is thus included in $Z$, as it is convex.

  \medskip
  \item The proof for $\G \F \neg Z$ is as in~\Cref{itm:1}, except
    that it remains to note that once an execution has left $Z$ by
    scheduling $\vec{m}$, if it keeps scheduling $\vec{m}$, then it
    will not re-enter $Z$.~\label{itm:3}

  \medskip
  \item The proof for $\F \G \neg Z$ is as
    in~\Cref{itm:3}. \hfill\qedhere
  \end{enumerate}
\end{IEEEproof}

\thmFGPhard*

\begin{IEEEproof}
  \emph{Fragment \fragmentB{\F}}. We reduce from linear programming
  feasibility. This problem asks whether a given zone $Z \subseteq
  \R^d$, described by a system of inequalities $\mat{A} \vec{x} \leq
  \vec{b}$, has a non-negative solution. The problem is P-complete
  even if zone $Z$ is bounded~\cite[Prob.~A.4.1]{GHR95}. Let $\M \defeq
  \{\vec{e}_i : i \in [1..d]\}$. It is readily seen that $Z \neq
  \emptyset$ iff $\vec{0} \models_\M \F Z$.

  \medskip\noindent\emph{Fragment \fragmentB{\G}}. We reduce from the
  monotone circuit-value problem (CVP)~\cite[Prob.~A.1.3]{GHR95}. Let
  $C(x_1, \ldots, x_n)$ be a boolean circuit with gates from $\{\land,
  \lor\}$, and let $w_1, \ldots, \allowbreak w_n \allowbreak \in
  \allowbreak \{0, 1\}$. We construct a $d$-dimensional MMS $\M$,
  $\vec{x} \in \R^d$ and a bounded zone $Z \subseteq \R^d$ such that
  $C(w) = 1$ iff $\vec{x} \models_\M \G Z$.

  Each gate $g$ is associated to a dimension $g$. We add two
  dimensions $\{\heartsuit, \overline{\heartsuit}\}$. Let $\vec{x}
  \defeq \vec{e}_\heartsuit + \sum_{i=1}^n w_i \cdot
  \vec{e}_{x_i}$. Let $Z$ be the zone defined by $c \in [0, 1]$ for
  every dimension $c$.

  We denote as $g_\text{out}$ the output gate of $C$. We associate a
  mode $\vec{m}_g$, to each gate $g = u \land v$, defined by
  $\vec{m}_g \defeq -\vec{e}_u - \vec{e}_v + \vec{e}_g$. We associate
  the modes $\vec{m}_g$ and $\vec{m}_g'$ to each gate $g = u \lor v$,
  defined by $\vec{m}_g \defeq -\vec{e}_u + \vec{e}_g$ and $\vec{m}_g
  \defeq -\vec{e}_v + \vec{e}_g$. We add two modes:
  $\vec{m}_\heartsuit \defeq -\vec{e}_{g_\text{out}} -
  \vec{e}_\heartsuit + \vec{e}_{\overline{\heartsuit}}$ and
  $\vec{m}_{\overline{\heartsuit}} \defeq \vec{e}_{g_\text{out}} +
  \vec{e}_\heartsuit - \vec{e}_{\overline{\heartsuit}}$.

  Let $g_1, \ldots, g_m$ be a topological ordering of the gates of
  $C$. Let $C_j$ be the subcircuit obtained by setting $g_j$ as the
  output gate. A routine induction shows that $C_j(w) = 1$ iff there
  exists $\vec{y}$ such that $\vec{x} \mreach{*}{Z} \vec{y}$ and
  $\vec{y}(g_j) > 0$. The claim implies that $C(w) = 1$ iff $\vec{x}
  \models_\M \G Z$. Indeed, any non-Zeno infinite schedule must use
  $\vec{m}_\heartsuit$. Moreover, upon reaching some $\vec{y}$ such
  that $\vec{y}(g_\text{out}) = \alpha > 0$, it is possible to
  alternate between $\alpha \vec{m}_\heartsuit$ and $\alpha
  \vec{m}_{\overline{\heartsuit}}$ indefinitely.
\end{IEEEproof}

\section{Missing proofs of \Cref{ssec:transu}}

\Cref{prop:reduc:Zsafe:exec} follows inductively from this lemma:

\begin{lemma}\label{lema:reduc:Zsafe:oneMode}
  Let $\rho (\alpha, \vec{m}) \rho' (\beta, \vec{m}) \rho''$ be a
  schedule. This holds:
  \begin{itemize}
  \item If $\vec{x} \mreach{\rho (\alpha, \vec{m}) \rho'
    (\beta,\vec{m}) \rho''}{Z} \vec{y}$, then $\vec{x} \mreach{\rho\,
    (\alpha, \vec{m})\, \frac{\alpha}{\alpha + \beta}[\rho'
    \rho'']}{Z} {}$,

  \item If $\vec{x} \mreach{\rho''(\beta, \vec{m})\rho'(\alpha,
    \vec{m})\rho}{Z} \vec{y}$, then ${} \mreach{\frac{\alpha}{\alpha +
      \beta}[\rho'' \rho']\, (\alpha, \vec{m})\, \rho}{Z} \vec{y}$.
  \end{itemize}
\end{lemma}

\begin{IEEEproof}
  We only prove the first item. The second follows symmetrically. We have
  $\vec{x} \mreach{\rho}{Z} \vec{u}$ for some $\vec{u}$. By convexity,
  there exist $\vec{v}, \vec{v}', \vec{w}$ such that:
  \[
  \vec{u} \mreach{\frac{\alpha}{\alpha+\beta}[(\alpha,
      \vec{m})\rho'(\beta, \vec{m})]}{Z} \vec{v}
  \mreach{\frac{\alpha}{\alpha+\beta}\rho''}{Z} \vec{w}
  \] and \[
  \vec{u} \mreach{(\alpha,
    \vec{m})\frac{\alpha}{\alpha+\beta}\rho'}{Z} \vec{v}'.
  \]
  Let us show that $\vec{v} = \vec{v}'$:
  \begin{align*}
    \vec{v}
    &= \vec{u} + \frac{\alpha}{\alpha+\beta}\left(\alpha\vec{m} +
    \effect{\rho'} + \beta\vec{m}\right) \\
    &= \vec{u} + \frac{\alpha}{\alpha+\beta}\left((\alpha +
    \beta)\vec{m} + \effect{\rho'}\right) \\
    &= \vec{u} + \alpha\vec{m} + \frac{\alpha}{\alpha+\beta}\effect{\rho'} \\
    &= \vec{v}'.
  \end{align*}
  From this, we get $\vec{x} \mreach{\rho}{Z} \vec{u} \mreach{(\alpha,
    \vec{m})\frac{\alpha}{\alpha+\beta}\rho'}{Z} \vec{v}
  \mreach{\frac{\alpha}{\alpha+\beta}\rho''}{Z} \vec{w}$.
\end{IEEEproof}

\lemaModeZSafeOneMode*

\begin{IEEEproof}
  We only prove the first item. The second follows symmetrically. By
  convexity, there exist $\vec{y}'$ and $\vec{w}$ such that
  \[
  \vec{x} \mreach{\rho(\frac{\alpha}{2},
    \vec{m})\frac{1}{2}\rho'}{Z} \vec{y'}
  \text{ and }
  \vec{x} \mreach{\rho(\alpha, \vec{m})\frac{1}{2}\rho'}{Z} \vec{w}.
  \]
  Moreover, we have $\vec{y'} \mreach{(\frac{\alpha}{2}, \vec{m})}{}
  \vec{w}$. Since $\vec{y}', \vec{w} \in Z$, by convexity, we conclude
  that
  \[
  \vec{x} \mreach{\rho(\frac{\alpha}{2}, \vec{m})\frac{1}{2}\rho'}{Z}
  \vec{y'} \mreach{(\frac{\alpha}{2}, \vec{m})}{Z} \vec{w}. \tag*{\qedhere}
  \]
\end{IEEEproof}

\propModeZSafeExec*

\begin{IEEEproof}
  By \Cref{lema:mode:Zsafe:oneMode}, we can pick
  \[
  \pi' \defeq \prod_{i = 1}^{|\pi|}\left(\frac{1}{2^i} \cdot \pi(i)\right),\
  \pi'' \defeq \prod_{i = 1}^{|\pi|}\left(\frac{1}{2^{(|\pi| - i +
      1)}} \cdot \pi(i)\right),
  \]
  and $\beta \defeq 2^{|\pi|} \cdot \lceil\tweight{\pi}\rceil$.
\end{IEEEproof}

\propZSafeLineConstruction*

\begin{IEEEproof}
  Let $\pi \defeq (\alpha_1, \vec{m}_1)(\alpha_2, \vec{m}_2) \cdots
  (\alpha_k, \vec{m}_k)$. Let $\vec{x}^0_0 \defeq \vec{x}$,
  $\vec{x}^k_{\beta k-1} \defeq \vec{y}$ and $\vec{x}^k_j \defeq
  \vec{x}^0_{j+1}$. Let
  \begin{align*}
    \vec{x}^i_j &\defeq \vec{x}^0_0 + j\sum_{h=1}^k\frac{\alpha_h \vec{m}_h}{\beta k} + \sum_{h=1}^i\frac{\alpha_h \vec{m}_h}{\beta k}\\
    &= \vec{x}^i_0 + \sum_{h=i+1}^k\frac{\alpha_h \vec{m}_h}{\beta k} + (j-1)\sum_{h=1}^k\frac{\alpha_h \vec{m}_h}{\beta k}+ \sum_{h=1}^i\frac{\alpha_h \vec{m}_h}{\beta k}\\
    &= \vec{x}^i_0 + \sum_{h=1}^k\frac{\alpha_h \vec{m}_h}{\beta k} + (j-1)\sum_{h=1}^k\frac{\alpha_h \vec{m}_h}{\beta k}\\
    &= \vec{x}^i_0 + j\sum_{h=1}^k\frac{\alpha_h \vec{m}_h}{\beta k}.
  \end{align*}
  Let $\mat{A}\vec{\ell} \leq \vec{b}$ be the system of inequalities
  that represents zone $Z$. By assumption, the following holds for all
  $i \in [1..k]$:
  \[
  \mat{A}\left(\vec{x}^0_0 + \frac{\alpha_i \vec{m}_i}{\beta}\right)
  \leq \vec{b}
  \text{ and }  
  \mat{A}\left(\vec{x}^k_{\beta k - 1} - \frac{\alpha_i
    \vec{m}_i}{\beta}\right) \leq \vec{b}.
  \]  
  The following holds:
  \[
  \vec{x}^0_0
  \mreach{\frac{1}{\beta k}\pi(1)}{Z} \vec{x}^1_0
  \mreach{\frac{1}{\beta k}\pi(2)}{Z} \vec{x}^2_0
  \mreach{\frac{1}{\beta k}\pi(3)}{Z} \cdots
  \mreach{\frac{1}{\beta k}\pi(k)}{Z} \vec{x}^k_0,
  \]
  since
  \begin{align*}
    \mat{A}\vec{x}^i_0
    &= \mat{A}\vec{x}^0_0 + \sum_{j=1}^i\mat{A}\frac{\alpha_j
      \vec{m}_j}{\beta k} \\
    &= \frac{k-i}{k}\mat{A}\vec{x}^0_0 + \frac{i}{k}\mat{A}\vec{x}^0_0
    + \frac{1}{k} \sum_{j=1}^i\mat{A}\frac{\alpha_j \vec{m}_j}{\beta} \\
    &= \frac{k-i}{k}\mat{A}\vec{x}^0_0 + \frac{1}{k} \sum_{j=1}^i
    \mat{A} \left(\vec{x}^0_0 + \frac{\alpha_j
      \vec{m}_j}{\beta}\right) \\
    &\leq \frac{k-i}{k}\vec{b} + \frac{1}{k} \vec{b} \\
    &= \vec{b}.
  \end{align*}
  Similarly, the following holds:
  \begin{align*}
  \vec{x}^0_{\beta k - 1} \mreach{\frac{1}{\beta k}\pi(1)}{Z}
  \vec{x}^1_{\beta k - 1} &\mreach{\frac{1}{\beta k}\pi(2)}{Z}
  \vec{x}^2_{\beta k - 1} \cdots
  \mreach{\frac{1}{\beta k}\pi(k)}{Z} \vec{x}^k_{\beta k - 1}.
  \end{align*}

  It remains to show that, for all $i \in [1..k]$, $j \in [1..(\beta
    k-1)]$, we have $\vec{x}^i_j = (1-\lambda)\vec{x}^i_0 + \lambda
  \vec{x}^i_{\beta k-1}$ where $\lambda = \frac{j}{\beta k-1}$.

  We have:
  \begin{align*}
    \vec{x}^i_j &= \vec{x}^i_0 + j\sum_{h=1}^k\frac{\alpha_h
      \vec{m}_h}{\beta k} \\
    &= \vec{x}^i_0 + \frac{\beta k-1}{\beta
      k-1} \cdot j \sum_{h=1}^k\frac{\alpha_h \vec{m}_h}{\beta k} \\
    &= \vec{x}^i_0 + \lambda(\beta k-1)\sum_{h=1}^k\frac{\alpha_h
      \vec{m}_h}{\beta k} \\
    &= (1 - \lambda)\vec{x}^i_0 + \lambda\left(\vec{x}^i_0 + (\beta
    k-1)\sum_{h=1}^k\frac{\alpha_h \vec{m}_h}{\beta k}\right) \\
    &= (1 - \lambda)\vec{x}^i_0 + \lambda \vec{x}^i_{\beta k-1}.
  \end{align*}
  Since $\lambda \in [0, 1]$, each $\vec{x}^i_j$ belongs to $Z$ by
  convexity.
\end{IEEEproof}

\section{Missing proofs of \Cref{ssec:lasso}}

\propZSafeExecOnLine*

\begin{IEEEproof}
  Let us prove the case where $\pi = \alpha m$ for some $\alpha \in
  \Rpos$ and mode $\vec{m} \in \M$. The general case follows by
  induction.  Note that $\vec{x}' = \vec{x} + \alpha \vec{m}$ and
  $\vec{x'} \in Z$. Let $\vec{z}' \defeq \vec{z} + \beta \alpha
  \vec{m}$. We clearly have $\vec{z} \trans{\beta \pi} \vec{z}'$. It
  remains to show that $\vec{z} \mreach{\beta \pi}{Z} \vec{z}'$. By
  convexity of $Z$, it suffices to show that $\vec{z}'$ is on the line
  passing through $\vec{x}' \in Z$ and $\vec{y} \in Z$:
  \begin{align*}
    \vec{z}'
    &=
    (\beta \vec{x} + (1-\beta) \vec{y}) + \beta \alpha \vec{m} \\
    &= \beta(\vec{x} + \alpha \vec{m}) + (1-\beta) \vec{y} \\
    &= \beta \vec{x}' + (1-\beta) \vec{y}. && \tag*{\qedhere}
  \end{align*}
\end{IEEEproof}

Within the proof of \Cref{prop:Dual}, we have claimed that $f(\vec{x},
\vec{y}) \geq \lambda$. Let us prove this.

\begin{IEEEproof}[Proof of \Cref{prop:Dual} (missing claim)]
  We first make two additional claims. Given $\vec{x} \in X'$ and
  $\vec{y} \in Y'$ such that $\vec{z} \mreach{\pi}{} \vec{x}
  \mreach{\pi'}{} \vec{y}$, it is the case that:
  \begin{enumerate}
  \item $f'(\vec{x}, \vec{y}) \geq \lambda$, and

  \item $f'(\vec{x}, \vec{y}) \leq \vec{v}_3^T(\vec{y} - \vec{x})$.
  \end{enumerate}

  \medskip\noindent To prove Claim~1, note that we have
  \begin{alignat*}{4}
    f'(\vec{x}, \vec{y})
    &=\ && -(\vec{v}_1^T\mat{A}_1(\vec{x} - \vec{z}) +
    \vec{v}_2^T\mat{A}_2(\vec{y} - \vec{z})) + {} \\    
    &&& \hspace*{20pt} \vec{v}_2^T\mat{A}_2(\vec{y} - \vec{x}) +
    \vec{v}_3^T(\vec{y} - \vec{x}) \\
    &=\ && -(\vec{v}_1^T\mat{A}_1 (\vec{x} - \vec{z}) +
    \vec{v}_2^T\mat{A}_2 (\vec{y} - \vec{z})) + {}\quad \\
    &&& \hspace*{20pt} (\vec{v}_2^T\mat{A}_2 \mat{M} +
    \vec{v}_3^T \mat{M}) \parikh{\pi'} \\
    &\geq\ && -(\vec{v}_1^T\mat{A}_1 (\vec{x} - \vec{z}) +
    \vec{v}_2^T\mat{A}_2 (\vec{y} - \vec{z}))
    && \text{(by~\eqref{eq:farkas:2})} \\
    &\geq\ && -(\vec{v}_1^T(\vec{b}_1 - \mat{A}_1\vec{z}) +
    \vec{v}_2^T(\vec{b}_2 - \mat{A}_2\vec{z}))
    && \\
    &&&&&\hspace{-48pt}\text{(by~$\vec{x} \in X', \vec{y} \in Y'$)} \\
    &=\ && \lambda.
  \end{alignat*}
  
  \medskip
  Next, to prove Claim~2, we have
  \begin{alignat*}{3}
    &&& f'(\vec{x}, \vec{y}) \\
    &=\ && -(\vec{v}_1^T\mat{A}_1(\vec{x}-\vec{z}) +
    \vec{v}_2^T\mat{A}_2(\vec{y} - \vec{z})) + {} \\    
    &&& \hspace*{20pt} \vec{v}_2^T\mat{A}_2(\vec{y} - \vec{x}) +
    \vec{v}_3^T(\vec{y} - \vec{x}) \\
    &=\ && -\left(\vec{v}_1^T\mat{A}_1(\vec{x} - \vec{z}) +
    \vec{v}_2^T\mat{A}_2(\vec{x} - \vec{z})\right) +
    \vec{v}_3^T(\vec{y}-\vec{x}) \\
    &=\ && -(\vec{v}_1^T\mat{A}_1 \mat{M} + \vec{v}_2^T\mat{A}_2 \mat{M})
    \parikh{\pi} + \vec{v}_3^T(\vec{y} - \vec{x}) \\
    &\leq\ && \vec{v}_3^T(\vec{y} - \vec{x}) &&
    \hspace{-20pt}\text{(by~\eqref{eq:farkas:1})}.
  \end{alignat*}

  \medskip
  Finally, by using Claims~2 and~1, we can prove the original claim:
  $f(\vec{x}, \vec{y})
  = \vec{v}_3^T(\vec{y} - \vec{x})
  \geq f'(\vec{x}, \vec{y})
  \geq \lambda$.
\end{IEEEproof}

\propLassoSameLine*

\begin{IEEEproof}
  
  By unrolling the definition of $\vec{x}_n$ and $\vec{y}_n$, we have
  \begin{alignat}{3}
    &&& \vec{x}_n \\
    &=\ && \vec{x}_0 + \sum_{i =
      0}^{n-1}\left(\lambda^i\left(\effect{\pi\pi'} +
    \frac{\effect{\rho} - \epsilon \effect{\pi\pi'}}{1+\epsilon}\right) +
    (1 - \lambda^i)\effect{\rho'\rho''}\right) \notag \\
    &=\ && \vec{x}_0 + \sum_{i = 0}^{n-1}\left(\lambda^i\left(\effect{\pi\pi'} +
    \frac{\effect{\rho} - \epsilon \effect{\pi\pi'}}{1+\epsilon}\right) +
    (1 - \lambda^i) \cdot \vec{0}\right) \notag\\
    &=\ && \vec{x}_0 + \sum_{i = 0}^{n-1}\lambda^i\left(\effect{\pi\pi'} +
    \frac{\effect{\rho} - \epsilon \effect{\pi\pi'}}{1+\epsilon}\right)
    \notag \\
    &=\ && \vec{x}_0 + \frac{\lambda^n - 1}{\lambda - 1}\left(\effect{\pi\pi'}
    + \frac{\effect{\rho} -
      \epsilon \effect{\pi\pi'}}{1+\epsilon}\right) \label{eq:geo} \\
    &=\ && \vec{x}_0 + \frac{\lambda^n -
      1}{-1 / (1 + \epsilon)}\left(\effect{\pi\pi'} +
    \frac{\effect{\rho} - \epsilon \effect{\pi\pi'}}{1+\epsilon}\right)
    \notag \\
    &=\ && \vec{x}_0 - (\lambda^n - 1)(1
    +\epsilon)\left(\effect{\pi\pi'} + \frac{\effect{\rho}
    - \epsilon \effect{\pi\pi'}}{1+\epsilon}\right) \notag \\    
    &=\ && \vec{x}_0 + (1 - \lambda^n)((1 + \epsilon)\effect{\pi\pi'} +
    \effect{\rho} - \epsilon \effect{\pi\pi'}) \notag \\
    &=\ && \vec{x}_0 + (1 - \lambda^n)\effect{\pi\pi'\rho} \notag \\
    &=\ && \lambda^n \vec{x}_0 + (1 - \lambda^n)(\vec{x_0}
    + \effect{\pi\pi'\rho}) \notag \\
    &=\ && \lambda^n\vec{x_0} + (1 - \lambda^n)\vec{x}_f, \notag
  \end{alignat}
  where~\eqref{eq:geo} follows from $\sum_{i = 0}^{n-1} x^{n-1} = (x^n
  - 1) / (x - 1)$. Furthermore, we have
  \begin{align}
    \vec{y}_n &= \vec{x}_n + \lambda^n\effect{\pi} + (1 - \lambda^n)\effect{\rho'} \notag\\
    &= \lambda^n \vec{x}_0 + (1 - \lambda^n) \vec{x}_f + \lambda^n\effect{\pi} + (1 - \lambda^n)\effect{\rho'} \label{eq:prop:Lasso:XSameLine}\\
    &= \lambda^n (\vec{x}_0 + \effect{\pi}) + (1 - \lambda^n)(\vec{x}_f + \effect{\rho'}) \notag\\
    &= \lambda^n \vec{y}_0 + (1 - \lambda^n) \vec{y}_f, \notag
  \end{align}
  where \eqref{eq:prop:Lasso:XSameLine}
  follows from the fact that $\vec{x}_n = \lambda^n \vec{x}_0 + (1 -
  \lambda^n)\vec{x}_f$ for all $n \in \N$, as proven above.
\end{IEEEproof}

\section{Missing proofs of \Cref{ssec:GZ}}

\propfarkasGZ*

\begin{IEEEproof}
  We define $\mat{M}$ as the matrix such that each column is a mode
  from $\M$. Observe that the following set of constraints
  $\mathcal{S}$ is equivalent to $\exists \pi, \vec{z}' : \vec{z}
  \preach{\parikh{\pi}}{} \vec{z'} \land \mat{A}\vec{z'} \leq
  \mat{A}\vec{z} \land \norm{\parikh{\pi}} \geq 1$:
  \begin{align*}
    \exists \vec{u} \geq \vec{0}:
    \begin{bmatrix}
      \mat{A}\mat{M}\\
      \vec{-1}^T
    \end{bmatrix}\vec{u} 
    \leq 
    \begin{bmatrix}
      \vec{0}\\
      -1
    \end{bmatrix}.
  \end{align*}  
  For the sake of contradiction, suppose that $\vec{z} \models \G Z$
  and that $\mathcal{S}$ has no solution. By Farkas' lemma, the
  following system $\mathcal{S}'$ has a solution:
  \begin{align*}
    &\exists \vec{v}_1 \geq \vec{0}, v_2 \geq 0:\\
    &\begin{bmatrix}
      \mat{M}^T\mat{A}^T & \vec{-1}
    \end{bmatrix}    
    \begin{bmatrix}
      \vec{v}_1\\
      v_2
    \end{bmatrix} 
    \geq 
    \vec{0},\,
    \begin{bmatrix}
      \vec{0}^T
      -1
    \end{bmatrix}
    \begin{bmatrix}
      \vec{v}_1\\
      v_2
    \end{bmatrix} 
    < 0.
  \end{align*}
  The latter can be rewritten equivalently as follows:
  \begin{alignat*}{3}
    &&&
    \exists \vec{v}_1 \geq \vec{0}, v_2 \geq 0:
      \mat{M}^T\mat{A}^T\vec{v}_1 - \vec{1}v_2 \geq \vec{0},\, -v_2 < 0\\
    &\iff\ &&
      \exists \vec{v}_1 \geq \vec{0}, v_2 \geq 0:
      \vec{v}_1^T\mat{A}\mat{M} \geq \vec{1}^Tv_2,\, v_2 > 0\\
    &\iff\ &&
    \exists \vec{v}_1 \geq \vec{0}, v_2 \geq 0:
      \vec{v}_1^T\mat{A}\mat{M} \geq \vec{1}^Tv_2  > \vec{0}^T.
  \end{alignat*}

  Since $\vec{z} \models \G Z$, there exists a non-Zeno infinite
  schedule $\pi$ such that $\vec{z} \mreach{\pi}{Z} {}$. Let $n \in
  \N$, $\pi_n \defeq \FromTo{\pi}{}{n}$ and $\vec{z}_n \defeq \vec{z}
  + \effect{\pi_n}$. We have $\vec{z}_n \in Z$ and hence $\mat{A}
  \vec{z}_n \leq \vec{b}$. Thus,
  \begin{align}
    v_2\vec{1}^T\parikh{\pi_n}
    &\leq \vec{v}_1^T\mat{A}\mat{M}\parikh{\pi_n} \notag \\
    &= \vec{v}_1^T\mat{A}(\vec{z}_n - \vec{z}) \notag \\
    &\leq \vec{v}_1^T(\vec{b} - \mat{A}\vec{z}).\label{eq:bounded:lambda}
  \end{align}
  Since $\pi$ is non-Zeno, we further have $\lim_{n \to \infty}
  v_2\vec{1}^T\parikh{\pi_n} = \infty$, which
  contradicts~\eqref{eq:bounded:lambda}.
\end{IEEEproof}

\propStillSafe*

\begin{IEEEproof}
  We consider the case case where $\rho = \alpha \vec{m}$. The general
  case follows inductively. Since $\vec{z} \mreach{\rho}{Z}$, we have
  $\mat{A}\vec{z} + \mat{A}\alpha \vec{m} \leq \vec{b}$. Thus,
  $\mat{A}\vec{z}' + \mat{A}\alpha \vec{m} \leq \mat{A}\vec{z} +
  \mat{A}\alpha \vec{m} \leq \vec{b}$.
\end{IEEEproof}

\section{Missing proofs of \Cref{subsec:manyzones}}

\begin{IEEEproof}[Proof of the construction in \Cref{lem:n:to:two}]

$\Rightarrow)$ Let $\pi$ be a non-Zeno infinite schedule that
  satisfies $\induced{\pi}{\x} \models \varphi$. We construct a
  non-Zeno infinite schedule $\pi'$ such that $\induced{\pi'}{(\x,
      \ldots, \x)} \models \varphi'$. By definition of $\varphi$, for
  each $i \in [1..n]$, there exist $\tau_{i, 0} < \tau_{i, 1} < \cdots
    \in \Rnon$ and $\x_{i, 0}, \x_{i, 1}, \ldots \in \R^d$ such that
  $\tau_{i, 0} = 0$, $\x_{i, 0} = \x$ and
  \begin{align*}
    \x_{i, j} \mreach{\FromTo{\pi}{\tau_{i, j}}{\tau_{i, j+1}}}{Z_0} \x_{i,
    j+1} \in Z_i &  & \text{ for all } j \in \N.
  \end{align*}
  Without loss of generality, we may assume that $\tau_{i, j} \leq
    \tau_{1, j} \leq \tau_{i, j+1}$ for all $i \in [1..n]$ and $j \in
    \N$. Indeed, the values can be chosen this way since each zone $Z_i$
  is visited infinitely often.

  Let $\pi_i$ denote the schedule obtained from $\pi$ by replacing
  each mode $\vec{m} \in \M$ with $\vec{m}_i \in \M'$. Let $\pi'
    \defeq u_0 v_0 u_1 v_1 \cdots$ be the infinite schedule, where
  \begin{alignat*}{4}
    u_j &\defeq\ &&
    \pi_1[\tau_{1,j}..\tau_{1,j+1}] &&
    \pi_2[\tau_{1,j}..\tau_{2,j+1}] &&
    \cdots
    \pi_n[\tau_{1,j}..\tau_{n,j+1}], \\
    v_j &\defeq\ &&
    \pi_1[\tau_{1,j+1}..\tau_{1,j+1}] &&
    \pi_2[\tau_{2,j+1}..\tau_{1,j+1}] &&
    \cdots
    \pi_n[\tau_{n,j+1}..\tau_{1,j+1}].
  \end{alignat*}
  By definition, we have
  \begin{alignat*}{3}
    (\x, \ldots, \x)
                                   & \mreach{u_0}{Z}
    (\x_{1, 1}, \ldots, \x_{n, 1}) &                 & \in X \\
                                   & \mreach{v_0}{Z}
    (\x_{1, 1}, \ldots, \x_{1, 1}) &                 & \in Y \\
                                   & \mreach{u_1}{Z}
    (\x_{1, 2}, \ldots, \x_{n, 2}) &                 & \in X \\
                                   & \mreach{v_1}{Z}
    (\x_{1, 2}, \ldots, \x_{1, 2}) &                 & \in Y \\
                                   & \mreach{u_2}{Z}
    \cdots.
  \end{alignat*}
  Thus, it follows that $(\x, \ldots, \x) \models_{\M'} \varphi'$.

  $\Leftarrow)$ Let $\pi'$ be a non-Zeno infinite schedule that
  satisfies $\induced{\pi'}{(\x, \ldots, \x)} \models \varphi'$. There
  exist $0 = \tau_0 < \gamma_1 < \tau_1 < \gamma_2 < \tau_2 < \cdots
    \in \Rnon$ and $\x_1, \y_1, \x_2, \y_2, \ldots \in \R^{nd}$ such
  that
  \begin{alignat*}{3}
    (\x, \ldots, \x)
         & \mreach{\FromTo{\pi'}{\tau_0}{\gamma_1}}{Z}
    \x_1 &                                             & \in X \\
         & \mreach{\FromTo{\pi'}{\gamma_1}{\tau_1}}{Z}
    \y_1 &                                             & \in Y \\
         & \mreach{\FromTo{\pi'}{\tau_1}{\gamma_2}}{Z}
    \x_2 &                                             & \in X \\
         & \mreach{\FromTo{\pi'}{\gamma_2}{\tau_2}}{Z}
    \y_2 &                                             & \in Y \\
         & \mreach{\FromTo{\pi'}{\tau_2}{\gamma_3}}{Z}
    \cdots
  \end{alignat*}

  For every $i \in [1..n]$, let $\pi'_i$ denote the schedule obtained
  from $\pi'$ by keeping only the modes from $\{\vec{m}_i : \vec{m}
    \in \M\}$, and changing each occurrence of $\vec{m}_i$ to
  $\vec{m}$. Let
  \begin{alignat*}{3}
    \pi
     & \defeq\  &  &
    \FromTo{\pi_1}{\tau_0}{\gamma_1}\
    \FromTo{\pi_1}{\gamma_1}{\tau_1}         \\
     &          &  &
    \FromTo{\pi_2}{\tau_1}{\gamma_2}\
    \FromTo{\pi_2}{\gamma_2}{\tau_2}         \\
     &          &  &
    \cdots                                   \\
     &          &  &
    \FromTo{\pi_n}{\tau_{n-1}}{\gamma_n}\
    \FromTo{\pi_n}{\gamma_n}{\tau_n}         \\
     &          &  &
    \FromTo{\pi_1}{\tau_n}{\gamma_{n+1}}\
    \FromTo{\pi_1}{\gamma_{n+1}}{\tau_{n+1}} \\
     &          &  &
    \cdots
  \end{alignat*}
  Recall that $\y_j[i] = \y_j[i']$ for all $i, i' \in [1..n]$. So, by
  definition, we have
  \begin{alignat*}{4}
    \x
     &\mreach{\FromTo{\pi_1}{\tau_0}{\gamma_1}}{Z_0}\
     &\x_1[1]\
     &\mreach{\FromTo{\pi_1}{\gamma_1}{\tau_1}}{Z_0}\
     &\y_1[1]\
     &= \y_1[2]                                                            \\
     &\mreach{\FromTo{\pi_2}{\tau_1}{\gamma_2}}{Z_0}\
     &\x_2[2]\
     &\mreach{\FromTo{\pi_2}{\gamma_2}{\tau_2}}{Z_0}\
     &\y_2[2]\
     &= \y_2[3]                                                            \\
     & \cdots                                                   &  & \cdots \\
     &\mreach{\FromTo{\pi_n}{\tau_{n-1}}{\gamma_n}}{Z_0}\
     & \x_n[n]\
     & \mreach{\FromTo{\pi_n}{\gamma_n}{\tau_n}}{Z_0}\
     & \y_n[n]\
     & = \y_n[1]                                                            \\
     &\mreach{\FromTo{\pi_1}{\tau_n}{\gamma_{n+1}}}{Z_0}\
     & \x_{n+1}[1]\
     & \mreach{\FromTo{\pi_1}{\gamma_{n+1}}{\tau_{n+1}}}{Z_0}\
     & \y_{n+1}[1]\
     & = \y_{n+1}[2]                                                        \\
     & \cdots                                                   &  & \cdots
  \end{alignat*}
  Recall that $\x_j[i] \in Z_i$ for all $i \in [1..n]$ and $j \in
    \N$. Thus, each zone $Z_i$ is visited infinitely often, and so
  $\induced{\pi}{\x} \models \varphi$.
\end{IEEEproof}

\section{Missing proofs of \Cref{sec:np}}

\begin{IEEEproof}[Proof of \Cref{thm:fa:NPhard} (correctness of reduction)]
  Let us show that $\vec{0} \models_\M \varphi$ iff there is a
  solution $V$ to the SUBSET-SUM instance $(S, t)$. Recall that
  $\vec{0} \models_\M \varphi$ holds iff $\M$ has a non-Zeno infinite
  schedule $\pi$ such that $\induced{\pi}{\vec{0}} \models \varphi$.

  $\Leftarrow)$ Let $V \subseteq S$ be such that $\sum_{v \in V} v =
  t$. We define a schedule $\pi$ that satisfies $\varphi$. Let $\pi
  \defeq \pi_1 \pi_2 \cdots \pi_n\, \vec{y}_1^\omega$, where
  \[
  \pi_i
  \defeq
  \begin{cases}
    \vec{y}_i\, \overline{\vec{y}}_i & \text{if } s_i \in V, \\
    \vec{n}_i\, \overline{\vec{n}}_i & \text{otherwise}.
  \end{cases}
  \]
  
  $\Rightarrow)$ Let $\pi$ be such that $\sigma \defeq
    \induced{\pi}{\vec{0}} \models \varphi$. By definition of $\varphi$,
  there exist $\tau_T, \tau_{Y_1}, \tau_{N_1}, \ldots, \tau_{Y_n},
    \tau_{N_n} \in \Rnon$ such that $\sigma(\tau_T) \in T$,
  $\sigma(\tau_{Y_i}) \in Y_i$ and $\sigma(\tau_{N_i}) \in N_i$ for
  all $i \in [1..n]$. Let all of these be minimal.

  Let $\sigma' \defeq \FromTo{\sigma}{0}{\tau_T}$. Since $\pi$ is a
  schedule for $\sigma$, there exists a schedule $\pi'$ such that
  $\sigma' = \induced{\pi'}{\vec{0}}$. We will show that:
  \begin{align}
    \weight{\pi'}{\vec{y}_i}, \weight{\pi'}{\vec{n}_i} \in \{0, 1\}
    \text{ for all } i \in [1..n].\label{eq:all:bin}\tag{*}
  \end{align}
  From \eqref{eq:all:bin}, we can finish the proof. Indeed, by
  definition of $\pi'$ and of the modes, we
  have \[\sigma(\tau_T)(c^*) = \sum_{i=1}^n \weight{\pi'}{\vec{y}_i}
    \cdot s_i.\] Additionally, $\sigma(\tau_T)(c^*) = t$ holds by
  definition of zone $T$. Since each $\weight{\pi'}{\vec{y}_i} \in \{0,
    1\}$, we obtain a solution $V \defeq \{s_i \colon
    \weight{\pi'}{\vec{y}_i} = 1\}$ to the SUBSET-SUM instance.

  \newcommand{\ci}[1]{c_{i,#1}}
  \newcommand{\pc}[1]{\vec{x}(c_{i,#1})}

  \newcommand{\psiweight}[1]{\weight{\psi}{#1}}
  \newcommand{\psipweight}[1]{\weight{\psi'}{#1}}

  It remains to show~\eqref{eq:all:bin}. We first make the following
  claims for every $i \in [1..n]$:
  \begin{enumerate}[label=(\arabic*)]
    \item $\tau_{Y_i} \leq \tau_T$ and $\tau_{N_i} \leq
            \tau_T$.

    \item $\weight{\pi'}{\vec{y}_i} + \weight{\pi'}{\vec{n}_i} = 1$ and

    \item[] $\weight{\pi'}{\overline{\vec{y}}_i} +
            \weight{\pi'}{\overline{\vec{n}}_i} = 1$.
  \end{enumerate}
  Let us prove these two claims.
  \begin{enumerate}[label=(\arabic*)]
    \setlength\itemsep{5pt}

    \item We only show that $\tau_{Y_i} \leq \tau_T$, as $\tau_{N_i}
            \leq \tau_T$ is symmetric. We proceed by proving that for any
          $\vec{x} \in T \setminus Y_i$, it is the case that $\vec{x}
            \not\models_\M \F Y_i$. By definition of $T$ and $Y_i$, we have
          $\pc{1} \in [-0.5, 0.5)$, $\pc{2} = 2$ and $\pc{3} = \pc{4} =
            1$. Note that the only modes affecting $\{c_{i,1}, \ldots,
            c_{i,4}\}$ are $\{\vec{y}_i, \vec{n}_i, \overline{\vec{y}}_i,
            \overline{\vec{n}}_i\}$. The only modes affecting $c_{i,1}$
          positively, and that could thus lead it to $0.5$, are
          $\vec{y}_i$ and $\overline{\vec{n}}_i$. However, both affect
          $c_{i,2}$ positively, and no mode decreases $c_{i,2}$. Hence,
          using either mode from $\vec{x}$ can never lead to a point in
          $Y_i$. Thus, $\vec{x} \not\models_\M \F Y_i$.

    \item By $\sigma(\tau_T) \in T$, we have $\sigma(\tau_T)(\ci{2}) =
            2$. Since $\vec{y}_i(\ci{2}) = \vec{n}_i(\ci{2}) =
            \overline{\vec{y}}_i(\ci{2}) = \overline{\vec{n}}_i(\ci{2}) =
            1$, and since no mode decreases $\ci{2}$, it is the case that
          $\weight{\pi'}{\vec{y}_i} + \weight{\pi'}{\vec{n}_i} +
            \weight{\pi'}{\overline{\vec{y}}_i} +
            \weight{\pi'}{\overline{\vec{n}}_i} = 2$. Since $T$ requires
          $\ci{3} = 1$, and no mode decreases $\ci{3}$, we have
          $\weight{\pi'}{\vec{y}_i} + \weight{\pi'}{\vec{n}_i} =
            1$. Similarly, as $T$ requires $\ci{4} = 1$, we have
          $\weight{\pi'}{\overline{\vec{y}}_i} +
            \weight{\pi'}{\overline{\vec{n}}_i} = 1$.
  \end{enumerate}

  It remains to use the above claims to prove~\eqref{eq:all:bin}. Note
  that $\tau_{Y_i} \neq \tau_{N_i}$ as the constraints of $Y_i$ and
  $N_i$ conflict on $\ci{1}$. Let us assume that $\tau_{Y_i} <
    \tau_{N_i}$ (the other case is symmetric). Let $\psi$ and $\psi'$ be
  schedules such that $\induced{\psi}{\vec{0}} =
    \sigma[0..\tau_{Y_i}]$ and $\induced{\psi'}{\sigma(\tau_{Y_i})} =
    \sigma[\tau_{Y_i}..\tau_{N_i}]$.

  By definition of $Y_i$ and $N_i$, we have
  $\sigma(\tau_{Y_i})(\ci{3}) = 1$ and $\sigma(\tau_{N_i})(\ci{3}) =
    1$. Since no mode decreases $\ci{3}$, modes $\vec{y}_i$ and
  $\vec{n}_i$ are not used in $\psi'$. Further, note that
  $\sigma(\tau_{Y_i})(\ci{1}) = 0.5$ and $\sigma(\tau_{N_i})(\ci{1}) =
    -0.5$. Therefore, it must be the case that
  \[
    0.5 -
    \psipweight{\overline{\vec{y}}_i} +
    \psipweight{\overline{\vec{n}}_i} = -0.5.
  \]
  Thus, $\psipweight{\overline{\vec{y}}_i} -
    \psipweight{\overline{\vec{n}}_i} = 1$. As $\psi'$ arises from
  $\pi'$, Claim~2 yields
  \[
    \psipweight{\overline{\vec{y}}_i} +
    \psipweight{\overline{\vec{n}}_i} \leq 1.
  \]
  So, we have $\psipweight{\overline{\vec{y}}_i} = 1$ and
  $\psipweight{\overline{\vec{n}}_i} = 0$. From Claim~2, we further
  derive $\psiweight{\overline{\vec{y}}_i} =
    \psiweight{\overline{\vec{n}}_i} = 0$.

  By definition of $Y_i$, we have $\sigma(\tau_{Y_i})(\ci{1}) =
    0.5$. Since $\vec{y}_i$ and $\vec{n}_i$ are the only modes possibly
  used in $\psi$ to change $\ci{1}$, we have $\frac{1}{2} \cdot
    \psiweight{\vec{y}_i} - \frac{1}{2} \cdot \psiweight{\vec{n}_i} =
    0.5$. By Claim~(2),
  \[\psipweight{\vec{y}_i} + \psipweight{\vec{n}_i} \leq 1.\]
  So, we have $\psiweight{\vec{y}_i} = 1$ and $\psiweight{\vec{n}_i} =
    0$, which, by Claim~(2), yields $\weight{\pi'}{\vec{y}_i} = 1$ and
  $\weight{\pi'}{\vec{n}_i} = 0$.
\end{IEEEproof}

\section{Missing proofs of \Cref{sec:undecidable}}

In \Cref{sec:undecidable}, we implicitly assume that Petri nets with
inhibitor arcs have no transition that consumes from, and produces in,
the same place. We can make this assumption without loss of
generality. Roughly, it is possible to split a transition $t$ that
consumes and produces in the same place into two transitions
$t_{\text{pre}}$ and $t_{\text{post}}$, while being equivalent with
respect to reachability. We let $t_{\text{pre}}$ consume from the
place and realize the effect of $t$ on all other places, and let
$t_{\text{post}}$ produce in the place.  We can ensure that when
$t_{\text{pre}}$, then immediately afterwards $t_{\text{post}}$ is
fired by adding a new place $p_t$, adding an arc from $t_{\text{pre}}$
to $p_t$ and an arc from $p_t$ to $t_{\text{post}}$, and adding an
inhibitor arc from $p_t$ to all other transitions, thus preventing any
transition other than $t_{\text{post}}$ from being fired until
$t_{\text{post}}$ was fired to consume the token from $p_t$.

Now, let us prove the statements from \Cref{sec:undecidable}.

\lemMmsAbc*

\begin{IEEEproof}
  Let $\vec{x}_A, \vec{x}_A' \in A$ and $\pi$ be as described.

  Let $\pi = (\alpha, \vec{m}) \pi'$. By definition of $A$, we must
  have $\vec{m} = \vec{a}_t$ for some $t \in T$. Since $\vec{x}_A(t_A)
  = 1$, we have $\alpha \in (0, 1]$. If $\alpha < 1$, then zone $A_t'
  \setminus A$ is reached, and the only mode that can be used is
  $\vec{a}_t$. Thus, we can assume w.l.o.g.\ that $\alpha = 1$.
  Let $\vec{x}_B \defeq \vec{x}_A + \vec{a}_t$. We have
  \[
  \vec{x}_A \mreach{\vec{a}_t}{A_t'} \vec{x}_B \text{ and }
  \vec{x}_B \in B_t.
  \]

  Let $\pi' = (\alpha', \vec{m}') \pi''$. By definition of $B_t$, we
  have $\vec{m}' = \vec{b}_t$ or $\vec{m}' = \vec{a}_s$ for some
  $s \neq t$. Since no zone allows for $t_B > 0$ and $s_B > 0$, we
  must have $\vec{m}' = \vec{b}_t$. Since $\vec{x}_B(t_B) = 1$, we
  have $\alpha' \in (0, 1]$. If $\alpha' < 1$, then zone $B_t'
  \setminus B_t$ is reached, and, again, the only mode that can be
  used is $\vec{b}_t$. So, we can assume w.l.o.g.\ that $\alpha' = 1$.
  Let $\vec{x}_C \defeq \vec{x}_B + \vec{b}_t$. We have
  \[
  \vec{x}_B \mreach{\vec{b}_t}{B_t'} \vec{x}_C \text{ and }
  \vec{x}_C \in C_t.
  \]

  Let $\pi'' = (\alpha'', \vec{m}'') \pi'''$. By definition of $C_t$,
  we have $\vec{m}'' = \vec{c}_t$ or $\vec{m}'' = \vec{a}_s$ for some $s
  \neq t$. Since no zone allows for $t_C > 0$ and $s_B > 0$, we must
  have $\vec{m}'' = \vec{c}_t$. Since $\vec{x}_C(t_C) = 1$, we have
  $\alpha'' \in (0, 1]$. If $\alpha'' < 1$, then zone $C_t' \setminus
  C_t$ is reached, and, again, the only mode that can be used is
  $\vec{c}_t$. So, we can assume w.l.o.g.\ that $\alpha'' = 1$.
  Let $\vec{y} \defeq \vec{x}_C + \vec{c}_t$. We have
  \[
  \vec{x}_C \mreach{\vec{c}_t}{C_t'} \vec{y} \text{ and }
  \vec{y} \in A.
  \]

  Since execution $\induced{\pi}{\vec{x}_A}$ does not contain
  intermediate points in $A$, we have $\vec{y} = \vec{x}_A'$ and hence
  $\pi'''$ is empty. Consequently, $\pi \equiv \vec{a}_t \vec{b}_t
  \vec{c}_t$.
\end{IEEEproof}

\propPniMmsUnbounded*

\begin{IEEEproof}
  We show the proposition with $\M$ and $\AP$ described above. Let
  $\vec{x}$ be the point such that $\vec{x}(t_A) \defeq 1$ for all $t
  \in T$, $\vec{x}(\vec{p}) \defeq \xsrc$, and $\vec{x}(j) \defeq 0$
  for any other $j$. Let $\vec{x}'$ be defined in the same way, but
  with $\xtgt$ rather than $\xsrc$.

  \Cref{itm:non:zero:cycle} follows from the fact that zones from
  $\AP$ are non-negative, and each mode of $\M$ decreases some
  dimension. It remains to show that $\xsrc \trans{+} \xtgt$ in $\pn$
  iff $\vec{x} \mreach{+}{\AP} \vec{x}'$ in $\M$.

  $\Rightarrow$) Let $\xsrc \trans{\pi} \xtgt$ in $\pn$, where $\pi =
  t_1 \cdots t_k$. Let $\pi' \defeq \vec{a}_{t_1} \vec{b}_{t_1}
  \vec{c}_{t_1} \cdots \vec{a}_{t_k} \vec{b}_{t_k} \vec{c}_{t_k}$. We
  have $\vec{x} \mreach{\pi'}{\AP} \vec{x}'$ in $\M$.

  $\Leftarrow$) Let $\vec{x} \mreach{\pi}{\AP} \vec{x}'$ in $\M$ where
  $\pi$ is nonempty. As $\vec{x}, \vec{x}' \in A$, repeated
  applications of \Cref{lem:mms:ABC} yield $t_1, \ldots, t_k \in T$
  and points $\vec{y}_{A, 1}$, $\vec{y}_{B, 1}$, $\vec{y}_{C, 1}$,
  $\ldots$, $\vec{y}_{A, k}$, $\vec{y}_{B, k}$, $\vec{y}_{C, k}$,
  $\vec{y}_{A, k+1}$ such that $\pi \equiv \vec{a}_{t_1} \vec{b}_{t_1}
  \vec{c}_{t_1} \cdots \vec{a}_{t_k} \vec{b}_{t_k} \vec{c}_{t_k}$, and
  for all $i \in [1..k]$:
  \begin{enumerate}
  \item $\vec{y}_{A, i} \in A$, $\vec{y}_{B, i} \in B_{t_i}$ and
    $\vec{y}_{C, i} \in C_{t_i}$,

  \item $\vec{y}_{A, 1} = \vec{x}$ and $\vec{y}_{A, k+1} = \vec{x}'$, and

  \item $\vec{y}_{A, i} \mreach{\vec{a}_{t_i}}{A_{t_i}'} \vec{y}_{B,
    i} \mreach{\vec{b}_{t_i}}{B_{t_i}'} \vec{y}_{C, i}
    \mreach{\vec{c}_{t_i}}{C_{t_i}'} \vec{y}_{A, i+1}$ in
    $\M$.\label{itm:exec:abc}
  \end{enumerate}
  By definition of the modes and zones, \Cref{itm:exec:abc} yields
  \begin{align*}
    \vec{y}_{A, i}(\vec{p}) \trans{t_i} \vec{y}_{A, i+1}(\vec{p})
    \text{ in } \pn
    && \text{for all } i \in [1..k].
  \end{align*}
  Thus, $\xsrc = \vec{y}_{A, 1}(\vec{p}) \trans{t_1 \cdots t_k}
  \vec{y}_{A, k+1}(\vec{p}) = \xtgt$ in $\pn$.
\end{IEEEproof}

\lemOrU*

\begin{IEEEproof}
  The two following equivalences hold for LTL formulas interpreted
  over infinite words:
  \begin{enumerate}
  \item $(\psi \lor \psi') \U \varphi \equiv (\psi \U \psi') \U
    ((\psi' \U \psi) \U \varphi)$,\label{itm:U:1}

  \item $\varphi \U (\psi \lor \psi') \equiv (\varphi \U \psi) \lor
    (\varphi \U \psi')$.\label{itm:U:2}
  \end{enumerate}
  By \Cref{prop:has:trace,prop:ltl:real:discrete}, these equivalences
  also hold for negation-free LTL formulas interpreted over executions.
  
  We proceed by induction on $n$. If $n = 1$, then the claim is
  trivial. Assume $n \geq 2$. Let $\psi' \defeq \psi_2 \lor \cdots
  \lor \psi_n$. We have:
  \begin{alignat}{3}
    & && (\psi_1 \lor \psi_2 \lor \cdots \lor \psi_n) \U \varphi
    \notag \\
    &\equiv\ && (\psi_1 \lor \psi') \U \varphi
    \notag \\
    &\equiv\ &&
    (\psi_1 \U \psi') \U ((\psi' \U \psi_1) \U \varphi)
    \label{eq:by:lem:U:1} \\
    &\equiv\ &&
    (\psi_1 \U \psi') \U (\theta \U \varphi)\quad
    \label{eq:chi:by:ind} \\
    &\equiv\ &&
    [(\psi_1 \U \psi_2) \lor \cdots \lor (\psi_1 \U \psi_n)] \U
     (\theta \U \varphi)
    \label{eq:by:lem:U:2} \\
    &\equiv\ &&
    \theta',
    \label{eq:chip:by:ind}
  \end{alignat}
  where \eqref{eq:by:lem:U:1} and \eqref{eq:by:lem:U:2} follow from
  \Cref{itm:U:1,itm:U:2}, and where \eqref{eq:chi:by:ind} and
  \eqref{eq:chip:by:ind} yield $\theta, \theta' \in \text{\fragmentB{\U}}$
  by induction hypothesis.
\end{IEEEproof}

\thmUGoUndec*
\begin{IEEEproof}[Proof for \fragmentB{\G, \lor}]
  Let $\pn$ be a Petri net with inhibitor arcs and let $\xsrc,
  \xtgt$. Let $\M$, $\vec{x}$, $\vec{x}'$ and $\AP$ be given by
  \Cref{prop:pni:mms}. We modify $\M$ and $\AP$ as follows. We add
  dimension $\heartsuit$ to indicate that $\vec{x}'$ was reached and
  extend the execution to an infinite one.

  Each mode $\vec{m} \in \M$ is extended with $\vec{m}(\heartsuit)
  \defeq 0$. Each zone of $\AP$ is extended with the constraint
  $\heartsuit = 0$. We add modes $\{\vec{a}_\heartsuit,
  \overline{\vec{a}_\heartsuit}\}$ and zone $A_\heartsuit$ defined by:
  \[
  \begin{array}{lp{2pt}rrr}
    \toprule
    j           && \vec{a}_\heartsuit(j) & \overline{\vec{a}_\heartsuit}(j) \\
    \midrule
    \heartsuit  && 1 & -1 \\
    \text{else} && 0 &  0 \\
    \bottomrule
  \end{array}
  \hspace{1cm}
  \begin{array}{lp{2pt}lll}
    \toprule
                  && A_\heartsuit \\
    \midrule
    \heartsuit    && \in [0, 1]   \\
    \textbf{rest} && = \vec{x}'   \\
    \bottomrule
  \end{array}
  \]

  Let $\M'$ and $\AP'$ be the resulting MMS and set of zones. Let
  $\varphi \defeq \G(\bigvee_{Z \in \AP'} Z)$. Let $\vec{y} \defeq (0,
  \vec{x})$. By \Cref{prop:pni:mms}, it suffices to show that $\vec{x}
  \mreach{*}{\AP} \vec{x}'$ in $\M$ iff $\vec{y} \models_{\M'}
  \varphi$.

  $\Rightarrow$) Let $\pi$ be a finite schedule such that $\vec{x}
  \mreach{\pi}{\AP} \vec{x}'$ in $\M$. Let $\pi' \defeq \pi\,
  \vec{a}_\heartsuit\, \overline{\vec{a}}_\heartsuit\,
  \vec{a}_\heartsuit\, \overline{\vec{a}}_\heartsuit\, \cdots$. We
  have $\vec{y} \mreach{\pi'}{\AP'}$ in $\M'$, and hence
  $\induced{\pi'}{\vec{y}} \models_{\M'} \varphi$, since $\pi'$ is
  non-Zeno.

  $\Leftarrow$) Let $\pi$ be an infinite non-Zeno schedule of $\M'$
  such that $\induced{\pi}{\vec{y}} \models \varphi$. Let $\sigma
  \defeq \induced{\pi}{\vec{y}}$. By \Cref{prop:pni:mms}, we have
  $\vec{y} \not\mreach{\pi}{\AP' \setminus \{A_\heartsuit\}}$.

  So, there exists $\tau \in \Rnon$ such that $\sigma(\tau) \in
  A_\heartsuit$. By definition of $\M'$, there is a minimal $\tau'
  \leq \tau$ with $\sigma(\tau') = (0, \vec{x}')$. Let $\pi'$ be a
  finite schedule such that $\induced{\pi'}{\vec{y}} =
  \FromTo{\sigma}{0}{\tau'}$. We have $\vec{y} \mreach{\pi'}{\AP'
    \setminus \{A_\heartsuit\}} (0, \vec{x}')$ in $\M'$. Hence,
  $\vec{x} \mreach{*}{\AP} \vec{x}'$ in $\M$.
\end{IEEEproof}

\end{document}